\documentclass[11pt,english]{article}
\usepackage{lmodern}

\usepackage[T1]{fontenc}
\usepackage[latin9]{inputenc}
\usepackage{geometry}
\usepackage{babel}
\usepackage{array}
\usepackage{verbatim}
\usepackage{units}
\usepackage{multirow}
\usepackage{amstext}
\usepackage{amssymb}
\usepackage{graphicx}
\usepackage{wasysym}
\usepackage[numbers,sort&compress]{natbib}
\usepackage[unicode=true,pdfusetitle,
 bookmarks=true,bookmarksnumbered=false,bookmarksopen=false,
 breaklinks=true,pdfborder={0 0 1},backref=false,colorlinks=false]
 {hyperref}

\makeatletter

\newcommand{\noun}[1]{\textsc{#1}}
\newcommand{\binom}[2]{{#1 \choose #2}}

\providecommand{\tabularnewline}{\\}
\newcommand{\lyxdot}{.}

\@ifundefined{date}{}{\date{}}
\@ifundefined{showcaptionsetup}{}{%
 \PassOptionsToPackage{caption=false}{subfig}}
\usepackage{subfig}
\makeatother

\begin{document}
\title{PAIReD jet: A multi-pronged resonance tagging strategy across all
Lorentz boosts}
\author{Spandan Mondal\thanks{Brown University, Providence, USA}\and Gaetano
Barone\footnotemark[1]{\addtocounter{footnote}{+2}}\and Alexander
Schmidt\thanks{RWTH Aachen University, Aachen, Germany}}
\maketitle
\begin{abstract}
We propose a new approach of jet-based event reconstruction that aims
to optimally exploit correlations between the products of a hadronic
multi-pronged decay across all Lorentz boost regimes. The new approach
utilizes clustered small-radius jets as seeds to define unconventional
jets, referred to as PAIReD jets. The constituents of these jets are
subsequently used as inputs to machine learning-based algorithms to
identify the flavor content of the jet. We demonstrate that this approach
achieves higher efficiencies in the reconstruction of signal events
containing heavy-flavor jets compared to other event reconstruction
strategies at all Lorentz boost regimes. Classifiers trained on PAIReD
jets also have significantly better background rejections compared
to those based on traditional event reconstruction approaches using
small-radius jets at low Lorentz boost regimes. The combined effect
of a higher signal reconstruction efficiency and better classification
performance results in a two to four times stronger rejection of light-flavor
jets compared to conventional strategies at low Lorentz-boosts, and
rejection rates similar to classifiers based on large-radius multi-pronged
jets at high Lorentz-boost regimes.
\end{abstract}
\thispagestyle{empty}

\newpage{}

\setcounter{page}{1}  

\section{Introduction\label{sec:Introduction}}

Jets are ubiquitous in collider physics experiments. A jet may originate
from a single quark/gluon generated in the hard scattering, or from
the hadronic decay of a heavy Lorentz-boosted object, such as a W,
Z, or Higgs (H) boson. In the former case, the jet is expected to
have a single-pronged structure\footnote{The top quark is an exception. It decays into a bottom quark and a
W boson before hadronizing, resulting in top quark jets having a multi-pronged
structure.}, while in the latter, it is expected to have a multi-pronged structure.
While looking at a jet as a single object does not let us distinguish
between these categories in general, investigation into jet substructures
\citep{Thaler_2011,Adams:2015hiv,Kogler:2018hem} provides additional
handles to predict the nature of the particle that resulted in the
jet. More recent developments use machine learning architectures to
achieve unprecedented accuracies in jet identification tasks \citep{Guest:2016iqz,Kogler:2018hem,Guest:2018yhq,LARKOSKI20201}.

The choice of one or more jet reconstruction algorithms in analysis
of collision data depends on several factors, such as the composition
of the expected final state and the Lorentz-boosts of physics objects.
For example, a search for the Higgs boson decaying to heavy-flavor
(bottom or charm) quarks may be divided into two distinct reconstruction-strategy
regimes \citep{Sirunyan:2020aa,VHcc}: (i) a resolved-jet regime,
when the Higgs boson has a low Lorentz boost (transverse momentum,
$p_{\text{T}}\apprle$ 200 GeV) and the two quarks arising from the
Higgs boson have a large angular separation (opening angle), giving
rise to two separate small-radius, single-pronged jets, and (ii) a
merged-jet or boosted regime, when the Higgs boson has a high Lorentz
boost ($p_{\text{T}}\apprge$ 200 GeV), and the decay products of
the Higgs boson can be reconstructed with either two small-radius
jets or one large-radius, multi-pronged jet, owing to a smaller angular
separation. Heavy-flavor decays of the Higgs boson can then be selected
using jet flavor identification algorithms. In case of single-pronged
small-radius jets, jet identification usually involves identifying
the flavor of the quark or gluon that initiated the jet (flavor tagging)
\citep{Sirunyan:2017ezt}. Hence, flavor-tagging classifiers that
discriminate between bottom (b), charm (c) and light-flavor (up, down,
strange, gluon) jets are employed. On the other hand, for multi-pronged
jets, either flavor identification or heavy object identification
algorithms are employed \citep{CMS:2020poo}. In the discussed example,
one might use a multiclassifier that discriminates between H(H$\rightarrow$$\text{b\ensuremath{\bar{\text{b}}}}$),
H(H$\rightarrow$$\text{c}\bar{\text{c}}$), and light-flavor jets.

The choice of using one large-radius jet instead of two small-radius
jets in the boosted regime is motivated by several factors. When a
heavy object identification or jet flavor identification algorithm
is used to identify a large-radius jet, it can exploit not only the
information of the two (or more) subjets contained in the jet, but
also the correlations that exist between the kinematics of the hadronization
products of the two (or more) quarks. In addition, the large-radius
jet potentially encompasses soft radiations that may lie spatially
between the subjets. The presence and kinematics of these particles
may provide additional handles to discriminate resonant subjet pairs
against non-resonant ones, as the former are expected to exhibit electrical
charge and Quantum Chromodynamics (QCD) color recombination. Small-radius
jets can also be used to reconstruct events in the boosted regime,
but each jet must be flavor-tagged individually. This means that the
flavor identification algorithm has access to the substructures of
the small-radius jets individually and cannot take advantage of correlations
between the constituents of the two jets, which inevitably results
in loss of information. On the other hand, the large-radius jet approach
is not suitable at low Lorentz-boosts of the Higgs boson as the decay
products of the two quarks are expected to have a large angular separation.
The event is therefore reconstructed with two small-radius jets. It
can be noted that the resolved regime usually caters to a large fraction
of potentially interesting physics processes (e.g. nearly 95\% of
the signal in Ref. \citep{VHcc}) and hence the small-radius jet approach
has the advantage of probing a large part of the signal phase space.

We refer to the choice of using either two small-radius (e.g. AK4)
jets \citep{Aad:2022aa}, or one large-radius (e.g. AK15) jet \citep{VHcc}
per event as a choice of the event reconstruction strategy. This is
different from choosing a jet tagging algorithm (e.g. ParticleNet
\citep{Qu:2019gqs}, Particle Transformer \citep{Qu:2022mxj}, ParticleTransformerAK4
\citep{CMS-DP-2022-050} used in the CMS experiment, GN1 \citep{ATL-PHYS-PUB-2022-027}
used in the ATLAS experiment, etc.) to identify the jet content. In
this paper, we examine several reconstruction strategies and evaluate
their performance relative to a fixed tagging algorithm. We propose
a new event reconstruction strategy called \textbf{P}article \textbf{A}ngular-separation
\textbf{I}ndependent \textbf{Re}sonant \textbf{D}i-jet (PAIReD jet)
tagging, with the goal of optimally exploiting the available information
in pairs of small-radius jets to identify resonant decays of heavy
particles irrespective of the angular separation of the two decay
products. This allows modern flavor-tagging algorithms to exploit
as much information as large-radius jet tagging algorithms typically
have access to, without being restricted to a narrow range of the
parent particle momentum.

Custom jets derived from reconstructed standard jets have been studied
and utilized in experiments previously, for example, in the form of
\emph{megajets} in supersymmetry searches \citep{CMS:2011xie,CMS:2014rcs}
and reclustering large-radius jets by using properties of small-radius
jets \citep{ATLAS:2013qzt,Nachman:2014kla,ATLAS-CONF-2017-062}. Methods
to improve measurements involving dijet resonances using novel observables
have also been proposed \citep{Izaguirre:2014ira}. 

Section \ref{sec:The-PAIReD-jet} outlines the PAIReD jet tagging
strategy. Section \ref{sec:Simulation} describes the simulation of
several physics processes which are used to compare the performances
of different event reconstruction strategies. Sections \ref{sec:Flavor-tagging-algorithm}
and \ref{sec:Neural-network-training} describe the architecture and
implementation of machine learning algorithms that are used to train
classifiers to identify heavy-flavor Higgs boson decays. A mock ZH(H$\rightarrow\text{c}\bar{\text{c}}$)
analysis is outlined in Sec. \ref{sec:Constructing-a-mock} as an
example and the reconstruction efficiencies and classifier performances
in the context of reconstructing H$\rightarrow\text{c}\bar{\text{c}}$
decays are compared across various event reconstruction strategies.
Section \ref{sec:Outlook} provides an outlook and Sec. \ref{sec:Conclusion}
provides a summary and conclusion of the topics discussed in the paper.

\section{The PAIReD jet strategy\label{sec:The-PAIReD-jet}}

A PAIReD jet is defined by the union of all the reconstructed particles
that lie within a radius of $\Delta R=R_{0}$ around the centers of
two clustered small-radius jets present in the event, where $R_{0}$
is the distance parameter used to cluster the small-radius jets and
$\Delta R=\sqrt{\Delta\eta^{2}+\Delta\phi^{2}}$ is the angular distance
calculated in pseudorapidity--azimuth ($\eta$--$\phi$) coordinates,
with $\phi$ being in radians. Thus, for an event with $n$ ($n\ge2$)
clustered AK4 jets\footnote{For the sake of simplicity, we use ``AK4 jets'' and ``small-radius
jets'' interchangeably henceforth. The PAIReD jet strategy can nevertheless
be extended to jets clustered with any clustering algorithm, with
distance parameters suitable for single-pronged decays. }, one can define $\binom{n}{2}$ PAIReD jets in the event, where $\binom{n}{2}$
denotes the number of ways one can choose 2 objects from a set of
$n$ objects.

This is practically implemented by calculating distances ($\Delta R_{j_{1}}$,
$\Delta R_{j_{2}}$) of each reconstructed particle in the event from
the centers of the two small-radius jets, $j_{1}$ and $j_{2}$; the
particles satisfying $\min(\Delta R_{j_{1}},\Delta R_{j_{2}})<R_{0}$
constitute the list of constituents of the PAIReD jet. As some tagging
algorithms may require ordered sets of particles, a reordering of
the list of constituents, e.g. in descending order of particle $p_{\text{T}}$,
may be necessary. It can be noted that this is distinct from simply
concatenating the lists of constituents clustered within the two AK4
jets. As shown in Table \ref{tab:Mean-percentage} and discussed in
the following paragraphs, this definition allows the PAIReD jet to
reconstruct a larger fraction of the particles arising from the Higgs
boson decay. On the other hand, the advantages that commonly used
sequential clustering algorithms provide, such as infrared and collinear
(IRC) safety \citep{Banfi:2004yd}, are not realized in this approach.
A more detailed discussion about the effect of using sequential clustering
algorithms has been presented in Appendix \ref{sec:Effects-of-(not)}.
An alternative variant of the PAIReD approach, called the PAIReDClustered
approach, that uses only the particles clustered in the two constituent
AK4 jets, is discussed in Appendix \ref{subsec:PAIReD-jets-with}.
The PAIReDClustered approach achieves a marginally lower performance
than the baseline PAIReD approach.

Since a PAIReD jet is composed of the hadronization products of both
quarks, a tagging algorithm employed to identify the hadronic decay
of a heavy particle has access to correlations between any pair of
particles arising from the hadronization, even if two particles are
reconstructed in two different AK4 jets. Furthermore, the fact that
two jets arising from the same hadronic decay inherently possess distinct
properties, such as charge and color, is preserved when analyzing
the entire PAIReD jet collectively; this information is lost when
each AK4 jet is processed separately. However, unlike large-radius
jets, the PAIReD jet still lacks information of the reconstructed
particles that lie spatially between the two AK4 jets. This leads
to the introduction of the second variant of the PAIReD jet strategy,
which we refer to as the PAIReDEllipse strategy.

\begin{figure}
\begin{centering}
\includegraphics[width=0.5\textwidth]{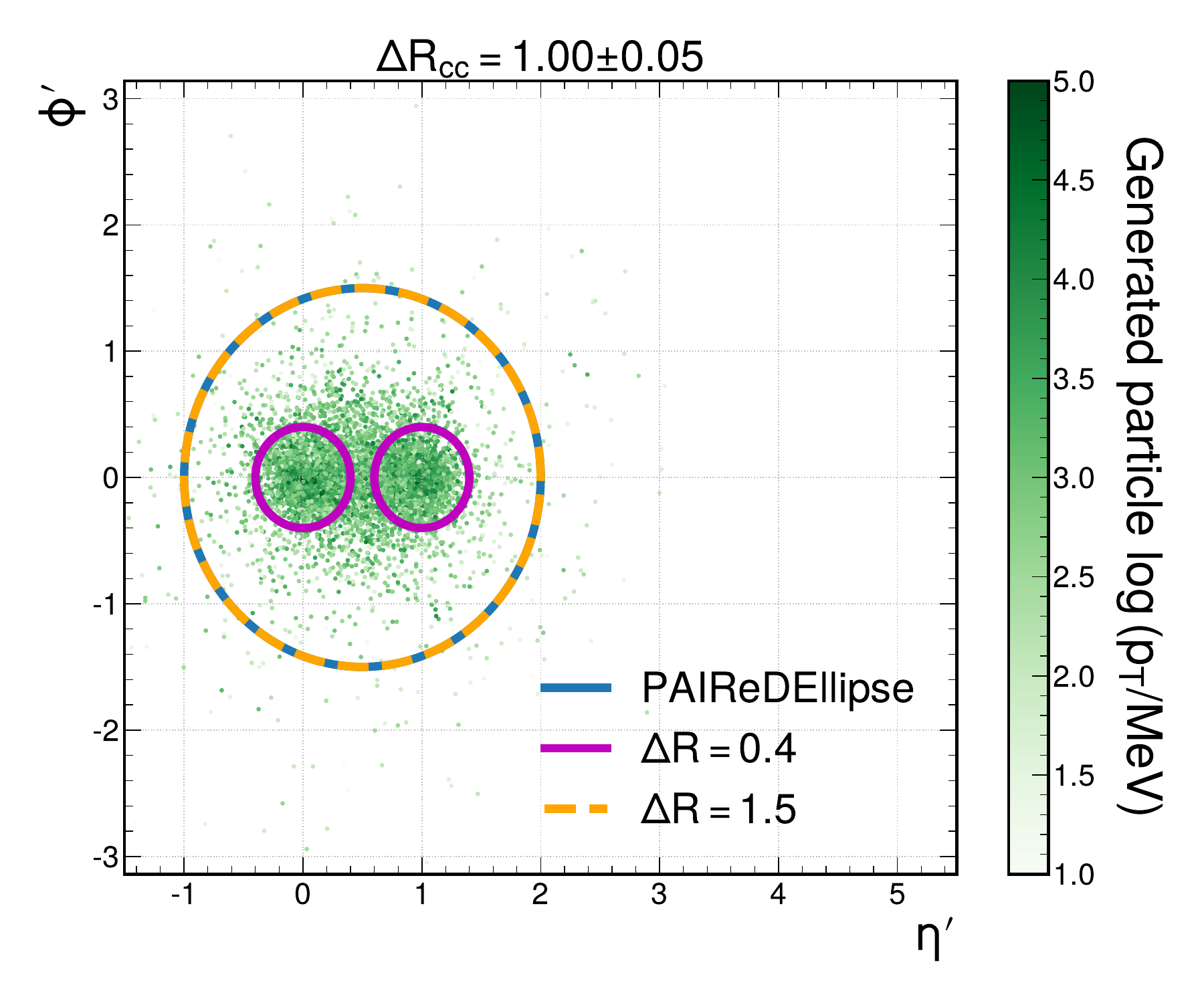}\includegraphics[width=0.5\textwidth]{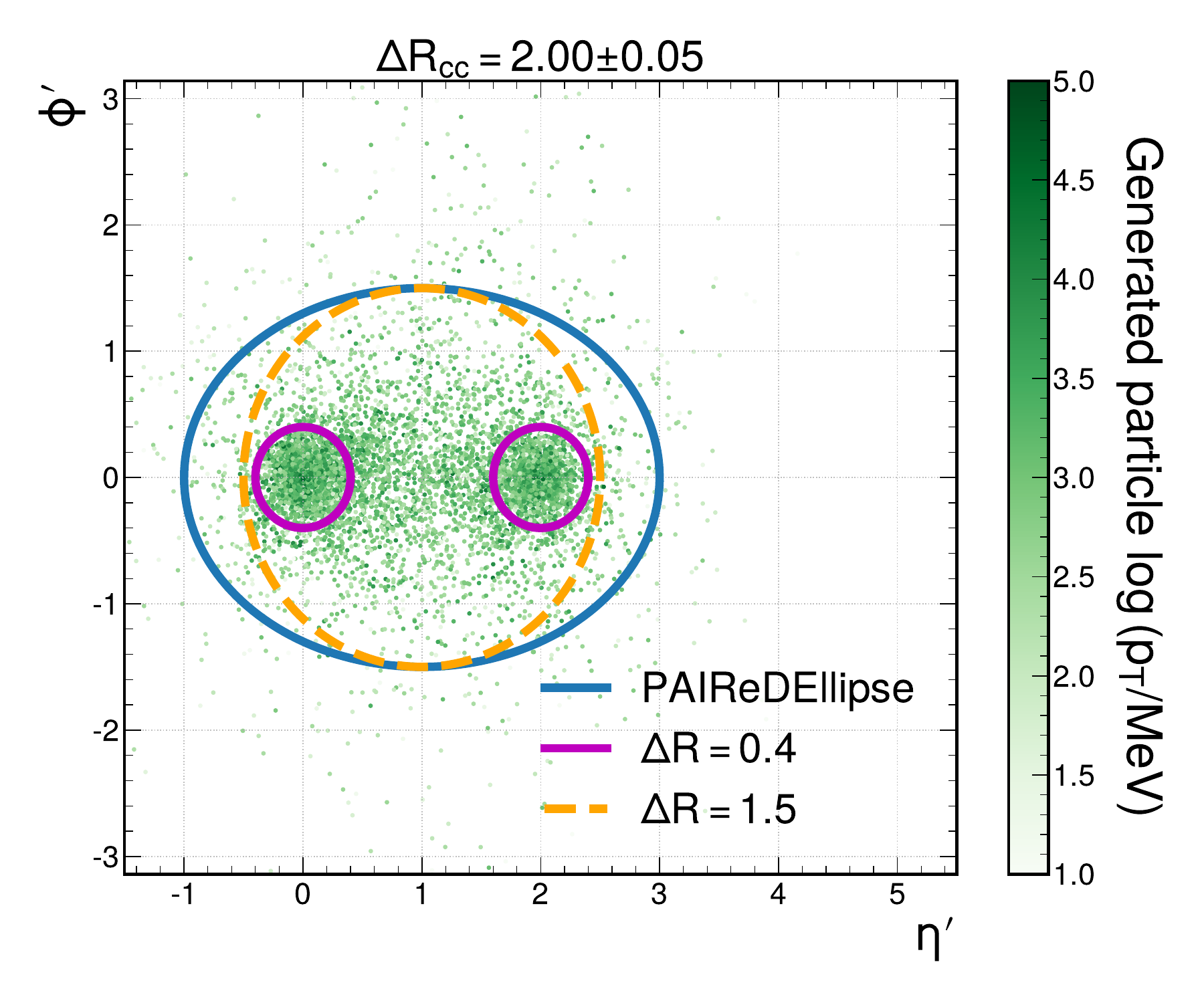}
\par\end{centering}
\begin{centering}
\includegraphics[width=0.5\textwidth]{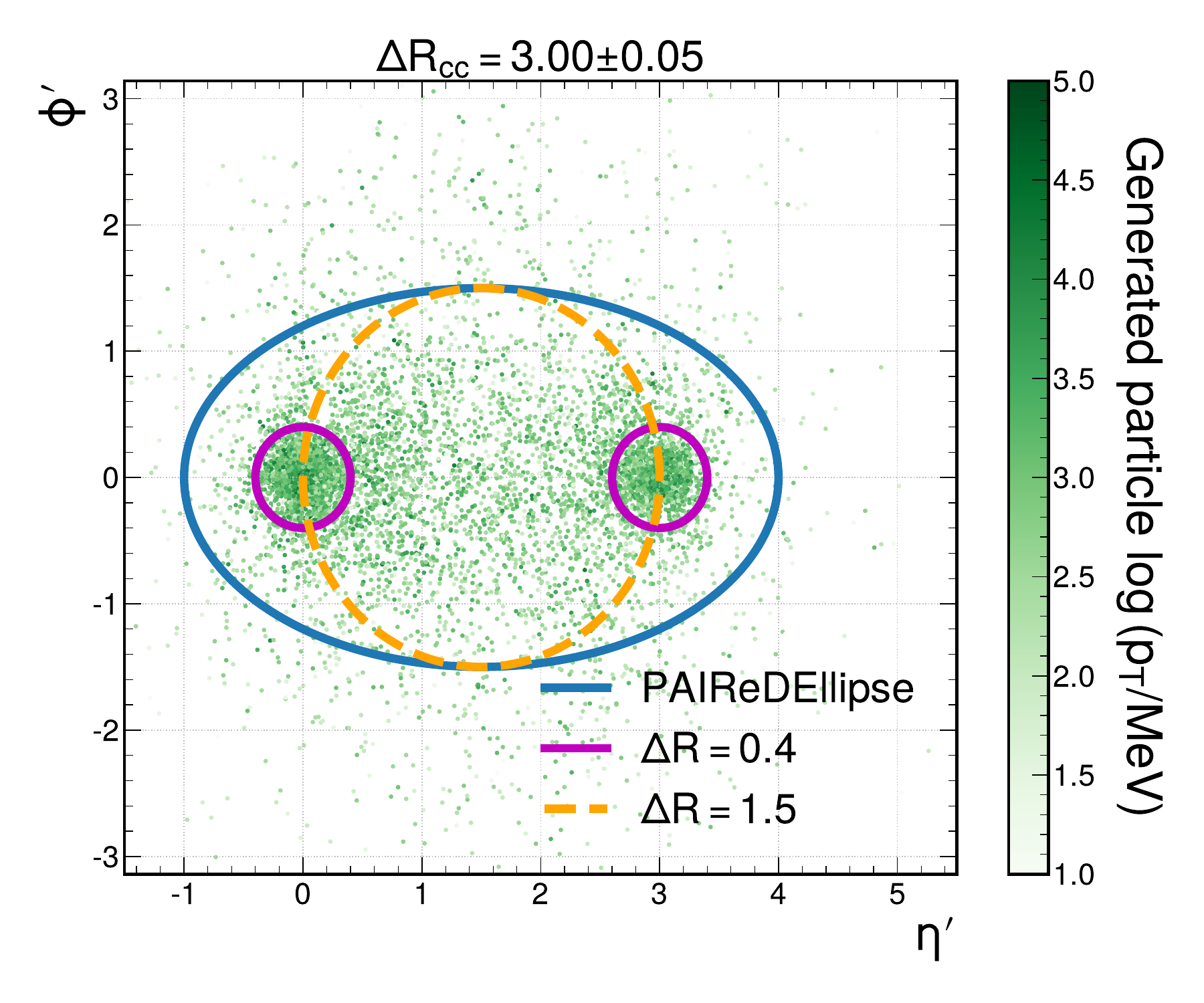}\includegraphics[width=0.5\textwidth]{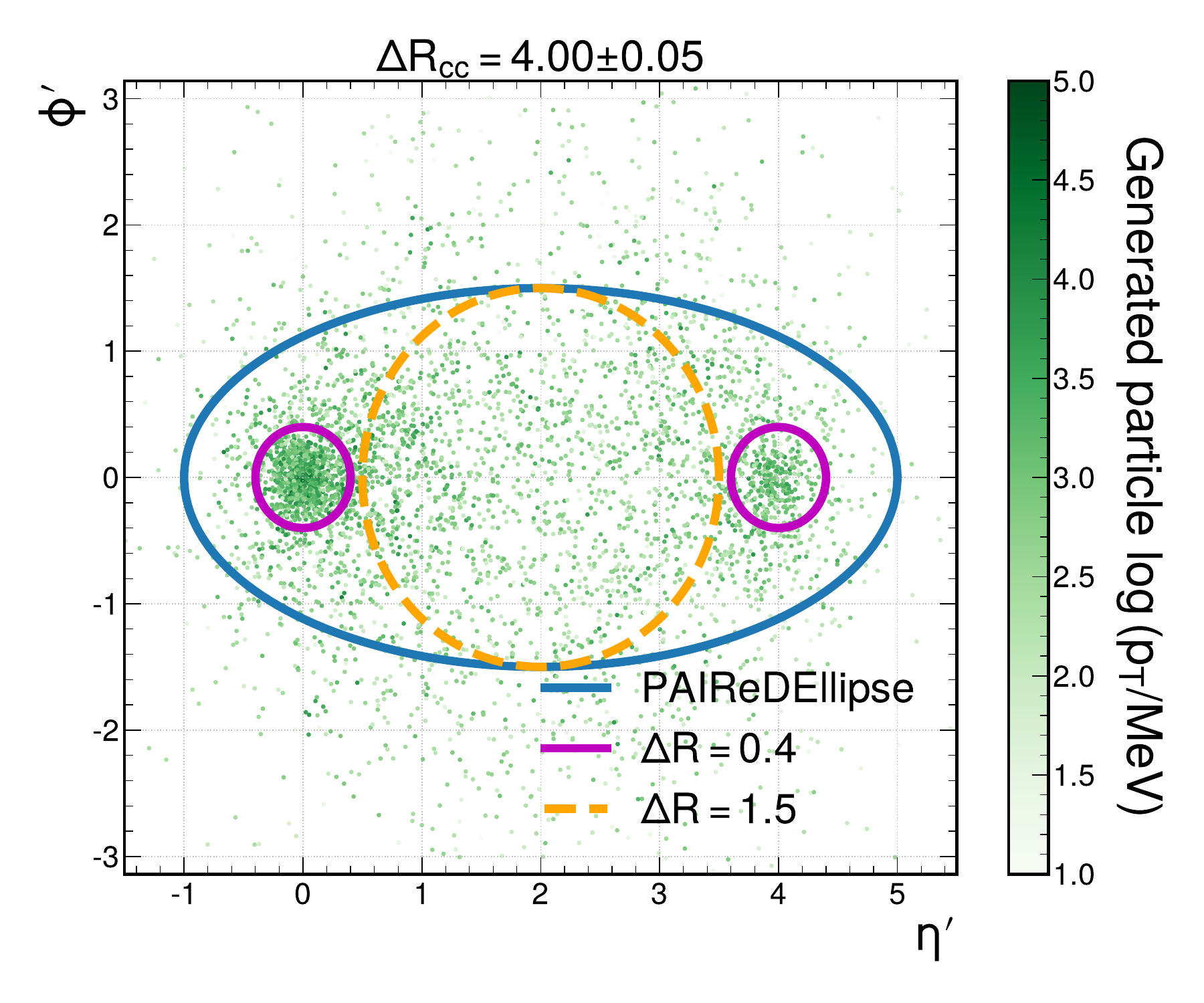}\caption{\label{fig:Maps}Simulated ZH(H$\rightarrow\text{c}\bar{\text{c}}$)
events with the $\eta$--$\phi$ coordinate system translated and
rotated in a way such that the c quark with the higher $p_{\text{T}}$
is located at $(\eta\prime,\phi\prime)=(0,0)$, while the second c
quark lies on the $\eta\prime$ axis ($\phi\prime=0$). Each subfigure
shows generated particles in 200 events, allowing particles across
events to overlap. The shade of green denoting each particle is chosen
according to the (logarithm of the) particle $p_{\text{T}}$, shown
on the colorbar to the right of each plot. The upper-left, upper-right,
lower-left, and lower-right subfigures show events with the angular
separation $\Delta R_{\text{cc}}$ between the two c quarks lying
in a window of 0.05 unit around 1, 2, 3, and 4, respectively. Each
of the magenta circles represents the area of a potential AK4 jet
that might be clustered centering one of the quarks. The two magenta
circles together represent the net area covered by a jet defined using
the PAIReD jet strategy. The blue ellipse represents the area of a
jet defined using the PAIReDEllipse strategy. The dashed orange line
represents the typical area a hypothetical AK15 jet would cover, had
it been constructed at the center of the 2-quark system (see discussion
in text).}
\par\end{centering}
\end{figure}

The PAIReDEllipse strategy defines a non-conventional jet using two
clustered AK4 jets as ``seeds''. The area of the PAIReDEllipse jet
is defined by an elliptical region in the $\eta$--$\phi$ plane.
The ellipse is defined such that the centers of the two seed AK4 jets
lie on the major axis of the ellipse. The length $b$ of the minor
axis of the ellipse is fixed to 3 units in the $\eta$--$\phi$ plane,
motivated by the largest jet radius ($\Delta R=1.5$) used in recent
physics analyses \citep{VHcc}. With the distance between the centers
of the two AK4 jets denoted by $\Delta R_{jj}$, the length of the
major axis is defined as
\[
a=\max(3,\Delta R_{jj}+2)
\]
units in the $\eta$--$\phi$ plane, so that the ellipse includes
particles lying up to 1 unit away from the centers of each small-radius
jet in the direction away from the center of the ellipse. As shown
in Fig. \ref{fig:Maps} and discussed in the following paragraphs,
this choice covers a large fraction of particles generated from the
decay of a Higgs boson. We have tested that increasing these numbers
does not significantly alter the final performance metrics of the
PAIReDEllipse strategy.

All the particles reconstructed within the perimeter of the ellipse
thus defined are considered constituents of the PAIReDEllipse jet.
Thus, for a pair of AK4 jets, one can uniquely define one PAIReDEllipse
jet. It can be noted that the definition of the PAIReDEllipse jet
does not involve any sequential clustering algorithm, except for the
ones necessary to define the centers of the seed AK4 jets, similar
to the PAIReD jet scenario.

In presence of additional jets arising from initial- and final-state
radiations, standard resolved-jet approaches typically involve sorting
all AK4 jets in the event by certain predetermined parameters ($p_{\text{T}}$,
b/c tagger score) to select the pair of jets that are most likely
to arise from the decay of a heavy particle. Analogously in the PAIReD(Ellipse)
jet strategy, the PAIReD(Ellipse) jet with the highest ``bb/cc-jet''
score can be treated as the decay of interest.

Figure \ref{fig:Maps} demonstrates the effective areas in the $\eta$--$\phi$
plane covered by the different jet construction strategies, using
simulated ZH(H$\rightarrow\text{c}\bar{\text{c}}$) events. The details
of the simulation are presented in Sec. \ref{sec:Simulation}. It
can be seen from the figure that the AK15 reconstruction strategy
covers nearly the entire Higgs boson decay when the angular separation
$\Delta R_{\text{cc}}$ between the two generated c quarks from the
Higgs boson decay is \textasciitilde 2 units or smaller. At higher
separations, a potential AK15 jet would be reconstructed around only
one of the quarks (or two separate AK15 jets around two quarks). On
the other hand, the AK4 reconstruction strategy can be implemented
at all angular separation regimes. However, it can be seen from Fig.
\ref{fig:Maps} that a significant fraction of the generated particles
arising from the Higgs boson decays lie outside the effective combined
area covered by the two AK4 jets. The PAIReD approach, which simply
concatenates the list of particles lying in a 0.4 radius around the
centers of two AK4 jets, would analogously fail to cover a significant
fraction of the generated particles. Finally, the PAIReDEllipse approach
covers a larger area and contains most of the generated particles
arising from the Higgs boson decay at all angular separations.

Table \ref{tab:Mean-percentage} shows the mean percentage of generated
particles arising from the decay of the Higgs boson that are clustered/contained
within the jet(s) for each reconstruction scenario at different values
of $\Delta R_{\text{cc}}$. The numbers reported are obtained after
pileup simulation and pileup subtraction (details in Sec. \ref{sec:Simulation}).
As expected, the fraction of particles clustered within single AK8
and AK15 jets decrease with increasing angular separation. In all
scenarios, using the PAIReD approach reconstructs a larger fraction
of the Higgs boson decay than using 2 AK4 jets. This motivates the
use of all particles lying within a characteristic angular distance
from the AK4 jet centers over using particles clustered within the
AK4 jets. Finally, the PAIReDEllipse approach reconstructs the maximum
fraction of Higgs boson decay products in all scenarios.

\begin{table}
\begin{centering}
\renewcommand{\arraystretch}{1.5}
\begin{tabular}{c|>{\centering}p{0.12\textwidth}>{\centering}p{0.12\textwidth}>{\centering}p{0.12\textwidth}>{\centering}p{0.12\textwidth}>{\centering}p{0.12\textwidth}}
\multirow{2}{*}{$\Delta R_{\text{cc}}$} & \multicolumn{5}{c}{Efficiency of clustering particles from Higgs boson decay (in \%)}\tabularnewline
\cline{2-6} \cline{3-6} \cline{4-6} \cline{5-6} \cline{6-6} 
 & AK4 & AK8 & AK15 & PAIReD & PAIReD\-Ellipse\tabularnewline
\hline 
$1.00\pm0.05$ & 51.7 & 35.8 & 54.9 & 88.4 & 98.7\tabularnewline
$2.00\pm0.05$ & 45.3 & 21.1 & 34.2 & 71.4 & 90.9\tabularnewline
$3.00\pm0.05$ & 44.6 & 18.7 & 31.6 & 65.7 & 84.3\tabularnewline
$4.00\pm0.05$ & 34.2 & 12.9 & 26.3 & 39.5 & 62.1\tabularnewline
\hline 
\end{tabular}
\par\end{centering}
\caption{\label{tab:Mean-percentage}Mean percentage of generated particles
arising from the decay of the Higgs boson in ZH(H$\rightarrow\text{c}\bar{\text{c}}$)
events clustered in each reconstruction scenario at various values
of $\Delta R_{\text{cc}}$. Only generated jets with $|\eta|$ < 2.5
are considered. In case of AK4, PAIReD, and PAIReDEllipse scenarios,
events with at least two AK4 generated jets each with $p_{\text{T}}$
> 20 GeV are taken into account. In case of the AK8 (AK15) strategy,
events with at least 1 AK8 (AK15) jet with $p_{\text{T}}$ > 100 (50)
GeV are considered. In all cases, the percentage corresponding to
the jet(s) reconstructing the highest fraction of Higgs decay products
in each event is reported. In case of AK4, AK8, and AK15 approaches,
the list of constituents clustered within generated jets are considered,
while in case of PAIReD and PAIReDEllipse approaches, the distances
of each generated particle from the center of the two constituent
AK4 jets are considered. }

\end{table}

Even though the higher area covered by the PAIReDEllipse jet strategy
allows capturing the entire Higgs boson decay regardless of the angular
separation of the decay products, it comes at the cost of capturing
additional unwanted particles, such as those arising from pileup interactions,
underlying event (UE), and other unrelated hadronic activities. However,
with the advent of high-level, attention-based \citep{NIPS2017_3f5ee243}
networks used for jet identification, it is expected that such spurious
particles would be identified and assigned less importance while performing
classification tasks downstream. Additionally, the inclusion of IRC-unsafe
observables, unclustered particles, and soft radiations as inputs
to the tagging algorithms potentially introduces a larger dependence
of the tagger performance on the modeling of the parton shower (PS)
and hadronization, modeling of the UE, and simulation-specific artifacts
that do not model the collision data correctly. While a systematic
study with different generators and PS models and their effects on
modern ML-based tagging algorithms is beyond the direct scope of this
study, a discussion about the effect of clustering algorithms and
an ablated variant excluding unclustered particles are presented in
Appendix \ref{sec:Effects-of-(not)}. The increased model dependency
of these taggers also sheds light on the importance of calibrating
the outputs of the tagger using collision data, to mitigate these
dependencies as well as potential mismodelings in simulation. 

In the case of PAIReD(Ellipse) jets, the momentum and other kinematic
features of the jet may be calibrated in two ways. Since experiments
routinely derive jet energy scale and resolution corrections for small-radius
jets \citep{Khachatryan:2016kdb,ATLAS:2020cli} for use in a broad
range of physics analyses, one may repurpose the corrected kinematics
of the constituent small-radius jets to define the effective momentum
and invariant mass of the PAIReD(Ellipse) jets. Thus, adding the corrected
four-momenta of two constituent AK4 jets to define the effective four-momentum
of the PAIReD(Ellipse) jet may be considered as the first, simple
approach towards calibrating energy scales of PAIReD(Ellipse) jets.
However, a more precise mass (see Appendix \ref{sec:Mass-regression})
and momentum\footnote{Whether regressing the momentum of PAIReD(Ellipse) jets provides a
substantial advantage over using the sum of the four-momenta of two
constituent AK4 jets has not been studied in this paper.} estimate of the initiating particle may be achieved by using mass
and momentum regression techniques using all constituents of the PAIReD(Ellipse)
jets as inputs. In order to calibrate the regressed (predicted) momentum,
one may use event samples enriched in leptonically-decaying Z bosons
recoiling against two small-radius jets (or equivalently, one PAIReD(Ellipse)
jet), a standard method used for calibrating energy scales of individual
small-radius jets \citep{Khachatryan:2016kdb,ATLAS:2020cli}. Since
the momentum of a leptonic Z boson can be reconstructed with high
precision, a system of Z boson recoiling against a PAIReD(Ellipse)
jet lets one measure the momentum of the jet with high precision and
hence derive corrections to the regressed momentum. The regressed
mass may be calibrated using standard techniques to calibrate masses
of large-radius jets \citep{VHcc}. This includes, for instance, reconstructing
hadronically-decaying W bosons with PAIReD(Ellipse) jets in events
enriched in semileptonic top quark-antiquark pair production ($\text{t}\bar{\text{t}}$),
and comparing the mass resolution in simulation and data.

The flavor tagging efficiency of PAIReD(Ellipse) jets can similarly
be compared in simulation and data using standard calibration methods
used for large-radius jets. These methods involve using the kinematics
of soft-muons or secondary vertices contained inside the jets to select
a subset of QCD multijet events in data that mimic $\text{b\ensuremath{\bar{\text{b}}}}$
or $\text{c}\bar{\text{c}}$ jets in terms of tagger score distributions
\citep{Sirunyan:2017ezt}. More recent methods \citep{CMS-DP-2022-005,CMS-PAS-BTV-22-001}
make use of Boosted Decision Trees (BDTs) \citep{FREUND1997119,Breiman:1984aa}
to achieve the same goal for highly performant taggers. Similar methods
may be used to isolate a subset of QCD multijet events in data that
mimic $\text{b\ensuremath{\bar{\text{b}}}}$ or $\text{c}\bar{\text{c}}$
PAIReD(Ellipse) jets in terms of tagger score distributions, followed
by a fit to data to extract correction factors at different efficiencies.
Calibration is a challenge that experiments must overcome before PAIReD(Ellipse)
jets can be used in physics analyses. Detailed demonstration of specific
calibration methods is beyond the scope of this paper.

\section{Simulation\label{sec:Simulation}}

We simulate proton--proton collision events to train, validate, and
compare the performance metrics across various event reconstruction
strategies. We focus on the 2-lepton channel of H$\rightarrow\text{b\ensuremath{\bar{\text{b}}}}/\text{c}\bar{\text{c}}$
searches, where the Higgs boson is produced in association with a
vector boson (Z/W) \citep{ATLAS:2018kot,CMS:2018nsn,Aad:2022aa,VHcc}
and the Z boson decays to a pair of electrons or muons. We refer to
the ZH(H$\rightarrow\text{b\ensuremath{\bar{\text{b}}}}/\text{c}\bar{\text{c}}$)
process as the signal process. The major background of these searches
are Z+jj events, where the Z boson decays to a pair of electrons or
muons and is generated in association with two additional jets. Additionally,
the semileptonic $\text{t}\bar{\text{t}}$ process forms a major background
specifically in the W-associated production mode of the Higgs boson.
Only Z+jj and $\text{t}\bar{\text{t}}$ backgrounds have been simulated
for this study.

All processes are generated with \textsc{MadGraph5\_aMC@NLO} (MG5)
\citep{Alwall:2014hca} at leading order accuracy in QCD, with up
to 3 additional partons in the event. The center of mass energy used
is $\sqrt{s}=13$ TeV. The Higgs boson decay into heavy-flavor quarks,
along with parton showering and hadronization, is simulated in \textsc{Pythia8}
\citep{Sjostrand:2014zea}. The \textsc{MLM} prescription \citep{Alwall:2007fs}
is used to match jets from the matrix element calculations and the
PS description. Finally, the detector response is simulated in \textsc{Delphes
\citep{deFavereau:2013fsa}} using the default CMS configuration available
in \textsc{Delphes}.

The effect of pileup is simulated within \textsc{Delphes} with a mean
pileup value set to 50. Particles are reconstructed using the E-Flow
algorithm in \textsc{Delphes}. Jets are clustered using the AK algorithm
implemented in the \textsc{FastJet} package \citep{Cacciari:2011ma}.
Distance parameters of 0.4 (AK4), 0.8 (AK8), and 1.5 (AK15) are used,
and only jets with $p_{\text{T}}$ greater than 20, 100, and 50 GeV,
respectively, are considered. Charged hadrons coming from pileup vertices
are discarded and the residual event pileup density is used along
with the jet area to perform pileup subtraction on jets \citep{deFavereau:2013fsa}.
The true flavor of jets is defined differently for different strategies:
\begin{itemize}
\item \textbf{AK4:} If at least one b quark is found within the jet radius,
the jet is labeled as a true b jet. If no b quark is found\footnote{Labeling a jet c, only if no b quark is found, ensures that a jet
with a c quark arising from a b hadron decay is labeled as a b jet
rather than a c jet.}, but at least one c quark is found, the jet is labeled c. If neither
a b nor a c quark is found, the jet is labeled udsg or light-flavor.
\item \textbf{AK8 and AK15:} If both generated b or c quarks arising from
the Higgs boson decay lie within a radius of 0.8 (1.5) units around
the jet center, the AK8 (AK15) jet is labeled as a true bb or true
cc jet. Otherwise, it is labeled as an ll jet.
\item \textbf{PAIReD jet:} If both generated b or c quarks arising from
the Higgs boson decay lie within a radius of 0.4 units around the
center of either of the two constituent AK4 jets, the PAIReD jet is
labeled as a true bb or cc jet. Otherwise, it is labeled as an ll
jet.
\item \textbf{PAIReDEllipse jet:} If both generated b or c quarks arising
from the Higgs boson decay lie within the perimeter of the ellipse
defined by the jet, the PAIReDEllipse jet is labeled as a true bb
or cc jet. Otherwise, it is labeled as an ll jet.
\end{itemize}
In addition to the standard Higgs boson samples discussed above, alternative
samples with a variable Higgs boson mass are also generated. In these
``flat-mass'' samples, the mass of the Higgs boson ($m_{\text{H}}$)
is set to integer values between 10 and 500 GeV with a uniform distribution,
while the Higgs boson decay width is set to 4 MeV as expected from
the SM \citep{PDG2022}. These samples are used in the classifier
trainings in order to achieve mass decorrelation \citep{Dolen:2016aa,Louppe:2016ylz,PhysRevD.96.074034,ATL-PHYS-PUB-2018-014,CMS-DP-2020-002,Bradshaw:2019ipy,PhysRevLett.125.122001,Benkendorfer:2020gek,Kitouni:2021aa,Klein:2022hdv}.
The technique of using flat-mass sample to achieve decorrelation has
previously been implemented in hadron collider experiments \citep{CMS-DP-2020-002,CMS-DP-2021-017}.
Using mass decorrelated taggers ensures a fair comparison among different
reconstruction strategies, since algorithms used in some strategies
(e.g. AK4-based reconstruction) do not have access to the Higgs boson
mass information.

\begin{figure}
\begin{centering}
\includegraphics[width=0.5\textwidth]{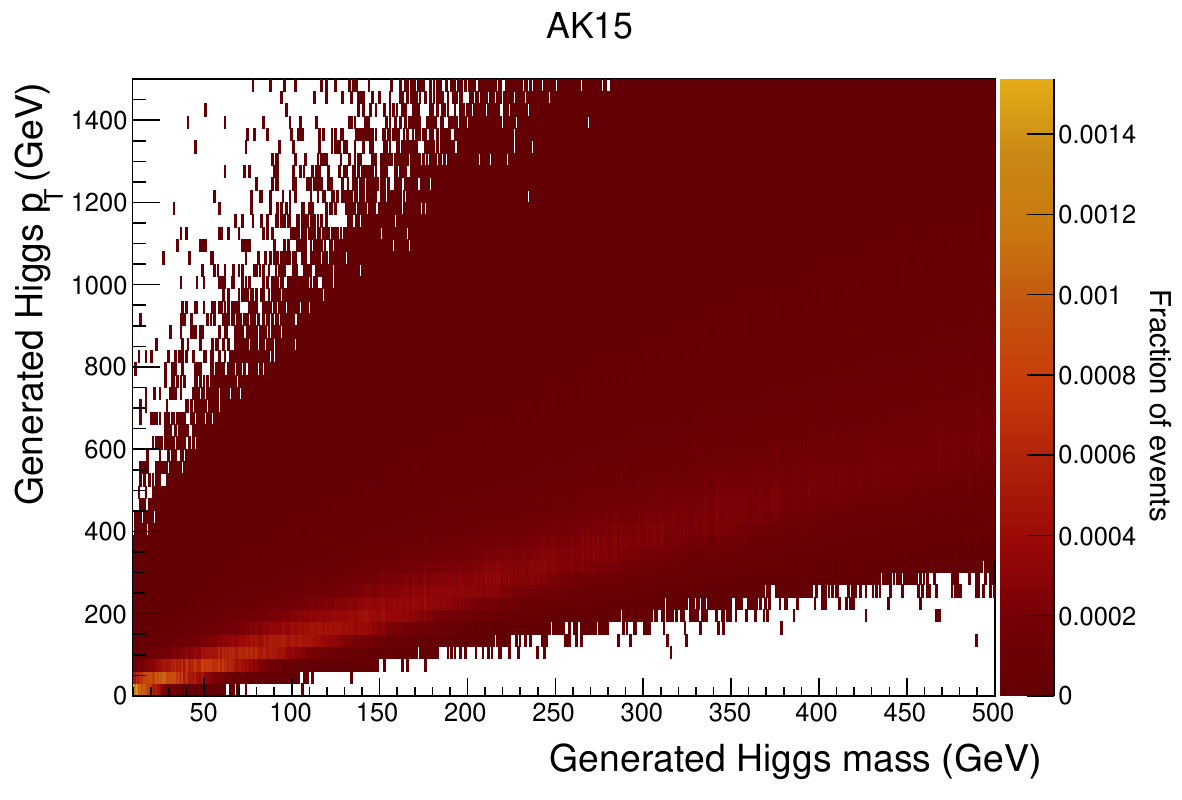}\includegraphics[width=0.5\textwidth]{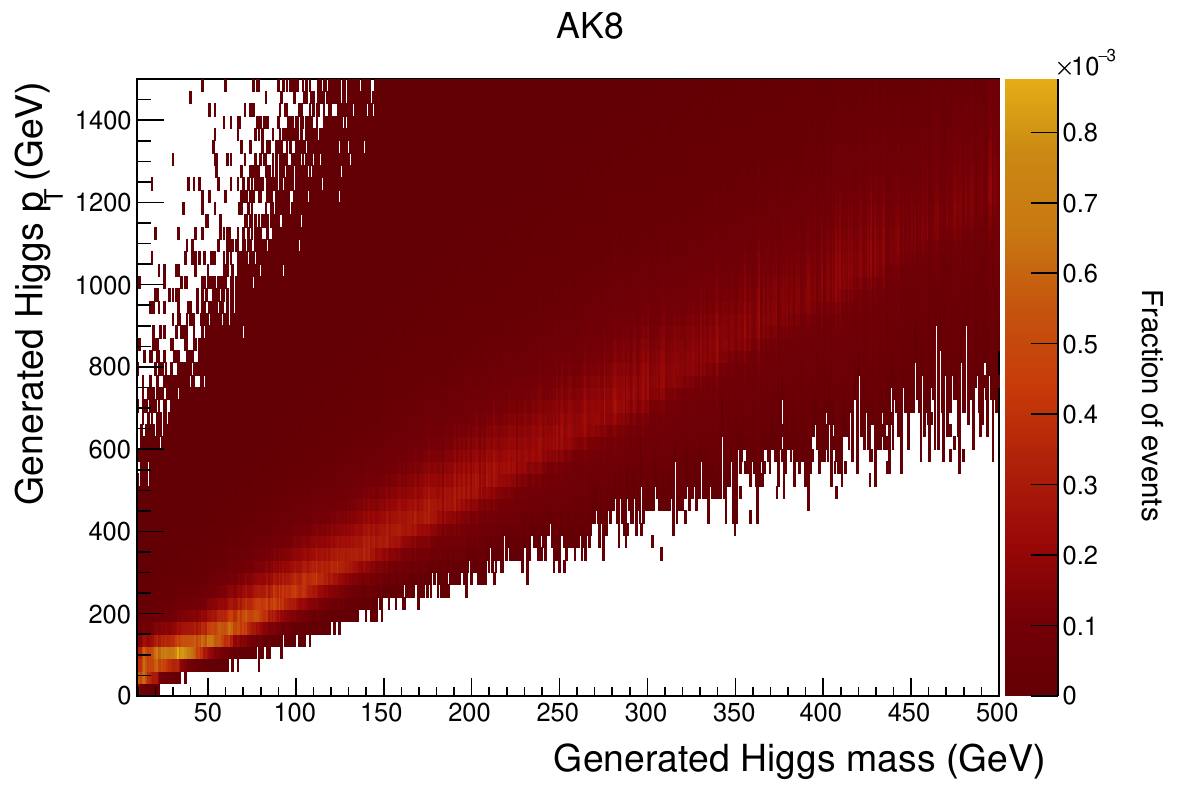}
\par\end{centering}
\centering{}\caption{\label{fig:Tricuts}The fraction of AK15 (left) and AK8 (right) cc
jets sampled at different values of the generated Higgs boson mass
(x-axis) and $p_{\text{T}}$ (y-axis) from the flat-mass ZH(H$\rightarrow\text{c}\bar{\text{c}}$)
sample.}
\end{figure}

An additional complexity arises in the flat-mass Higgs boson samples.
The $p_{\text{T}}$ spectrum of the Higgs boson is dependent on the
generated mass of the Higgs boson; the fraction of events containing
at least one bb/cc AK8 or AK15 jet diminishes at high masses of the
Higgs boson. To mitigate this effect, an empirical Higgs boson mass-dependent
selection criterion is applied\footnote{The parameter \texttt{pt\_min\_pdg} in MG5 is used. The resulting
Higgs boson $p_{\text{T}}$ spectrum, however, starts at a value slightly
below the applied selection criterion.} on the minimum $\text{\ensuremath{p_{\text{T}}}}$ of the generated
Higgs boson ($p_{\text{T}}^{\text{H}}$) in the physics processes
used to sample AK8 and AK15 jets. This criterion ranges between $p_{\text{T}}^{\text{H}}>0$
GeV at Higgs boson mass $m_{\text{H}}=10$ GeV to $p_{\text{T}}^{\text{H}}>390$
GeV at $m_{\text{H}}=500$ GeV for AK15 jets, and between $p_{\text{T}}^{\text{H}}>0$
GeV at $m_{\text{H}}=10$ GeV to $p_{\text{T}}^{\text{H}}>900$ GeV
at $m_{\text{H}}=500$ GeV for AK8 jets. These selections ensure that
a sufficient number of AK15 and AK8 jets are sampled throughout the
Higgs boson mass range, without generating events that are unlikely
to have any AK15 or AK8 jets. The distributions of AK15 and AK8 jets
as functions of the generated Higgs boson mass and $p_{\text{T}}$
are shown in Fig. \ref{fig:Tricuts} after the empirical selection
criteria on $p_{\text{T}}^{\text{H}}$ are applied.

A fraction of the simulated dataset was made available publicly for
testing and development \citep{mondal_2024_11150993}.

\section{Flavor tagging algorithm\label{sec:Flavor-tagging-algorithm}}

The algorithm used in this paper to identify the flavor of jets defined
using different reconstruction strategies is the Particle Transformer
(ParT) algorithm \citep{Qu:2022mxj}. The choice of the tagging algorithm
is largely independent of the choice of event reconstruction strategy
and one can evaluate the performance of various reconstruction strategies
relative to any one, fixed tagging algorithm.

The ParT algorithm makes use of a number of features of jet constituents
as well as interactions corresponding to every pair of constituents.
The network comprises of several particle and class attentions blocks.
Detailed description of the network architecture can be found in Ref.
\citep{Qu:2022mxj}. As inputs, we use kinematics and trajectory displacement
information of particles, but not the particle identification information.
This choice is consistent with the standard methods for jet tagging
used in particle physics experiments which intend to minimize potential
biases in subsequent independent validation steps that exploit dedicated
soft lepton information \citep{CMS:2021scf}.

In addition to the classification tasks that have been discussed so
far, the ParT algorithm is also implemented for mass regression (Appendix
\ref{sec:Mass-regression}) with minor changes in the architecture.
For regression, the default settings were modified to use a PReLU
activation function \citep{He2015DelvingDI} and the logarithm of
the hyperbolic cosine (log cosh) \citep{Wang:2022aa} as the loss
function. In addition, a gradient clipping feature \citep{FITPT283}
was implemented to prevent gradients from becoming too large owing
to the output node containing mass values of $\mathcal{O}(100)$ in
units of GeV.

\section{Neural network training\label{sec:Neural-network-training}}

Several networks corresponding to the various reconstruction strategies
are trained using the weaver framework \citep{weaver} based on PyTorch
\citep{NEURIPS2019_bdbca288}, implementing the ParT architecture
\citep{Qu:2022mxj} with minor modifications as described in Sec.
\ref{sec:Flavor-tagging-algorithm}. Jets are classified into three
classes, namely bb, cc, and ll (b, c, and udsg for AK4 jets). The
bb (cc) jets used in the trainings are sampled from the flat-mass
ZH(H$\rightarrow\text{b\ensuremath{\bar{\text{b}}}}$) (ZH(H$\rightarrow\text{c}\bar{\text{c}}$))
processes. The ll jets are sampled from ZH(H$\rightarrow\text{b\ensuremath{\bar{\text{b}}}}$),
ZH(H$\rightarrow\text{c}\bar{\text{c}}$), and Z+jj processes. Wrong
combinations of jets, such as multi-pronged jets containing only one
b/c parton in the ZH processes, are also labeled ll (cf. Sec. \ref{sec:Simulation}).
The number of ll jets sampled from each of ZH(H$\rightarrow\text{b\ensuremath{\bar{\text{b}}}}$),
ZH(H$\rightarrow\text{c}\bar{\text{c}}$), and Z+jj processes ranges
between 25--30 million. In case of AK4 jet trainings, all jets labeled
b, c, and udsg are used regardless of the physics process they were
sampled from.

Jets are assigned weights during the training such that each class
contains an equal number of jets. In case of AK8, AK15, PAIReD, and
PAIReDEllipse trainings, the jets are weighted such that a mass metric
has the same distribution for all labels, using the mass distribution
of the ll jets as the reference. For the purpose of defining a uniform
mass metric, mass regression techniques, discussed in Appendix \ref{sec:Mass-regression},
are used. Only jets with regressed masses ($m_{\text{reg}}$) of 50--400
GeV are used to train the classifier, in order to avoid any biases
in the prediction of $m_{\text{reg}}$ for jets lying at the extreme
edges of the range ($m_{\text{reg}}\in[10,50)\cup(400,500]$ GeV).
In case of AK4 jets, the jets are weighted such that the $p_{\text{T}}$
and $\eta$ of the jets in the b and c classes have the same distributions
as those in the reference udsg class. 

About a tenth of the jets are used as an independent \emph{validation
dataset} to check for overtraining. The initial learning rate (LR),
batch size, and number of epochs were adjusted to ensure optimal convergence
and balance between model performance and training efficiency. Finally,
the Ranger optimizer \citep{Ranger}, along with a uniform configuration
of batch size 2048 and initial LR of $5\times10^{-3}$, was chosen
across all trainings, and each training was run for 80 epochs. The
model from the epoch achieving highest classification accuracy on
the validation dataset is selected as the final classifier model.
The trainings are then evaluated on an independent \emph{test dataset}
comprising jets from ZH(H$\rightarrow\text{b\ensuremath{\bar{\text{b}}}}$)
and ZH(H$\rightarrow\text{c}\bar{\text{c}}$) processes with $m_{\text{H}}=125$
GeV, and an independent Z+jj sample. The performance comparison of
the various classifiers corresponding to different reconstruction
strategies is presented in Sec. \ref{sec:Performance-comparisons}.

\section{Performance evaluation on H$\rightarrow$$\text{c}\bar{\text{c}}$
events \label{sec:Constructing-a-mock}}

The various event reconstruction strategies outlined in this paper
can be compared and contrasted using H$\rightarrow\text{c}\bar{\text{c}}$
events. In this paper, we demonstrate a mock ZH(H$\rightarrow\text{c}\bar{\text{c}}$)
analysis as an example. The analysis strategies corresponding to the
various reconstruction strategies are outlined in Sec. \ref{subsec:Event-selection-strategies}.
In all cases, we ignore the reconstruction of the leptonic Z decay
and focus on reconstructing the hadronic Higgs boson decay. Another
example using Z$\rightarrow\text{c}\bar{\text{c}}$ events is presented
in Appendix \ref{sec:Validation-with-ZZ(Z)}. Only the Z+jj background
with a leptonically-decaying Z boson is studied in this section. Since
only the hadronic part is considered for the analysis, it is expected
that the results would be similar when discriminating signal events
from W+jj and Z+jj (Z decaying into neutrinos) in other channels,
and hence they are not studied separately. However, since the $\text{t}\bar{\text{t}}$
background is a major background in search for the W-associated Higgs
production and has a signature distinct from Z+jj events, it is studied
in Appendix \ref{sec:Performance-with-ttbar}, along with an extended
PAIReD(Ellipse) jet tagger retrained with additional $\text{t}\bar{\text{t}}$
classes.

\subsection{Performance metrics\label{subsec:Performance-metrics}}

The performances of jet flavor taggers are usually quantified using
metrics such as accuracy, area under the ROC curve (AUC), background
rejection at a given signal efficiency, etc. While these metrics convey
useful information when evaluating the performance of different algorithms
on the same type of reconstructed jets (e.g. public benchmark jet
datasets \citep{komiske_patrick_2019_3164691,kasieczka_gregor_2019_2603256,JetClass,Chen:2022aa})
they are not very interpretable when comparing across various event
reconstruction strategies and across various Lorentz-boost regimes.

We define end-to-end selection efficiency as the fraction of simulated
events that are selected and correctly reconstructed after all selection
criteria and classifier score requirements are applied. This includes
the requirement that a certain number of jets be reconstructed per
event, as well as any requirements that are applied to the jet tagger
scores to achieve a target signal efficiency. This quantity is, therefore,
dependent on the reconstruction strategy being used, as well as the
tagging performance achievable using that strategy. Events that cannot
be reconstructed using a given reconstruction strategy, and events
that are incorrectly reconstructed (e.g. wrong choice/selection of
jets) contribute to a decline in the end-to-end efficiency metric.
This is in contrast to the standard metrics that are used to evaluate
jet tagging algorithm performances---standard metrics focus on the
background rejection for a specific type of jets, disregarding the
fraction of events that contain such jets and the fraction of events
where the correct jet(s) is (are) chosen. This makes the end-to-end
efficiency a suitable metric to compare between event reconstruction
strategies and across Lorentz-boost regimes.

The end-to-end selection efficiency can be defined for both signal
and background events. They can be interpreted as the end-to-end true
positive and false positive rates, respectively. A more natural metric
for background events is the background rejection, which can be defined
as the inverse of the background selection efficiency. Thus, one can
define an end-to-end background rejection rate at a given end-to-end
signal selection efficiency. Additionally, a rough estimation of the
differential signal sensitivity as a function of the Higgs boson $p_{\text{T}}$
is shown in Appendix \ref{sec:Differential-signal-sensitivity}.

\subsection{Event selection strategies\label{subsec:Event-selection-strategies}}

\subsubsection{Analysis using AK4 jets}

We follow the strategy adopted by the CMS H$\rightarrow\text{c}\bar{\text{c}}$
search \citep{VHcc} in the resolved-jet regime. Only events with
at least 2 well-reconstructed AK4 jets with $p_{\text{T}}>20$ GeV
lying within $|\eta|<2.5$ are selected. In each event, jets are sorted
by the so-called CvsL score, which is defined as
\begin{equation}
\text{CvsL}=\frac{P(\text{c)}}{P(\text{c})+P(\text{udsg)}},\label{eq:CvsL}
\end{equation}
where $P(\text{c})$ and $P(\text{udsg})$ are the tagger scores obtained
at the c and udsg output nodes of the classifier prediction, respectively.
We verified that using $P(\text{c})$ instead of CvsL to sort the
jets did not change the results significantly. The two jets with the
highest CvsL values in each event are used to reconstruct the Higgs
boson candidate. The invariant mass $m_{jj}$ of the Higgs boson candidate
is required to lie within the range 50--200 GeV. The same procedure
is repeated for the background sample.

Only signal events where the correct pair of AK4 jets is chosen are
counted towards the ``pass'' events in the signal selection efficiency
calculation. To obtain the end-to-end background rejection rate at
a fixed (say, $\epsilon$) end-to-end signal selection efficiency
(in line with the strategy outlined in Sec. \ref{subsec:Performance-metrics}),
the $P(\text{c})$ score of the subleading jet in signal events for
which exactly $\epsilon$ fraction of the signal events pass all the
selection criteria is first determined. Next, the same requirement
on the $P(\text{c})$ score is applied on the subleading jet of the
background events. The fraction of background events thus selected
gives the end-to-end background mistag rate at an end-to-end signal
efficiency of $\epsilon$. The inverse of this quantity gives the
end-to-end background rejection rate.

\subsubsection{Analyses using AK8 and AK15 jets}

Events with at least one AK15 (AK8) jet with $p_{\text{T}}>50$ (100)
GeV and $m_{\text{reg}}\in(50,400)$ GeV, lying within the tracker
acceptance ($|\eta|<2.5$) are selected. The preselection on $m_{\text{reg}}$
is necessary since the classifiers are trained only in a limited $m_{\text{reg}}$
range. In each event, the jet with the highest $P(\text{cc})$ score
obtained from the classifier prediction is considered as the Higgs
boson candidate. As before, only signal events where the selected
jet contains both c quarks from the Higgs boson decay are counted
towards the ``pass'' events in the signal selection efficiency calculation.
Finally, an additional requirement of $m_{\text{reg}}\in(50,200)$
GeV is applied for the selected jet.

As before, a requirement on the $P(\text{cc})$ score for which a
certain fraction of the signal events pass the selection criteria
is applied to the selected jet of the background events to calculate
the end-to-end background rejection rate.

\subsubsection{Analyses using PAIReD and PAIReDEllipse jets}

The analyses using PAIReD and PAIReDEllipse jets follow the same strategy
as the ones with AK8 and AK15 jet. No minimum $p_{\text{T}}$ requirement
is applied for the jets, but the AK4 jets from which the PAIReD(Ellipse)
jets are defined are required to have $p_{\text{T}}>20$ GeV each.
Events with at least one PAIReD(Ellipse) jet with $m_{\text{reg}}\in(50,400)$
GeV lying within the tracker acceptance ($|\eta|<2.5$) are preselected.
In each event, the jet with the highest $P(\text{cc})$ score obtained
from the classifier prediction is considered as the Higgs boson candidate.
Finally, a more stringent requirement of $m_{\text{reg}}\in(50,200)$
GeV is applied for the selected jet.

Similarly as before, the same requirement on the $P(\text{cc})$ score
is applied to the selected jet of the background events and the end-to-end
background rejection rate is calculated.

\subsubsection{Summary}

\begin{figure}
\begin{centering}
\subfloat[\label{fig:Inclusive-in-Higgs/diparton}Signal (left) and background
(right) events, inclusively in Higgs/diparton $p_{\text{T}}$.]{\begin{centering}
\includegraphics[width=0.5\textwidth]{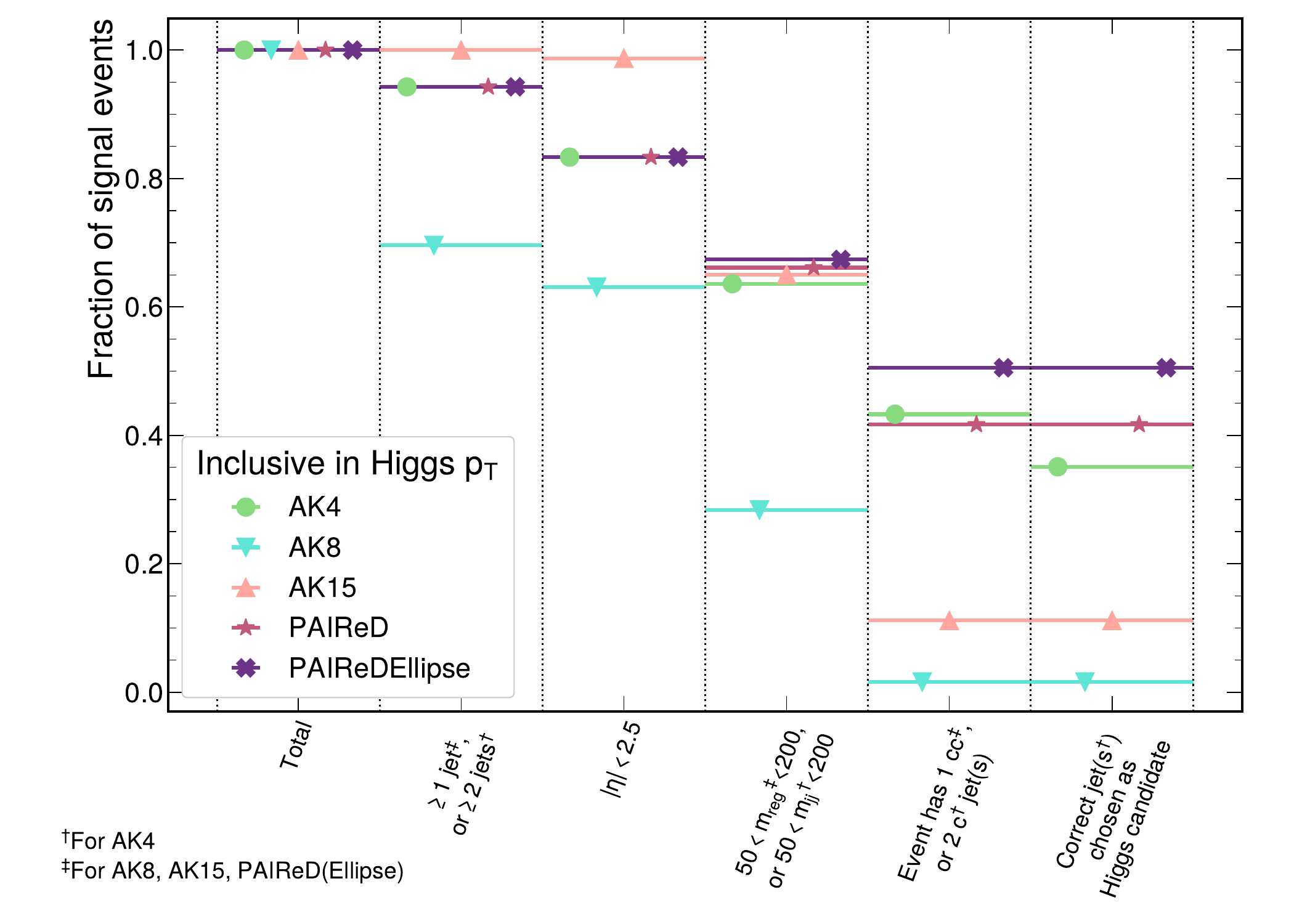}\includegraphics[width=0.5\textwidth]{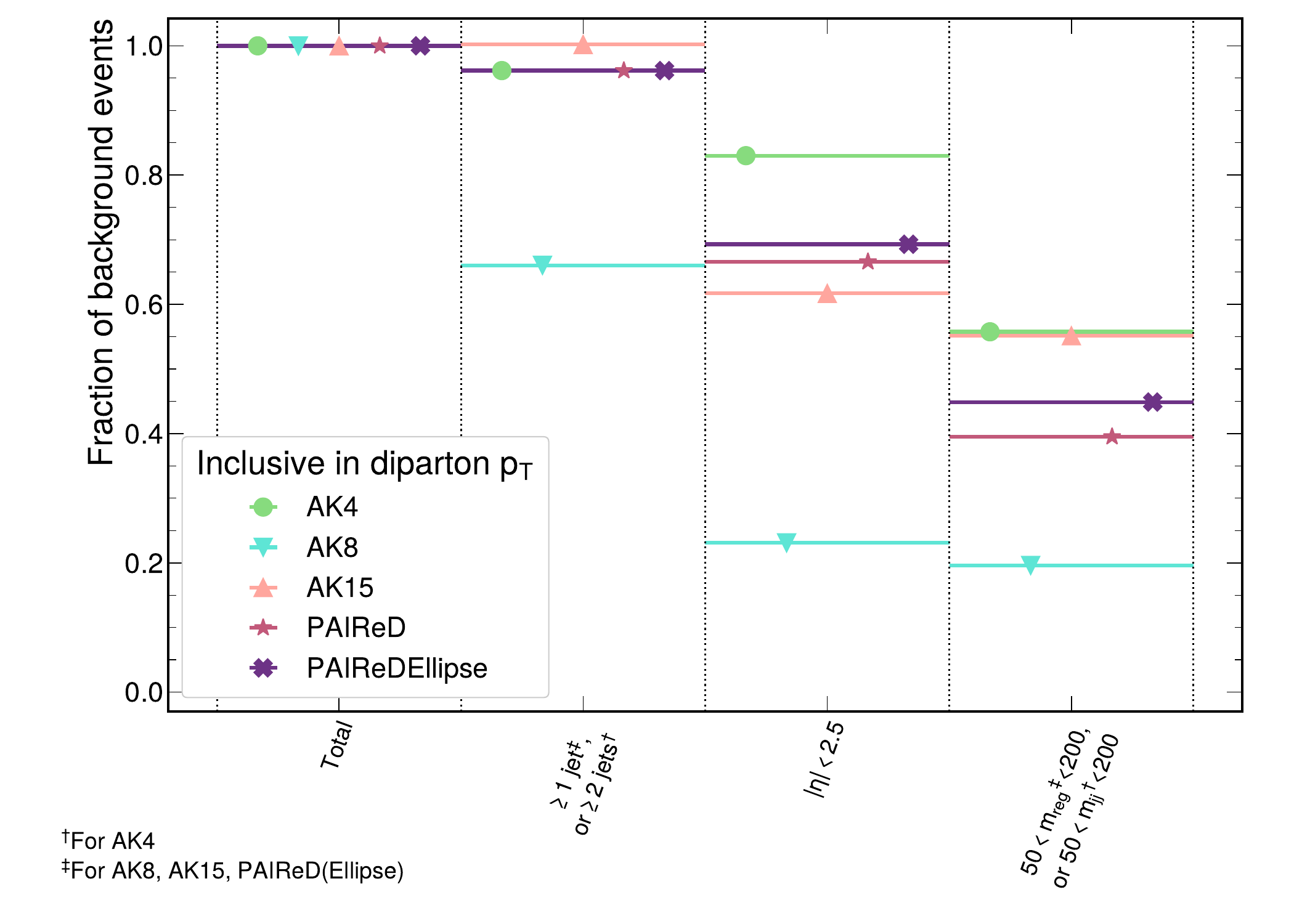}
\par\end{centering}
}
\par\end{centering}
\begin{centering}
\subfloat[\label{fig:Signal-events-at}Signal events at various ranges of the
Higgs boson $p_{\text{T}}$.]{\begin{centering}
\includegraphics[width=0.33\textwidth]{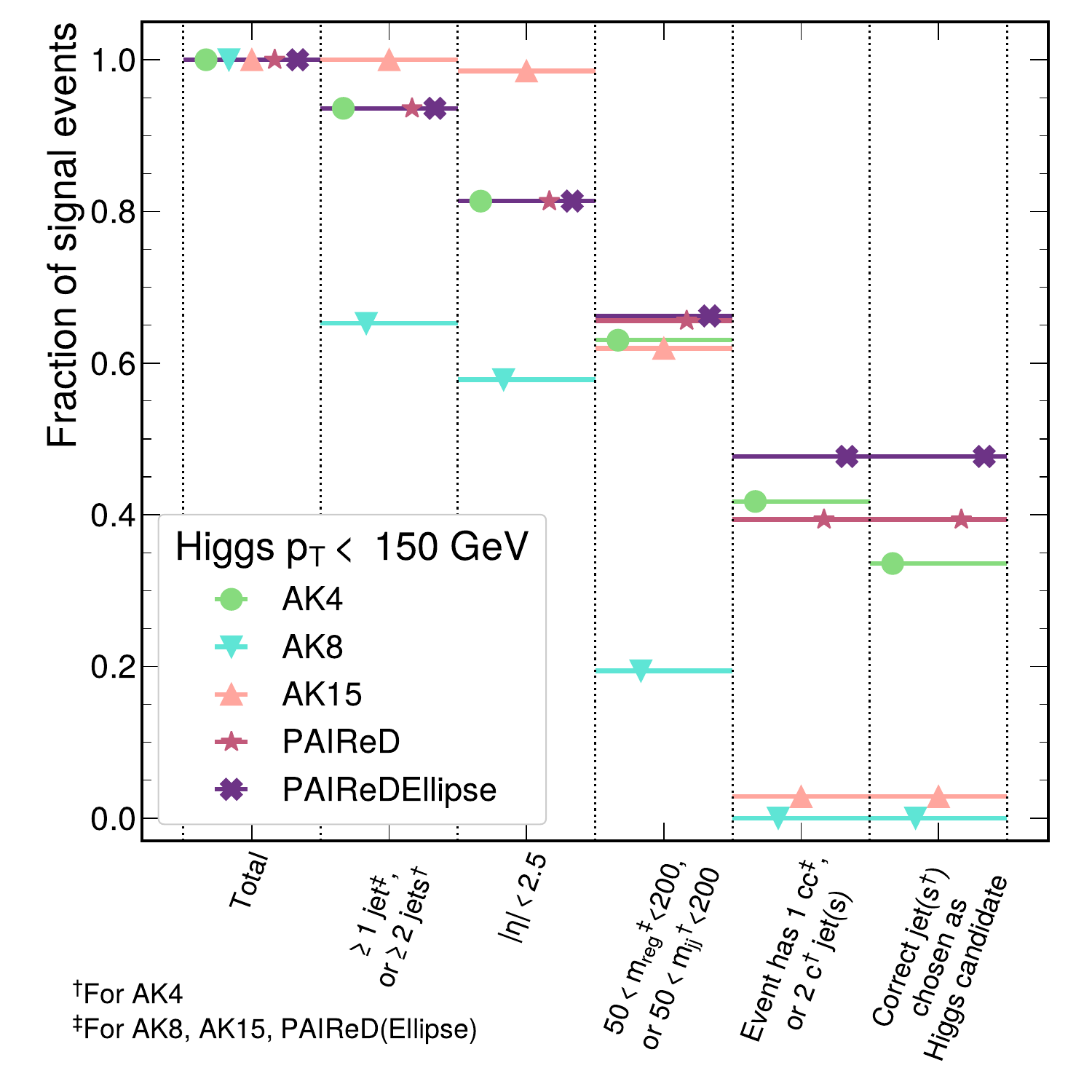}\includegraphics[width=0.33\textwidth]{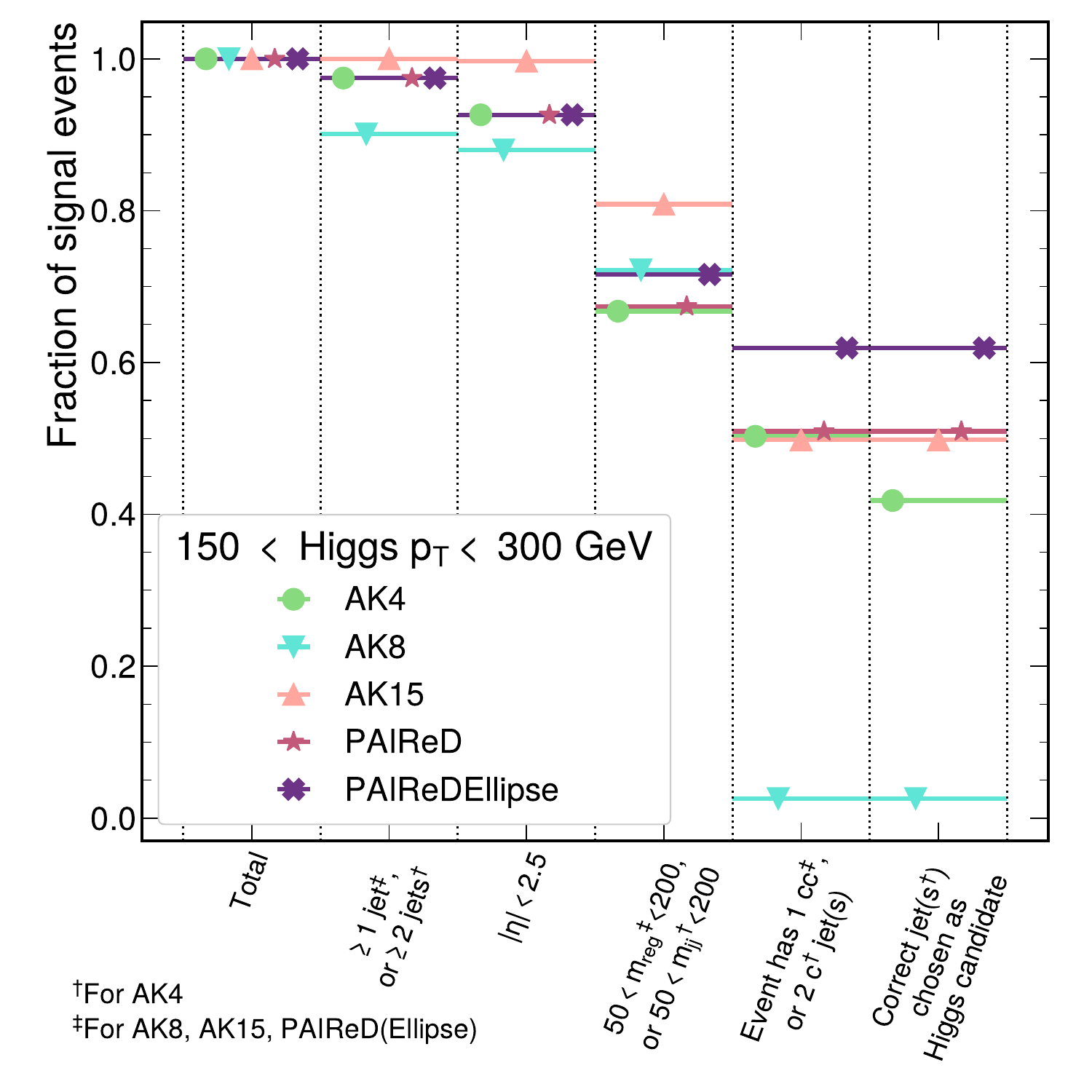}\includegraphics[width=0.33\textwidth]{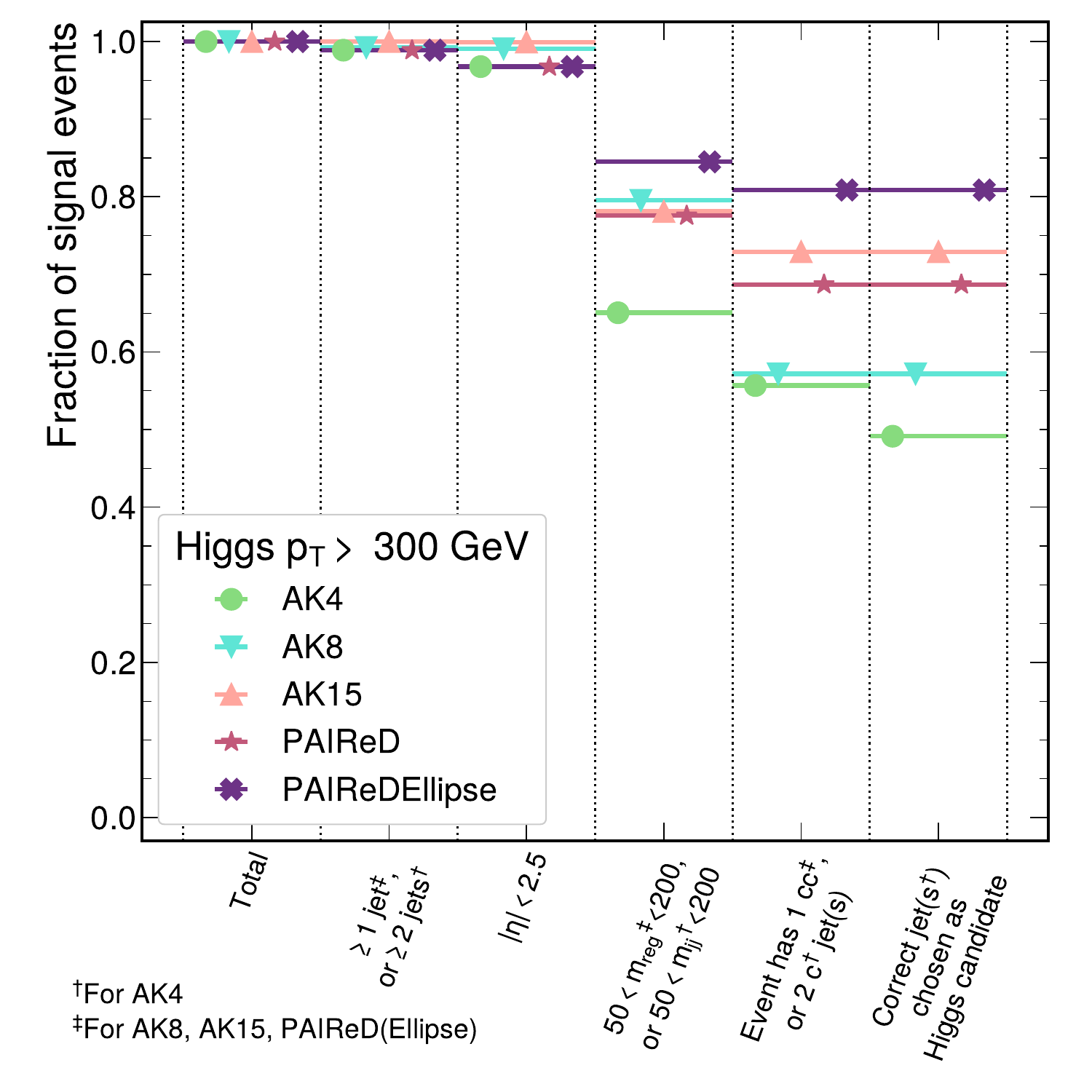}
\par\end{centering}
}
\par\end{centering}
\begin{centering}
\subfloat[Background events at various ranges of the diparton system $p_{\text{T}}$.]{\begin{centering}
\includegraphics[width=0.33\textwidth]{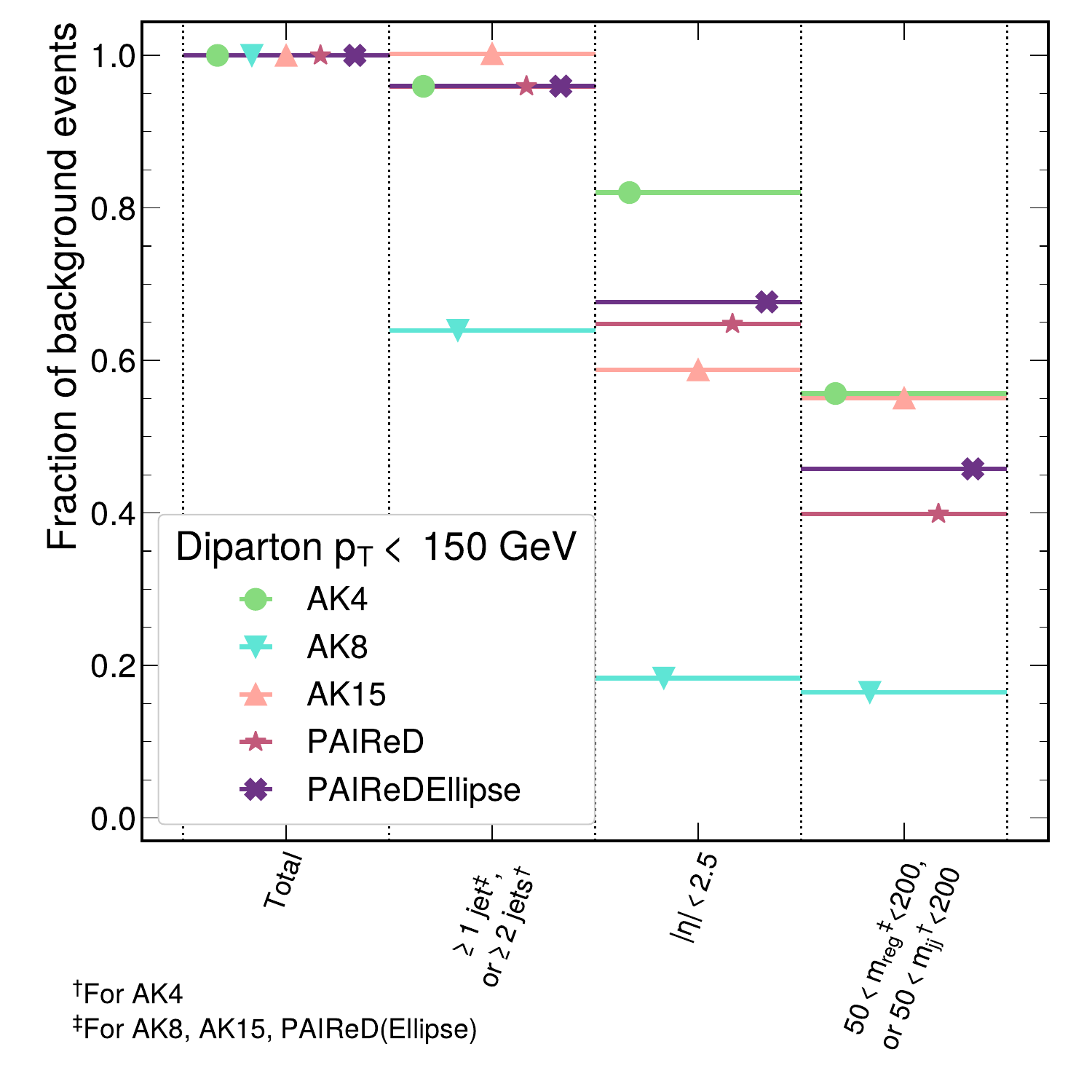}\includegraphics[width=0.33\textwidth]{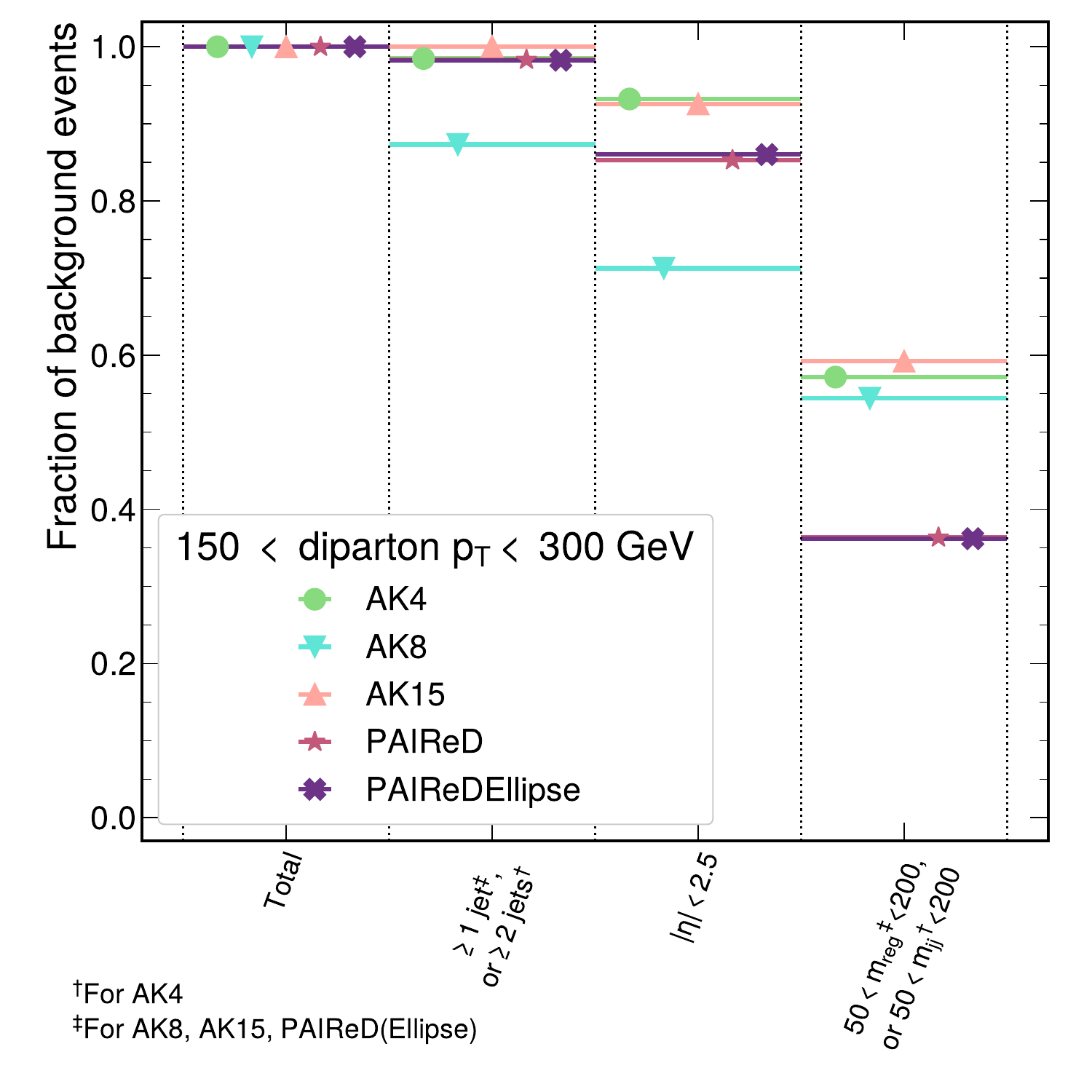}\includegraphics[width=0.33\textwidth]{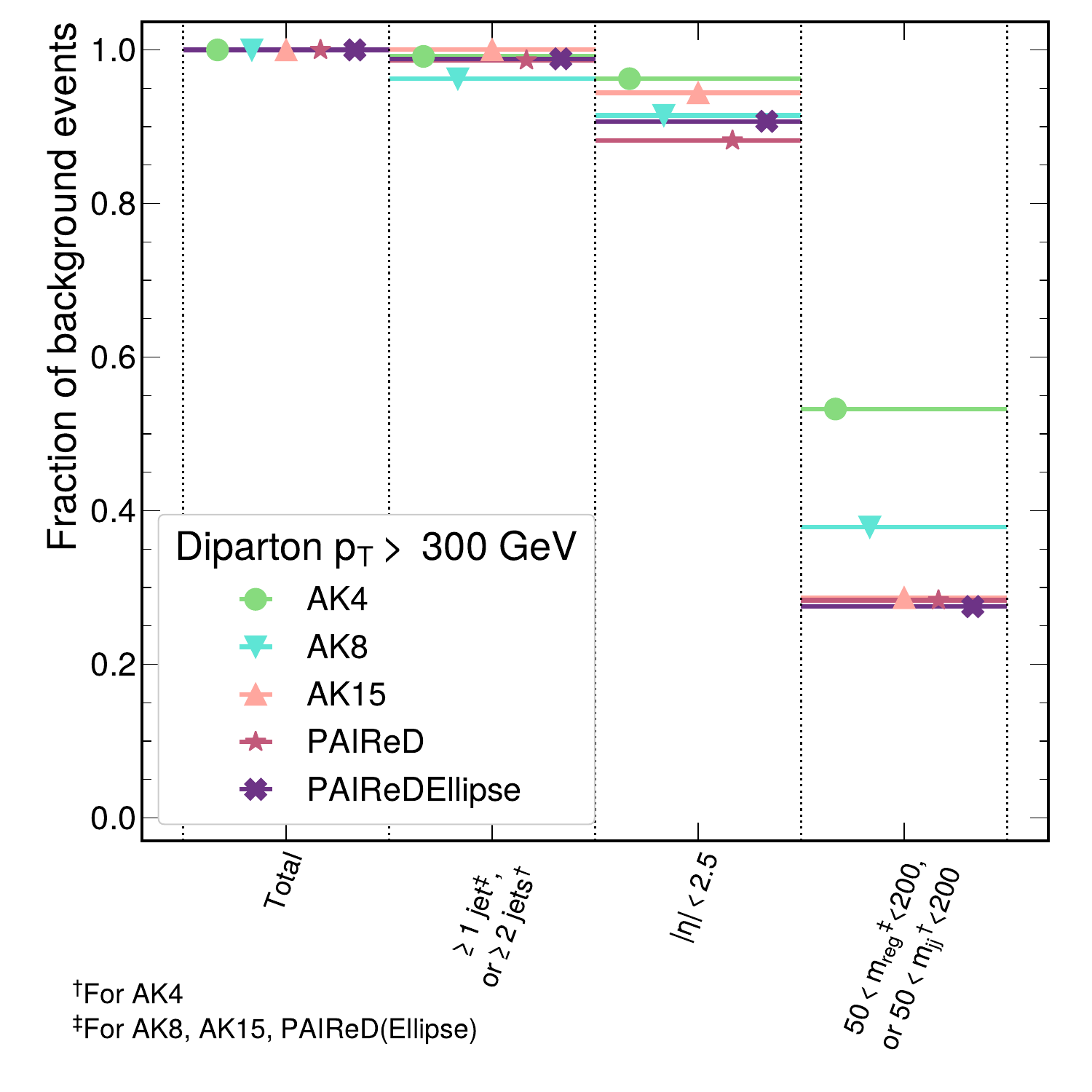}
\par\end{centering}
}
\par\end{centering}
\caption{\label{fig:Cutflow1}Plots showing fraction of signal or background
events obtained after each selection criterion is progressively applied.
The last two selections in Fig. \ref{fig:Inclusive-in-Higgs/diparton}
(left) and Fig. \ref{fig:Signal-events-at} involve using generator-level
flavor information in signal samples to investigate how many correctly
reconstructed events pass the selection criteria. Selections are,
in general, different for small-radius (AK4) and large-radius (AK8,
AK15, PAIReD, PAIReDEllipse) jets, and are marked using $\dagger$
and $\ddagger$, respectively. The masses are in units of GeV.}

\end{figure}

The hadronic part of the ZH(H$\rightarrow\text{c}\bar{\text{c}}$)
final state is reconstructed with 5 different approaches. The fraction
of signal and background events obtained after progressively applying
each selection criterion is shown in so-called cutflow plots at different
$p_{\text{T}}$ ranges of the Higgs boson (or diparton system) in
Fig. \ref{fig:Cutflow1}. These cutflow plots exclude the final requirement
on the classifier scores applied to obtain a fixed signal efficiency.

\subsection{Performance comparisons\label{sec:Performance-comparisons}}

The end-to-end selection efficiency metric can be factorized into
two components: (a) the reconstruction efficiency, and (b) the classifier
performance. The combined effect of the signal reconstruction efficiency
and the classifier performance determines the final analysis sensitivity.
For instance, a poorer classifier performance can offset any gains
achieved from having a high signal reconstruction efficiency. The
two efficiencies are investigated separately to study how each component
contributes to the final gains.

\subsubsection{Signal reconstruction efficiency\label{subsec:Signal-reconstruction-efficiency}}

The signal reconstruction efficiency can be defined as the ratio of
the maximum number of signal events that can be reconstructed using
a given strategy, to the total number of events generated. This involves
calculating the signal efficiency in each of the analyses without
applying the requirement on the classifier output score (cf. Sec.
\ref{sec:Constructing-a-mock}). This metric can also be considered
as the maximum signal efficiency that can be achieved using a particular
reconstruction strategy. 

\begin{figure}
\begin{centering}
\includegraphics[width=0.5\textwidth]{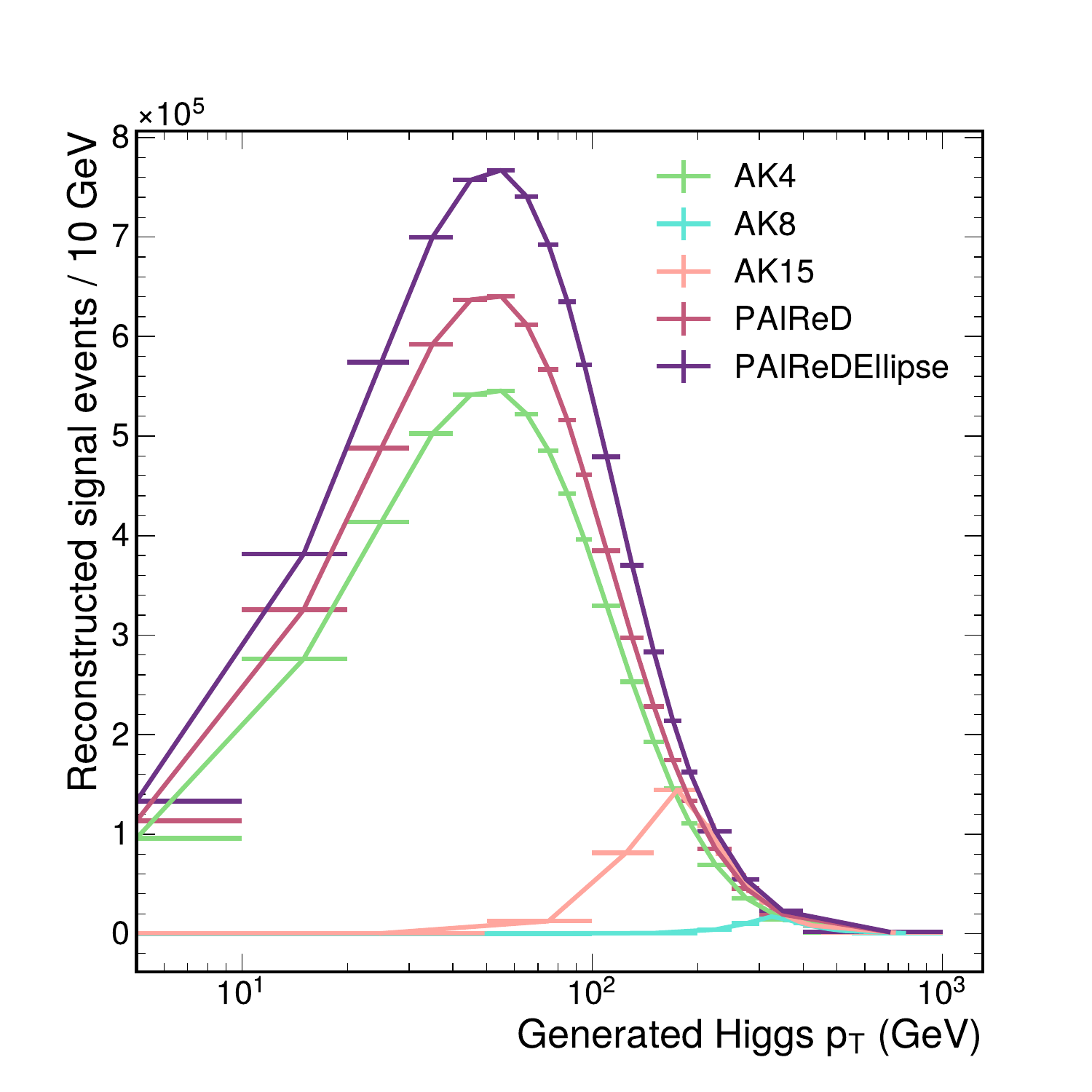}\includegraphics[width=0.5\textwidth]{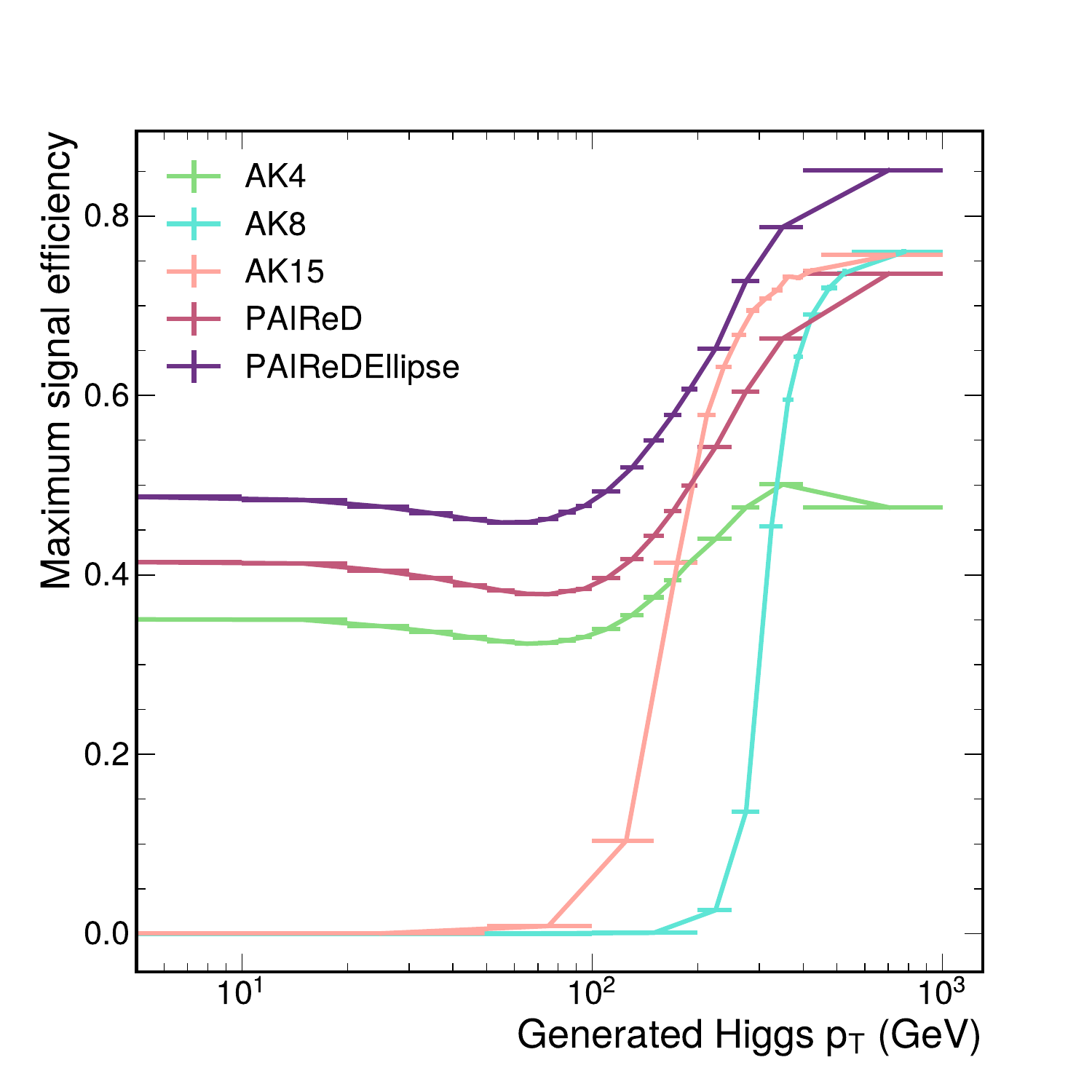}
\par\end{centering}
\caption{\label{fig:MaxSigEff}The number of reconstructed ZH(H$\rightarrow\text{c}\bar{\text{c}}$)
signal events out of the 10M signal events generated (left) and the
signal reconstruction efficiency (right) as functions of the generated
Higgs boson $p_{\text{T}}$, for various reconstruction strategies.}
\end{figure}

The signal reconstruction efficiency is evaluated using the ZH(H$\rightarrow\text{c}\bar{\text{c}}$)
process at various ranges of the generated Higgs boson $p_{\text{T}}$.
A comparison of the yield of reconstructed signal events and the signal
reconstruction efficiency across various reconstruction strategies
is presented in Fig. \ref{fig:MaxSigEff}. The PAIReDEllipse strategy
improves the signal reconstruction efficiency by about 40--50\% compared
to AK4-based event reconstruction at low Lorentz-boosts ($\apprle$200
GeV) of the Higgs boson. The efficiency of the PAIReDEllipse strategy
is slightly higher (\textasciitilde 2--14\%) compared to that of
the AK15-based strategy at higher boosts. The PAIReD strategy has
efficiencies intermediate to that of the AK4 and PAIReDEllipse approaches,
as expected. The AK4, PAIReD, and PAIReDEllipse approaches reconstruct
a significantly larger number of signal events compared to the AK8
and AK15-based approaches. The gain comes mainly from the low Lorentz-boost
regime.

\subsubsection{Classifier performance}

The ability of classifiers to distinguish between signal and background
jets is also expected to vary across different reconstruction strategies.
Classifiers trained using AK4 jets, for instance, evidently have access
to less information than classifiers trained with AK15 jets and hence
are expected to have lower background rejection rates.

The classifier performances are evaluated by calculating the AUC using
all the signal and background events that pass the selection criteria
(except the requirement on the classifier output score). Furthermore,
a proxy Higgs boson $p_{\text{T}}$ needs to be defined for the background
events in order to evaluate the AUC performances at different Lorentz-boosts
of the generated Higgs boson. For this purpose, a diparton $p_{\text{T}}$
is defined by adding the four-vectors of the two MG5-level partons
produced in association with a Z boson in the background Z+jj events.
The diparton $p_{\text{T}}$ is used as the proxy for Higgs boson
$p_{\text{T}}$ in the background events. The motivation behind this
choice is that the fake Higgs boson candidate defined in the background
events passing the selection criteria is most likely reconstructed
from the hadronization products of the two associated MG5-level partons
in the Z+jj events. Therefore, the generator-level diparton system
in background events is a suitable proxy for the generator-level Higgs
boson in signal events.

\begin{figure}
\begin{centering}
\includegraphics[width=0.5\textwidth]{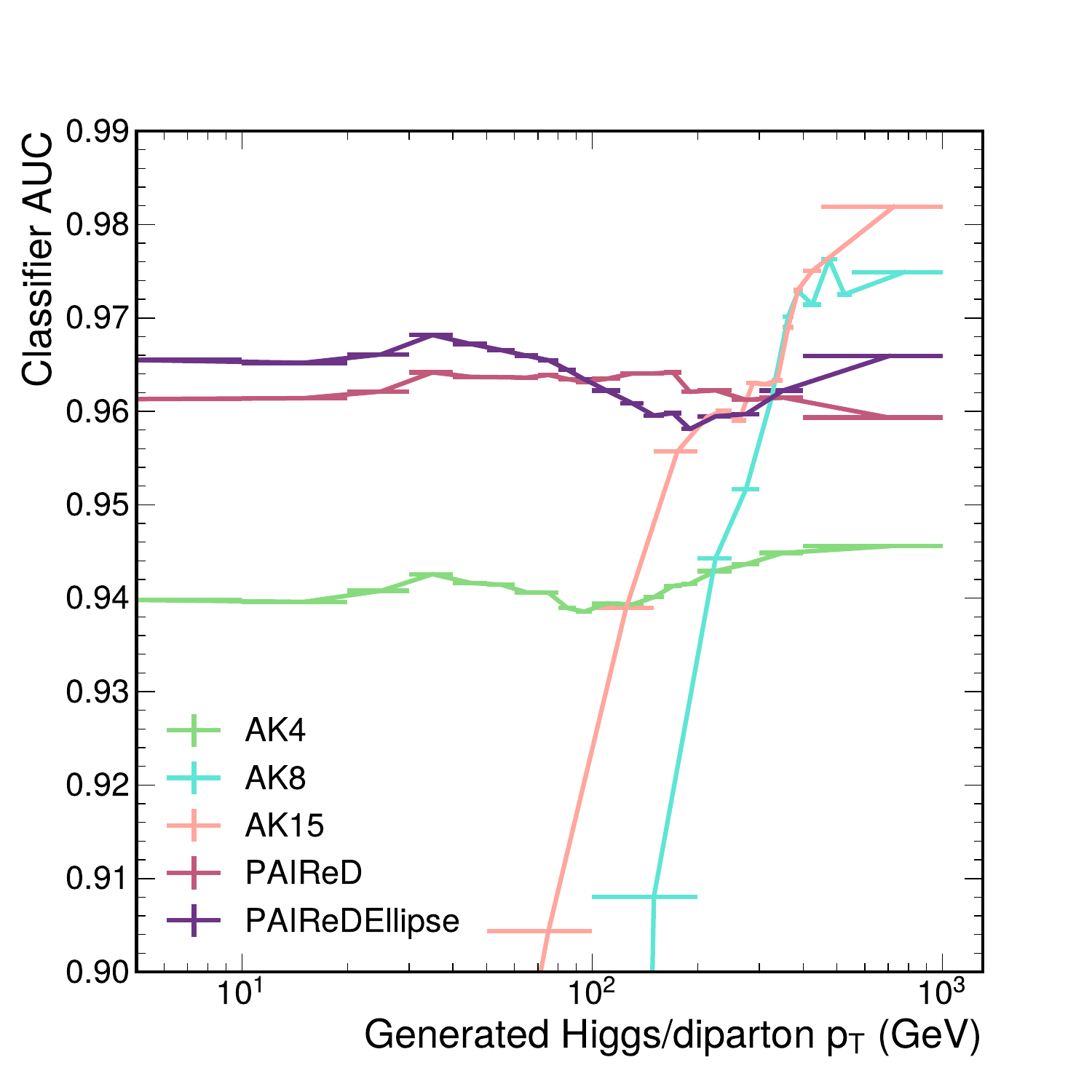}
\par\end{centering}
\caption{\label{fig:Classifier}The area under the ROC curves (AUC) obtained
when classifiers corresponding to various reconstruction strategies
are evaluated on the signal and background events passing the respective
selection criteria, plotted as a function of the Higgs boson $p_{\text{T}}$
(for signal events) or the diparton $p_{\text{T}}$ (for background
events). }
\end{figure}

The AUCs of the classifiers corresponding to the various reconstruction
strategies are compared in Fig. \ref{fig:Classifier}. The classifiers
corresponding to the PAIReD and PAIReDEllipse strategies have comparable
performances and outperform those corresponding to other strategies
in the low boost regimes. In the high boost regimes, the AK15- and
AK8-based classifiers outperform other classifiers. The improvement
compared to PAIReD and PAIReDEllipse approaches at high boosts can
be attributed to the fact that AK8 and AK15 jets are constructed using
IRC-safe sequential clustering algorithms and are less sensitive to
presence of pileup particles. This is discussed in more detail in
Appendix \ref{sec:Effects-of-(not)}.

\subsubsection{End-to-end efficiencies\label{subsec:End-to-end-efficiencies}}

\begin{figure}
\begin{centering}
\includegraphics[width=0.5\textwidth]{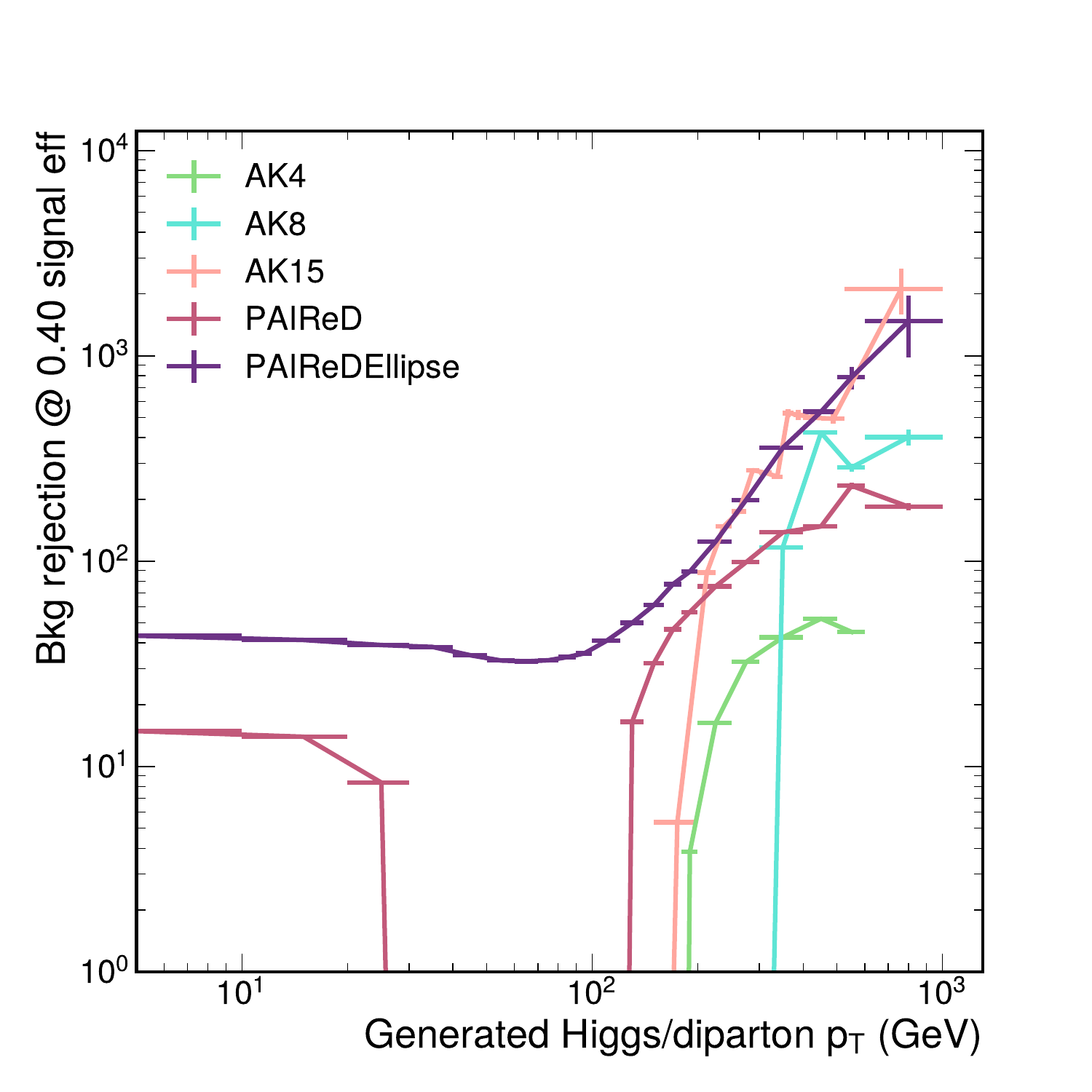}\includegraphics[width=0.5\textwidth]{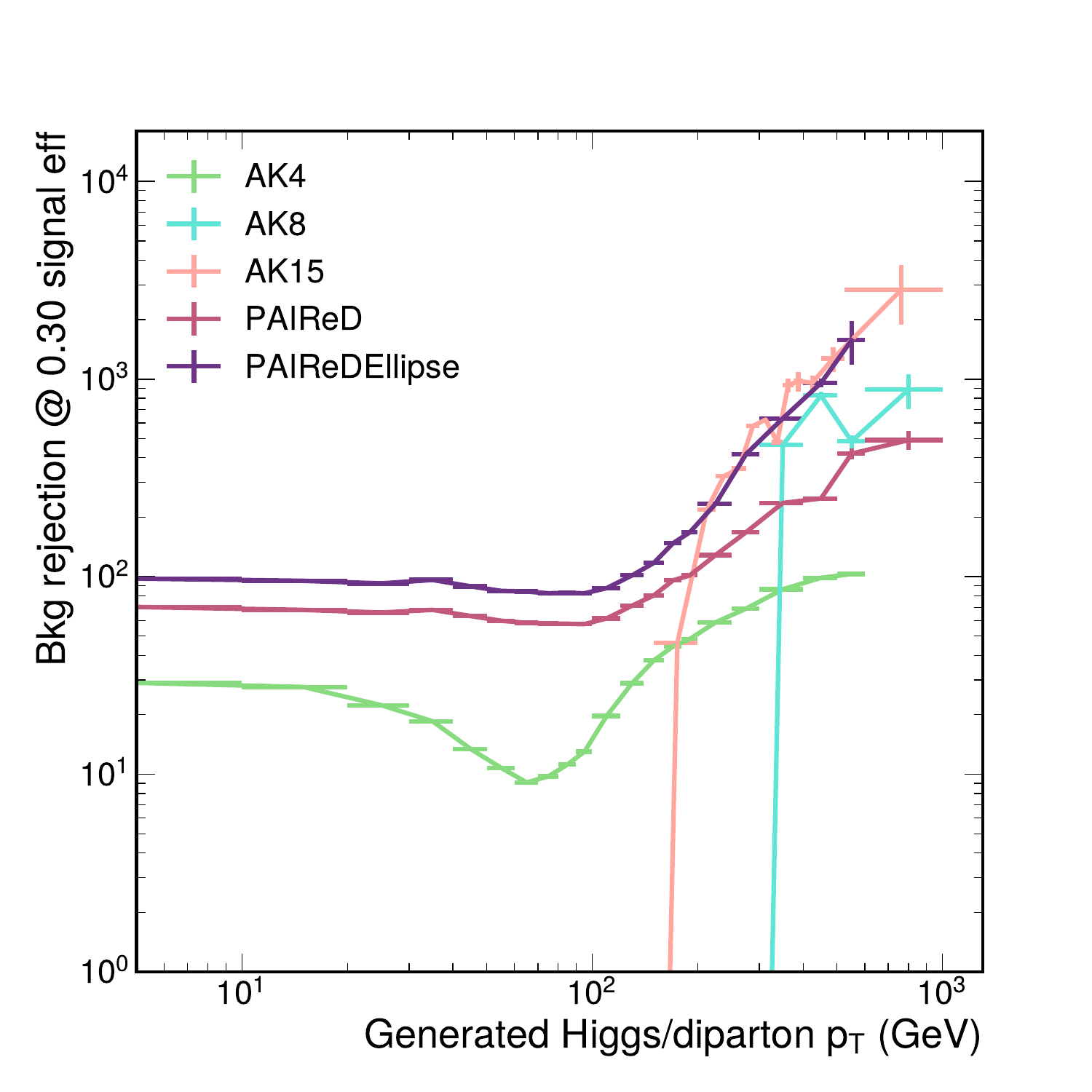}
\par\end{centering}
\begin{centering}
\includegraphics[width=0.5\textwidth]{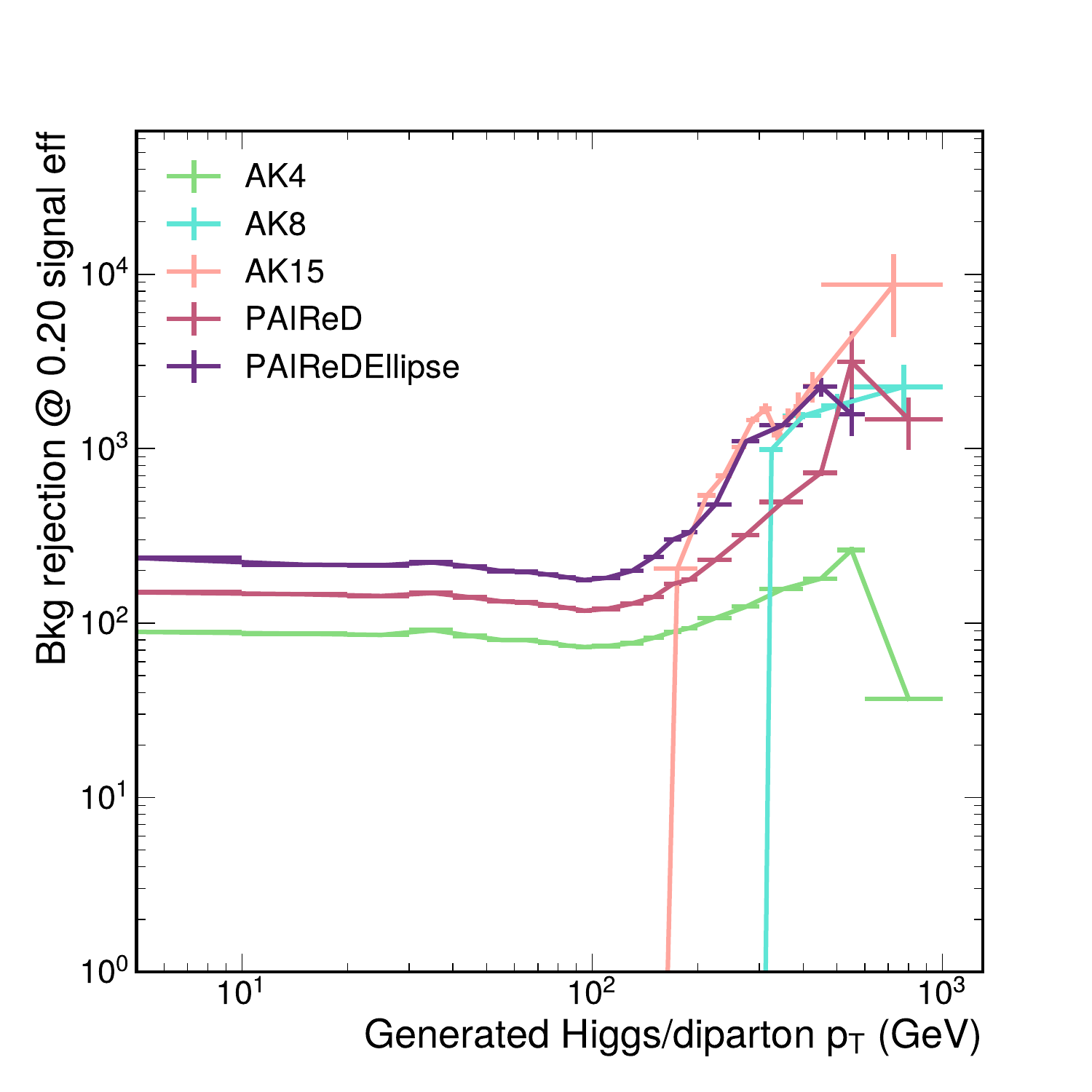}\includegraphics[width=0.5\textwidth]{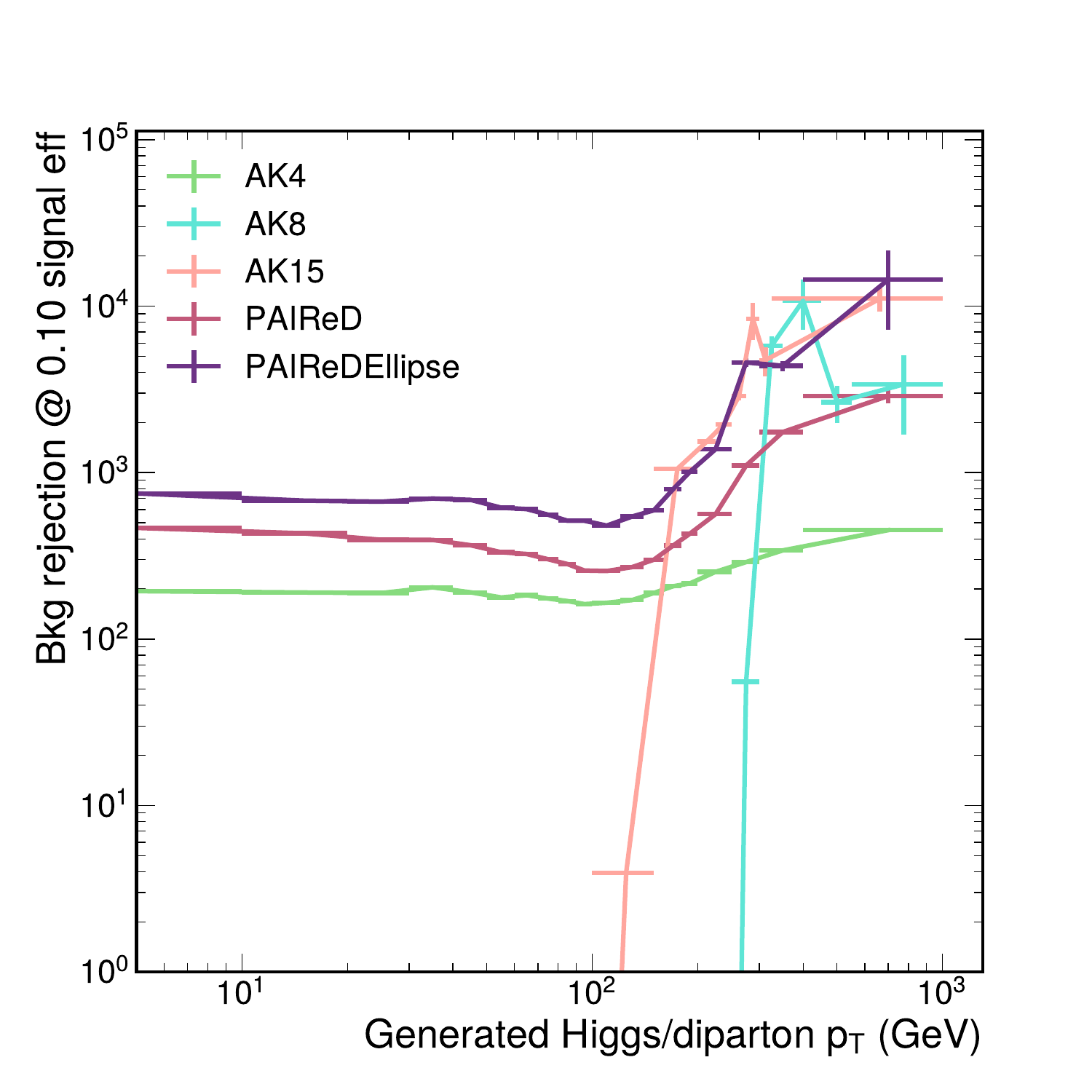}
\par\end{centering}
\caption{\label{fig:BkgRej}The end-to-end background rejection at end-to-end
signal efficiencies of 0.4 (upper-left), 0.3 (upper-right), 0.2 (lower-left),
and 0.1 (lower-right) as a function of the generated Higgs boson $p_{\text{T}}$
(for signal) or diparton $p_{\text{T}}$ (for background), plotted
for various reconstruction strategies. End-to-end signal efficiencies
and background rejections are calculated per bin. The background rejection
is set to 0 wherever the target signal efficiency cannot be achieved.
Bins with very low yield of background events are either merged with
adjacent bins or are omitted. }
\end{figure}

The end-to-end background rejection rates at a given target end-to-end
signal efficiency for different reconstruction strategies are discussed
in this subsection. Figure \ref{fig:MaxSigEff} (right) shows the
maximum signal efficiency achievable at different boosts of the generated
Higgs boson. As AK4 jets achieve a maximum signal efficiency of around
\textasciitilde 0.37 at low Lorentz-boosts, requiring a target signal
efficiency above \textasciitilde 0.37 is expected to yield no sensitivity
with the AK4-based reconstruction approach at low Lorentz-boosts.
Nevertheless, end-to-end signal efficiencies of 0.4, 0.3, 0.2, and
0.1 are explored and the corresponding background rejection rates
as a function of the Higgs/diparton $p_{\text{T}}$ are shown in Fig.
\ref{fig:BkgRej}. The end-to-end signal efficiencies and background
rejections are calculated per bin, i.e. exclusively in the Higgs/diparton
$p_{\text{T}}$ range covered by each bin. 

At low Lorentz-boosts of the Higgs boson, the PAIReDEllipse strategy
performs the best with around 2.5-4 times higher background rejection
compared to the AK4 strategy. The PAIReD strategy has a performance
intermediate to that of AK4 and PAIReDEllipse strategies, with around
2-3 times improvement compared to AK4-based reconstruction. Both AK4
and PAIReD strategies fail to achieve signal efficiencies of 0.4 at
Higgs boson boosts of \textasciitilde 25--150 GeV (cf. Fig. \ref{fig:MaxSigEff}
(right)) resulting in a dip in the observed background rejection at
a signal efficiency of 0.40 (Fig. \ref{fig:BkgRej}, upper-left);
this makes PAIReDEllipse the only strategy to achieve a non-zero background
rejection at 0.40 signal efficiency in this boost regime. At higher
boosts, the PAIReDEllipse and AK15 approaches have similar performances.
The AK8 approach has comparable performances at low signal efficiencies,
but performs slightly worse than PAIReDEllipse and AK15 at higher
signal efficiency requirements.

\subsubsection{Event-level classification\label{subsec:Event-level-classification}}

Going beyond using only jet tagger scores for signal vs. background
event classification, analyses generally make use of additional event-level
observables to improve signal-background separation. Jet kinematics,
Higgs boson candidate mass and $p_{\text{T}}$, lepton kinematics,
and Z boson candidate kinematics, e.g., may provide additional information
to differentiate ZH(H$\rightarrow\text{c}\bar{\text{c}}$) events
from Z+jj background. Similar to Refs. \citep{Sirunyan:2020aa,VHcc},
event-level BDTs are trained to distinguish signal events from background
events. The BDTs use the jet tagger scores along with various kinematic
quantities associated with events as inputs. Observables related to
the leptons and Z boson candidate are not used as inputs to the BDT
for simplicity. Five different BDTs, corresponding to the five different
event reconstruction approaches, are trained using different sets
of observables relevant to the corresponding event reconstruction
approach. Table \ref{tab:BDT inputs} shows the sets of event-level
observables used as inputs to the BDTs. It can be noted that unlike
performance metrics discussed until Sec. \ref{subsec:End-to-end-efficiencies},
the event-level BDT discussed here is not agnostic to the mass of
the Higgs boson candidate by design.

\begin{table}
\begin{centering}
\renewcommand{\arraystretch}{1.6}
\begin{tabular}{c|>{\centering}p{0.1\textwidth}>{\centering}p{0.1\textwidth}>{\centering}p{0.1\textwidth}>{\centering}p{0.1\textwidth}>{\centering}p{0.1\textwidth}}
\textbf{Observable} & \textbf{AK4} & \textbf{AK8} & \textbf{AK15} & \textbf{PAIReD} & \textbf{PAIReD\-Ellipse}\tabularnewline
\hline 
AK4 jet 1 kinematics & $\checkmark$ &  &  & $\checkmark$ & $\checkmark$\tabularnewline
AK4 jet 1 CvsL score & $\checkmark$ &  &  &  & \tabularnewline
AK4 jet 1 CvsB score & $\checkmark$ &  &  &  & \tabularnewline
AK4 jet 2 kinematics & $\checkmark$ &  &  & $\checkmark$ & $\checkmark$\tabularnewline
AK4 jet 2 CvsL score & $\checkmark$ &  &  &  & \tabularnewline
AK4 jet 2 CvsB score & $\checkmark$ &  &  &  & \tabularnewline
PAIReD/Fat jet kinematics &  & $\checkmark$ & $\checkmark$ &  & \tabularnewline
PAIReD/Fat jet CCvsLL &  & $\checkmark$ & $\checkmark$ & $\checkmark$ & $\checkmark$\tabularnewline
PAIReD/Fat jet CCvsBB &  & $\checkmark$ & $\checkmark$ & $\checkmark$ & $\checkmark$\tabularnewline
Higgs candidate mass (dijet mass) & $\checkmark$ &  &  &  & \tabularnewline
Higgs candidate mass (regressed) &  & $\checkmark$ & $\checkmark$ & $\checkmark$ & $\checkmark$\tabularnewline
Higgs candidate $p_{\text{T}}$ & $\checkmark$ & ($\checkmark$) & ($\checkmark$) &  & \tabularnewline
\end{tabular}
\par\end{centering}
\caption{\label{tab:BDT inputs}List of inputs to the various event-level BDTs.
Jet kinematics refer to $p_{\text{T}}$, $\eta$, $\phi$, and energy
of the jet. The Higgs candidate mass refers to the regressed mass
in case of fat and PAIReD(Ellipse) jets and dijet invariant mass in
case of AK4 jets. The Higgs candidate $p_{\text{T}}$ and fat jet
$p_{\text{T}}$ are equivalent in case of fat jets. CvsB is defined
as $\text{CvsB}=\frac{P(\text{c)}}{P(\text{c})+P(\text{b)}}$, in
the same fashion as in eq. \ref{eq:CvsL}. Similarly CCvsLL and CCvsBB
are defined as $\text{CCvsLL}=\frac{P(\text{cc)}}{P(\text{cc})+P(\text{ll)}}$
and $\text{CCvsBB}=\frac{P(\text{cc)}}{P(\text{cc})+P(\text{bb)}}$.}
\end{table}

The BDTs are implemented as Gradient Boosting Machines \citep{10.1214/aos/1013203451}
using the \noun{XGBoost} library \citep{Chen_2016} in Python \citep{10.5555/1593511},
while tools in the \noun{Scikit-Learn} package \citep{scikit-learn}
are used for data handling. Half (\textasciitilde 5M ZH(H$\rightarrow\text{c}\bar{\text{c}}$)
with $m_{H}=125$ GeV and \textasciitilde 5M Z+jj events) of the
generated dataset (referred to as \emph{test dataset} in Sec. \ref{sec:Neural-network-training})
is considered for training the BDTs. Of these events, only those passing
the event selections for a given event reconstruction approach are
used in training the corresponding BDT. Each model is trained to minimize
the fraction of misclassified instances (error rate) during training.

\begin{figure}
\begin{centering}
\includegraphics[width=0.5\textwidth]{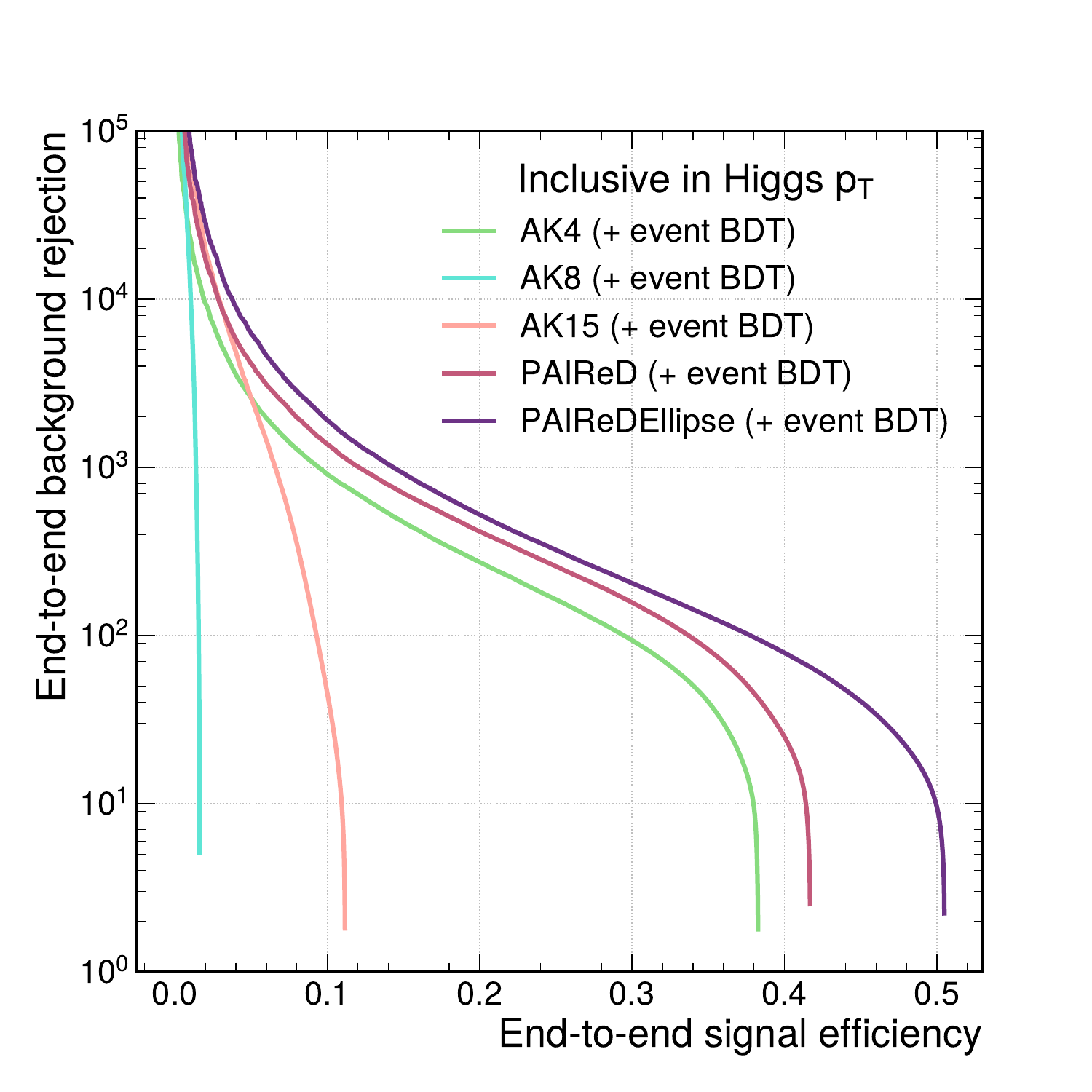}\includegraphics[width=0.5\textwidth]{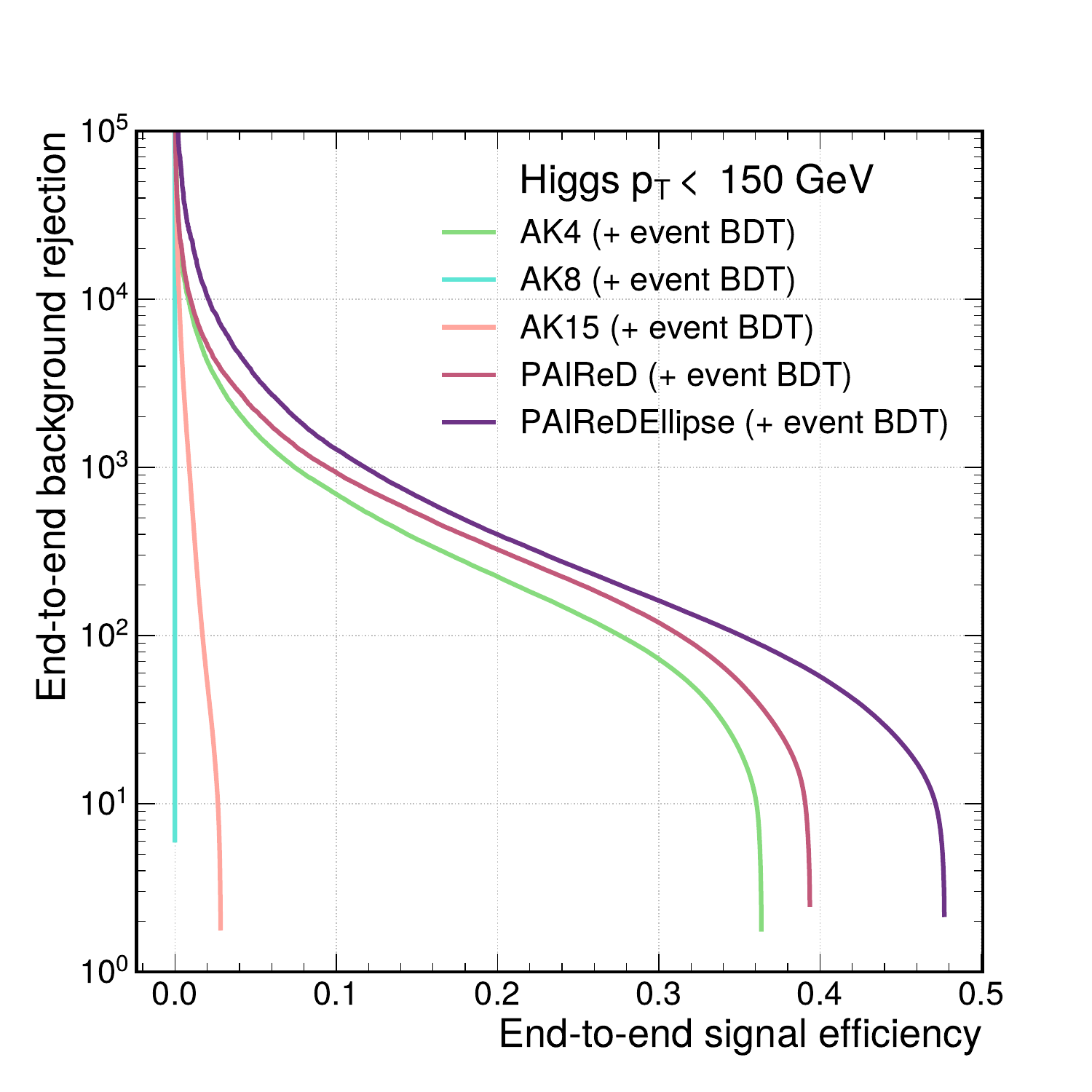}
\par\end{centering}
\begin{centering}
\includegraphics[width=0.5\textwidth]{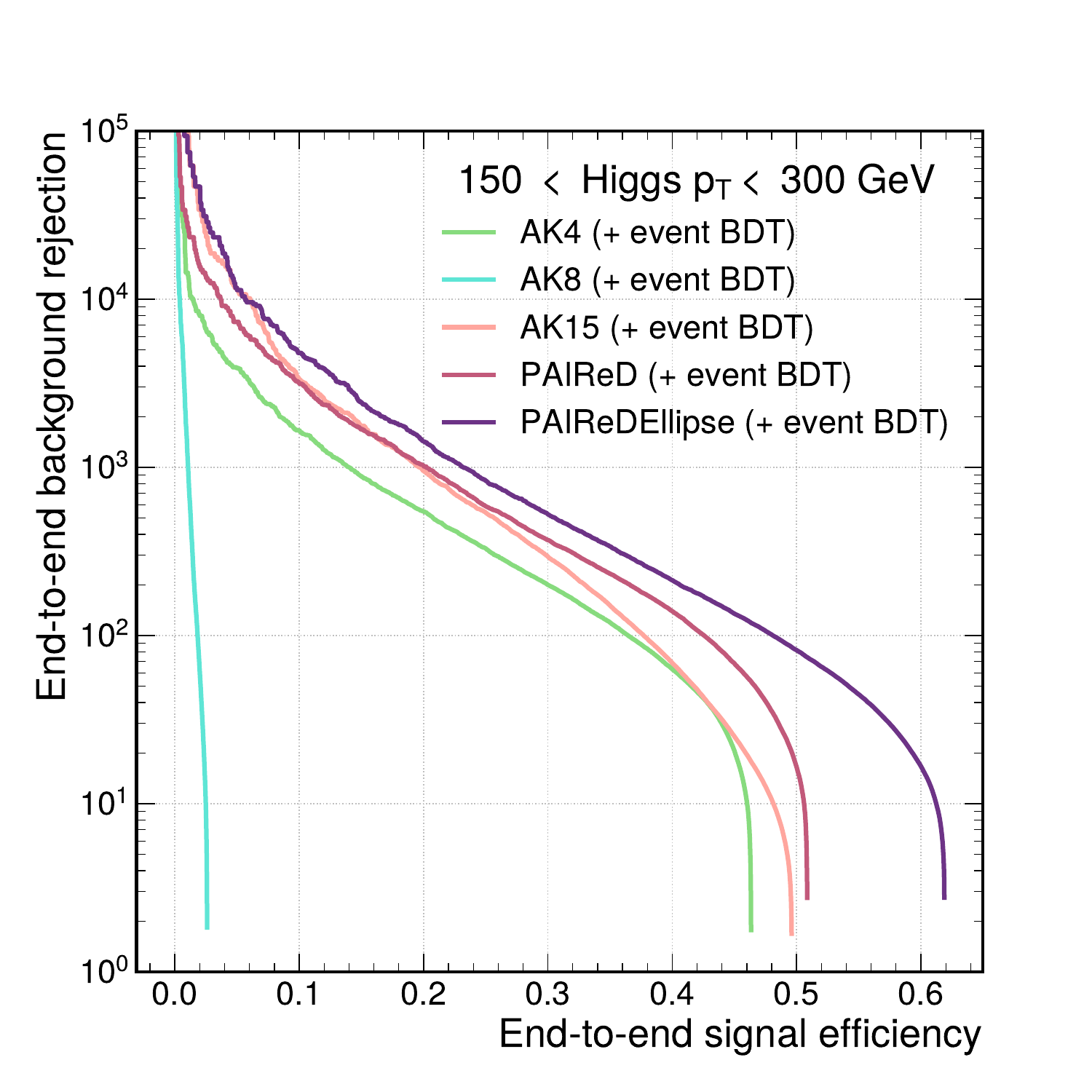}\includegraphics[width=0.5\textwidth]{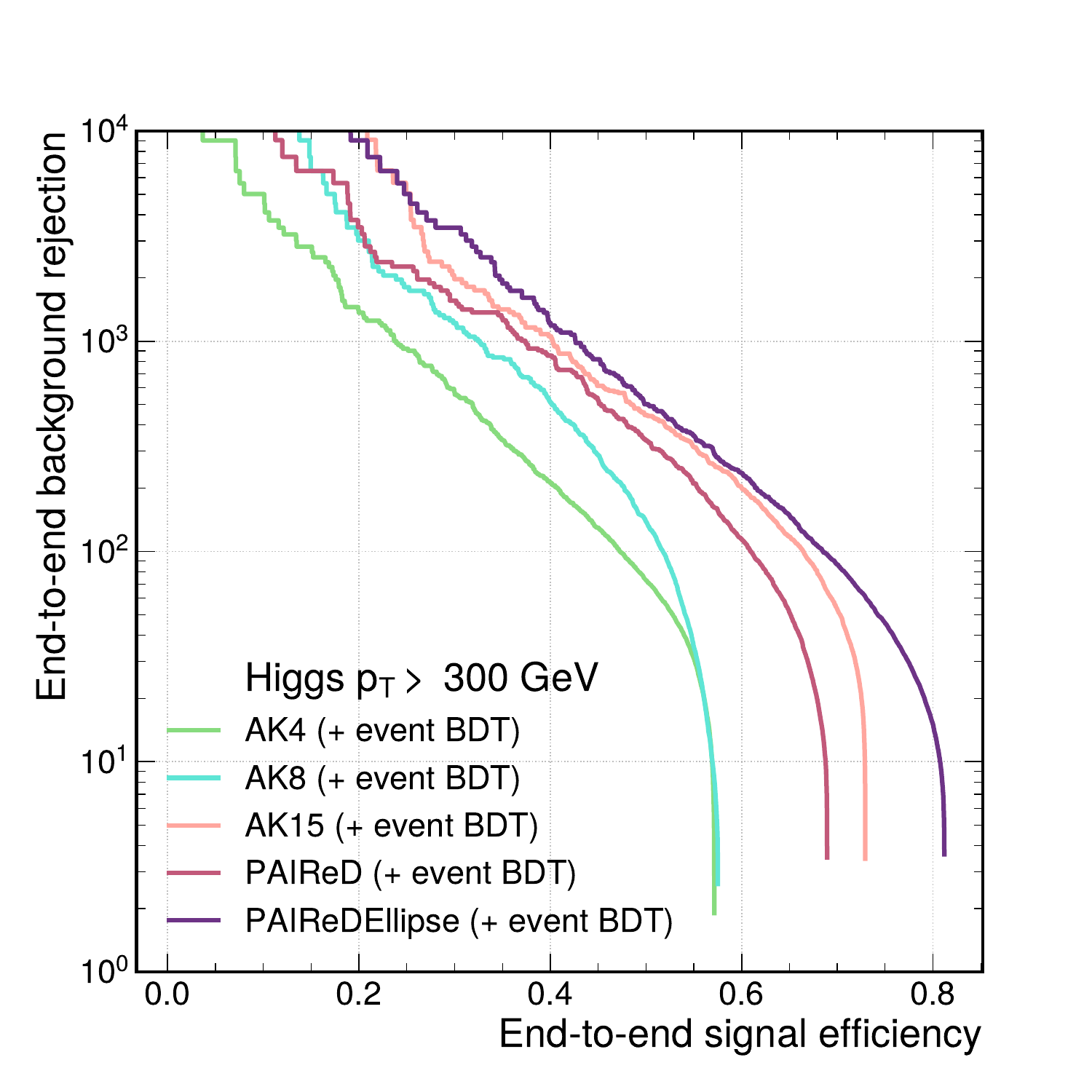}
\par\end{centering}
\caption{\label{fig:BDT}The end-to-end background event rejection rate as
a function of the end-to-end signal efficiency inclusively in Higgs
boson (or diparton) $p_{\text{T}}$ (upper-left), at low ($p_{\text{T}}$
< 150 GeV) $p_{\text{T}}$ ranges (upper-right), at medium (150 <
$p_{\text{T}}$ < 300 GeV) $p_{\text{T}}$ ranges (lower-left), and
at high ($p_{\text{T}}$ > 300 GeV) $p_{\text{T}}$ ranges (lower-right),
as predicted by the event-level BDT. The highest value along the x-axis
reached by each curve represents the maximum signal reconstruction
efficiency achievable using the respective event reconstruction strategy
(cf. Sec. \ref{subsec:Signal-reconstruction-efficiency}), while the
lowest value along the y-axis for each curve represents the trivial
background rejection achieved from reconstructing the event with reconstructed
physics objects satisfying quality criteria (i.e. without any selection
criterion on the BDT output).}
\end{figure}

The trained BDT models are then evaluated on the remaining half of
the\emph{ test dataset}. For each event reconstruction approach, the
end-to-end background event rejection rate as predicted by the corresponding
BDT is plotted as a function of the end-to-end signal efficiency.
Figure \ref{fig:BDT} shows the performance of these event-level BDTs
at various ranges of the Higgs $p_{\text{T}}$. Inclusively in $p_{\text{T}}$,
the PAIReDEllipse approach achieves 2--2.5 times better background
rejection compared to the AK4 approach, depending on the target signal
efficiency, while the PAIReD approach has an intermediate performance.
The AK15 and AK8 approaches do not play a significant role in low
$p_{\text{T}}$'s, but surpass the AK4-based approach at high $p_{\text{T}}$'s.
At all $p_{\text{T}}$ ranges, the BDT corresponding to the PAIReDEllipse
approach exhibits the best classification performance. The performance
of the BDTs as a function of the Higgs $p_{\text{T}}$ and at different
jet multiplicities, along with some discussions, is presented in Appendix
\ref{sec:Event-level-BDT-performance}.

An additional variant of the AK4-based BDT was trained using the kinematic
information of a third jet in the event (whenever reconstructed),
and a second Higgs mass input reconstructed from the four-vectors
of three AK4 jets. This was done with the motivation to capture a
possible third jet arising from final state radiations. However, this
BDT performed only marginally better (\textasciitilde 0.03\% improvement
in the accuracy metric) compared to the nominal AK4-based BDT and
hence was not used.

\section{Outlook \label{sec:Outlook}}

The PAIReD(Ellipse) approaches potentially leverage information across
a larger volume of the detector, especially at low boosts, while the
related observables are still calibratable using modified versions
of the standard jet calibration techniques used in hadron collider
experiments. The approach can be generalized to multi-pronged decays
as well, as long as suitable event samples can be identified in collision
data for calibration. A three-pronged tagger, for example, may be
calibrated using hadronically-decaying top quark events.

Furthermore, more complex final states, like 4b, 6b, 8b, and so on,
can potentially be factorized into two or more two-pronged PAIReD(Ellipse)
jets for reconstruction. The classifier output scores are expected
to be higher for the correct pairs of b quark resonances, which can
help identify the correct combinations of small-radius jets potentially
in conjunction with jet assignment algorithms \citep{Erdmann:2013rxa,Erdmann:2019evj,Shmakov:2021qdz}.
Tagger calibration for such complex final states can be achieved by
evaluating correction factors for simulated events as a function of
the efficiency correction factors of all PAIReD(Ellipse) jets reconstructed
in the event. The strategy may also be used to discriminate events
containing resonant heavy-flavor decays (H$\rightarrow$$\text{b\ensuremath{\bar{\text{b}}}}$,
for example) from those containing non-resonant decays (b and $\bar{\text{b}}$
from $\text{t}\bar{\text{t}}$ decay) at low boosts, without using
the invariant mass information. These extended use cases, however,
require additional studies and may be studied systematically for different
high-multiplicity final states in the future.

The method can also be suitably modified to identify, for example,
charged Higgs boson decays into c$\bar{\text{s}}$, by introducing
additional cs flavor nodes in the outputs. Such a classifier can be
calibrated using hadronically-decaying W bosons from $\text{t}\bar{\text{t}}$
events, for instance. Thus, the PAIReD and PAIReDEllipse approaches
may be extended to several kinds of physics analyses.

\section{Conclusion\label{sec:Conclusion}}

We propose a new method to reconstruct events containing hadronically-decaying
heavy particles. The two variants of the method, namely PAIReD and
PAIReDEllipse jet based reconstruction, define unconventional jets
using clustered small-radius jets as seeds. The former is limited
to the area defined by the small-radius jets, whereas the latter defines
an elliptical area in the $\eta$--$\phi$ plane based on the centers
of the small-radius jets and considers all reconstructed particles
lying in this area as jet constituents.

We have demonstrated this method in the context of di-pronged resonances,
where the PAIReD\-(Ellipse) jets are constructed from exactly two
AK4 jets. We have examined two rare physics processes, namely Higgs
bosons decaying to a bottom or charm quark-antiquark pair, where the
Higgs boson is produced in association with a Z boson (ZH(H$\rightarrow$$\text{b\ensuremath{\bar{\text{b}}}}$/$\text{c}\bar{\text{c}}$)).
We have used the ParticleTransformer jet tagging algorithm for regression
and classification tasks. We have trained mass-decorrelated multi-classifiers
to discriminate $\text{b\ensuremath{\bar{\text{b}}}}$, $\text{c}\bar{\text{c}}$,
and light-flavor jets (b, c and udsg, for AK4 jets).

The PAIReDEllipse strategy has the highest signal reconstruction efficiency
at all Lorentz-boosts of the Higgs boson. The AK4, PAIReD, and PAIReDEllipse
approaches reconstruct significantly more signal events than AK15
and AK8 approaches, while the AK15 and AK8 surpass the AK4-based approach
in terms of signal efficiency at high boosts. In terms of classifier
performance, the PAIReDEllipse and PAIReD perform better than AK4
at all boosts, while the AK15 and AK8 classifiers surpass all others
at high boosts. The combined effect results in the PAIReDEllipse approach
having a background rejection about 2.5--4 times higher than the
AK4-based approach at low boosts, while the PAIReD approach performs
about 2--3 times better than the AK4-based approach. At higher boosts,
the PAIReDEllipse approach has a performance at par with the AK15-based
approach.

The previous state-of-the-art analysis techniques involve using AK4-based
jet reconstruction at low Higgs boson boosts ($p_{\text{T}}^{\text{H}}\apprle$
300 GeV) and using AK15-based reconstruction at high boosts. Our proposed
PAIReDEllipse tagger improves the background rejection at low boosts
by up to 4 times compared to AK4-based taggers, while maintaining
the same performance as AK15-based taggers at high boosts. The main
gain in performance at low boosts comes from the ability of PAIReD(Ellipse)-based
classifiers to exploit correlations between the hadronization products
of the two quarks---a feat that was so far achieved only at high
boosts using the substructure information of large-radius jets. The
new method is expected to significantly improve the sensitivities
of physics searches in the low-boost regime and enable searches that
were previously considered inaccessible at low boosts.

\section*{Acknowledgements}

We are grateful to the authors of Ref. \citep{Qu:2022mxj}, H. Qu,
C. Li, and S. Qian for many helpful discussions. We are also grateful
to them for publishing and sharing codes which we adapted for this
work. We would also like to thank U. Heintz and J. Luo for many fruitful
discussions. This research was supported by the Deutsche Forschungsgemeinschaft
(DFG) under grant 400140256 - GRK 2497: The physics of the heaviest
particles at the LHC, and the U.S. Department of Energy, Office of
Science, Office of Basic Energy Sciences Energy Frontier Research
Centers program under Award Number DE-SC0010010. Part of this research
was conducted using computational resources and services at the Center
for Computation and Visualization, Brown University.

\bibliographystyle{lucas_unsrt}
\bibliography{bib}
\pagebreak{} 

\appendix

\section{Mass regression\label{sec:Mass-regression}}

Before flavor classifiers are trained (cf. Sec. \ref{sec:Neural-network-training}),
regression networks are trained to define a regressed jet mass for
AK8, AK15, PAIReD, and PAIReDEllipse jets. The same ParT architecture,
with some modifications (cf. Sec. \ref{sec:Flavor-tagging-algorithm}),
is used. Defining a jet mass is necessary so that the classification
networks can be trained to be uncorrelated with respect to this mass
quantity \citep{CMS-DP-2020-002}. While regular mass terms like SoftDropped
mass \citep{Larkoski:2014aa} can be used in the case of AK8 and AK15
jets, they are not well-defined for jets that do not have a characteristic
jet radius (the soft-drop condition requires a radius parameter $R_{0}$,
see Ref. \citep{Larkoski:2014aa}) such as the PAIReD and PAIReDEllipse
jets. Therefore, for a fair comparison in subsequent steps, the regressed
mass ($m_{\text{reg}}$) is devised as a mass term that can be uniformly
defined for all four kinds of jets. Moreover, using the same technique
to regress mass values for all reconstruction strategies allows us
to compare the accuracy and resolution of the jet mass across all
strategies.

A mass regression network is trained each for AK8, AK15, PAIReD, and
PAIReDEllipse jets. Jets with labels bb and cc (cf. Sec. \ref{sec:Simulation})
sampled from the flat-mass ZH(H$\rightarrow\text{b\ensuremath{\bar{\text{b}}}}$)
and ZH(H$\rightarrow\text{c}\bar{\text{c}}$) processes are used in
the training. The mass of the generated Higgs boson is set as the
target of the regression. The jets are further assigned weights to
ensure that the distribution of the generated Higgs boson mass is
flat in the entire range of 10--500 GeV. 

About 13--15 million jets with label bb, and a similar number of
jets with label cc, are used in the training process. The exact number
varies across reconstruction strategies. About 90\% of these jets
are used in the training while the remaining are used as an independent
validation dataset to avoid overtraining. Each training is run for
50 epochs, which is deemed enough for the training to converge. The
model from the epoch with the lowest validation loss is chosen as
the final regression model.

\begin{figure}
\begin{centering}
\includegraphics[width=0.33\textwidth]{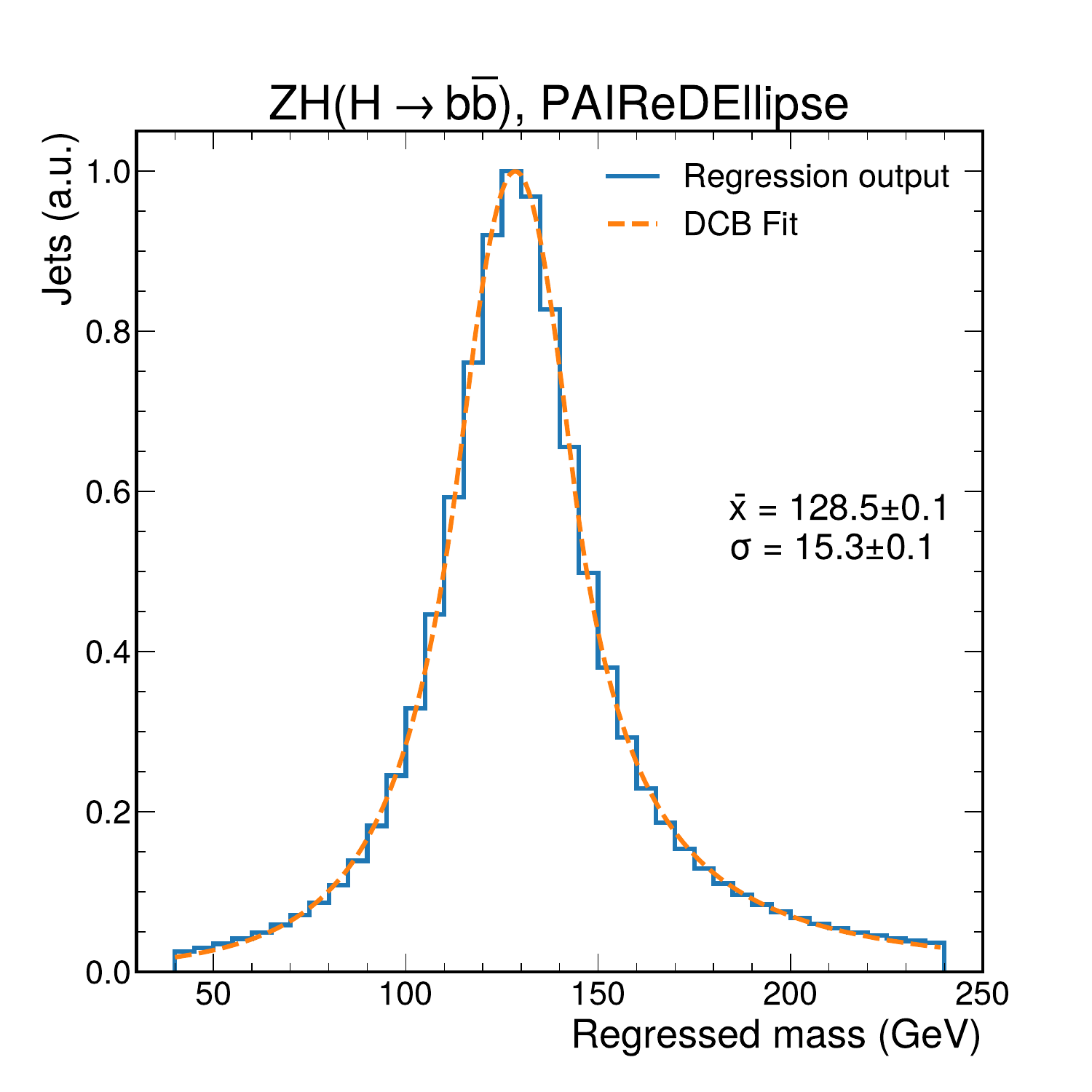}\includegraphics[width=0.33\textwidth]{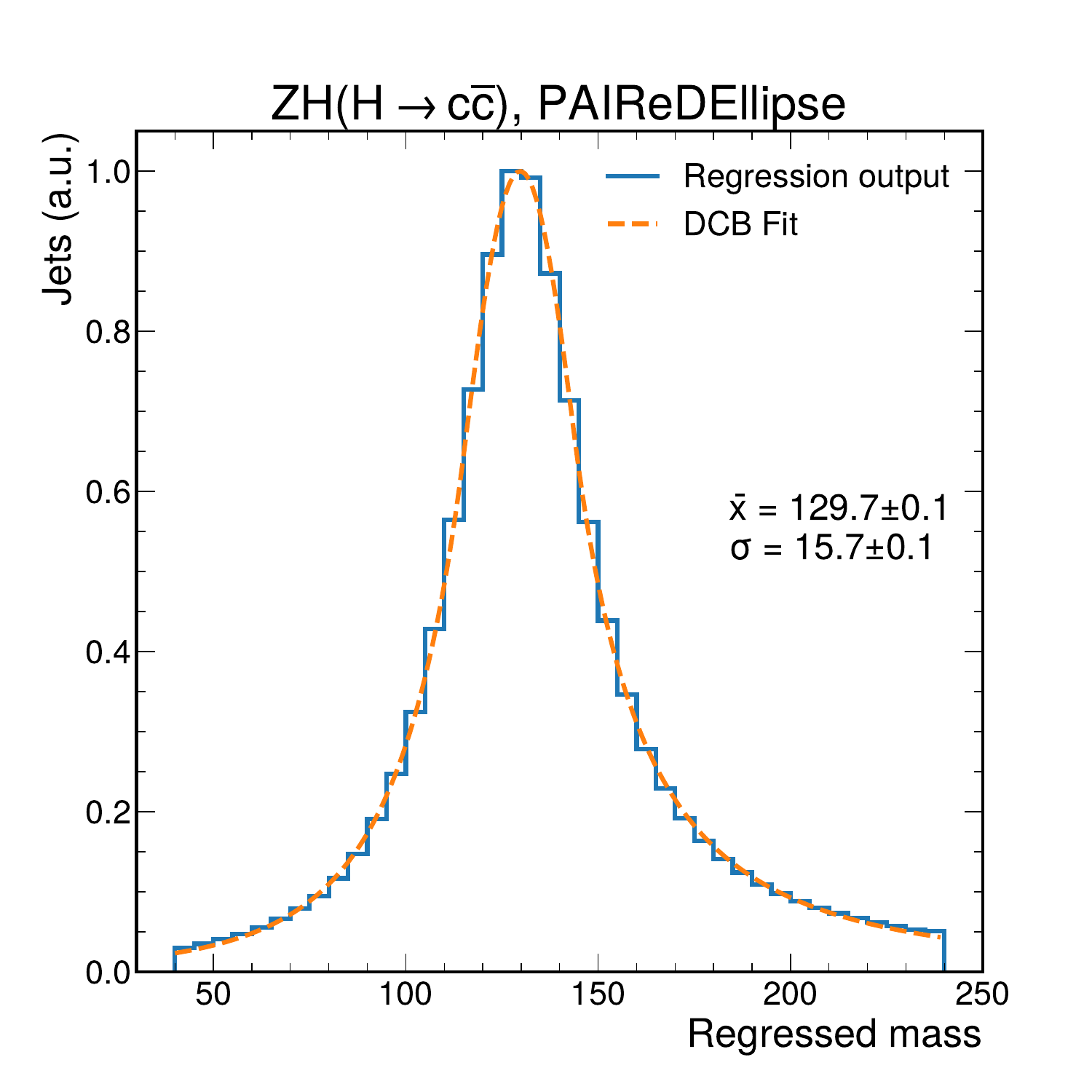}\includegraphics[width=0.33\textwidth]{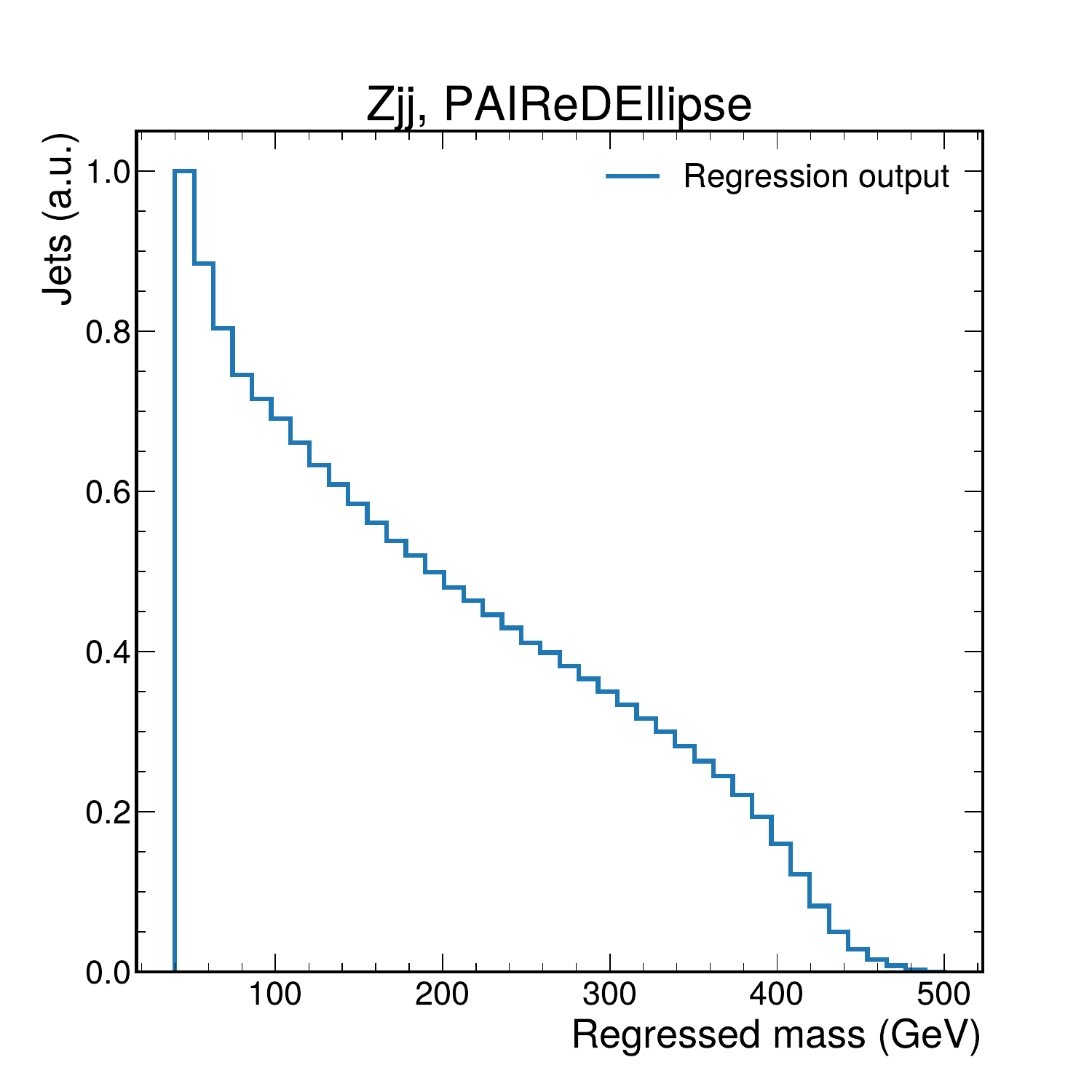}
\par\end{centering}
\caption{\label{fig:masspeaks}The distribution of the regressed mass obtained
from the PAIReDEllipse jet regression training evaluated on bb jets
from ZH(H$\rightarrow\text{b\ensuremath{\bar{\text{b}}}}$) (left),
cc jets from ZH(H$\rightarrow\text{c}\bar{\text{c}}$) (center), and
light-flavor jets from Z+jj (right) processes. The blue histograms
represent the network outputs, while the orange lines in the first
two plots show the results of a Double-sided Crystal Ball (DCB) fit.
The mean ($\bar{x}$) and spread ($\sigma$) parameters of the fit
are also displayed on the first two plots. The distributions are truncated
between 40--240 GeV for bb and cc jets, and 40-500 GeV for ll jets.}
\end{figure}

The trainings are evaluated on an independent test dataset comprising
jets from ZH(H$\rightarrow\text{b\ensuremath{\bar{\text{b}}}}$) and
ZH(H$\rightarrow\text{c}\bar{\text{c}}$) processes with $m_{\text{H}}=125$
GeV, and Z+jj processes. The evaluation results for the PAIReDEllipse
jet training are shown in Fig. \ref{fig:masspeaks} as an example.
As expected, the distribution of $m_{\text{reg}}$ peaks at around
the Higgs boson mass for the bb and cc samples. A Double-sided Crystal
Ball (DCB) fit is performed to extract the mean and spread of these
distributions. A mass resolution of about 15 GeV is obtained. The
distribution of $m_{\text{reg}}$ assumes a smoothly falling shape
for ll jets taken from Z+jj processes, as expected.

\begin{figure}
\begin{centering}
\includegraphics[width=0.5\textwidth]{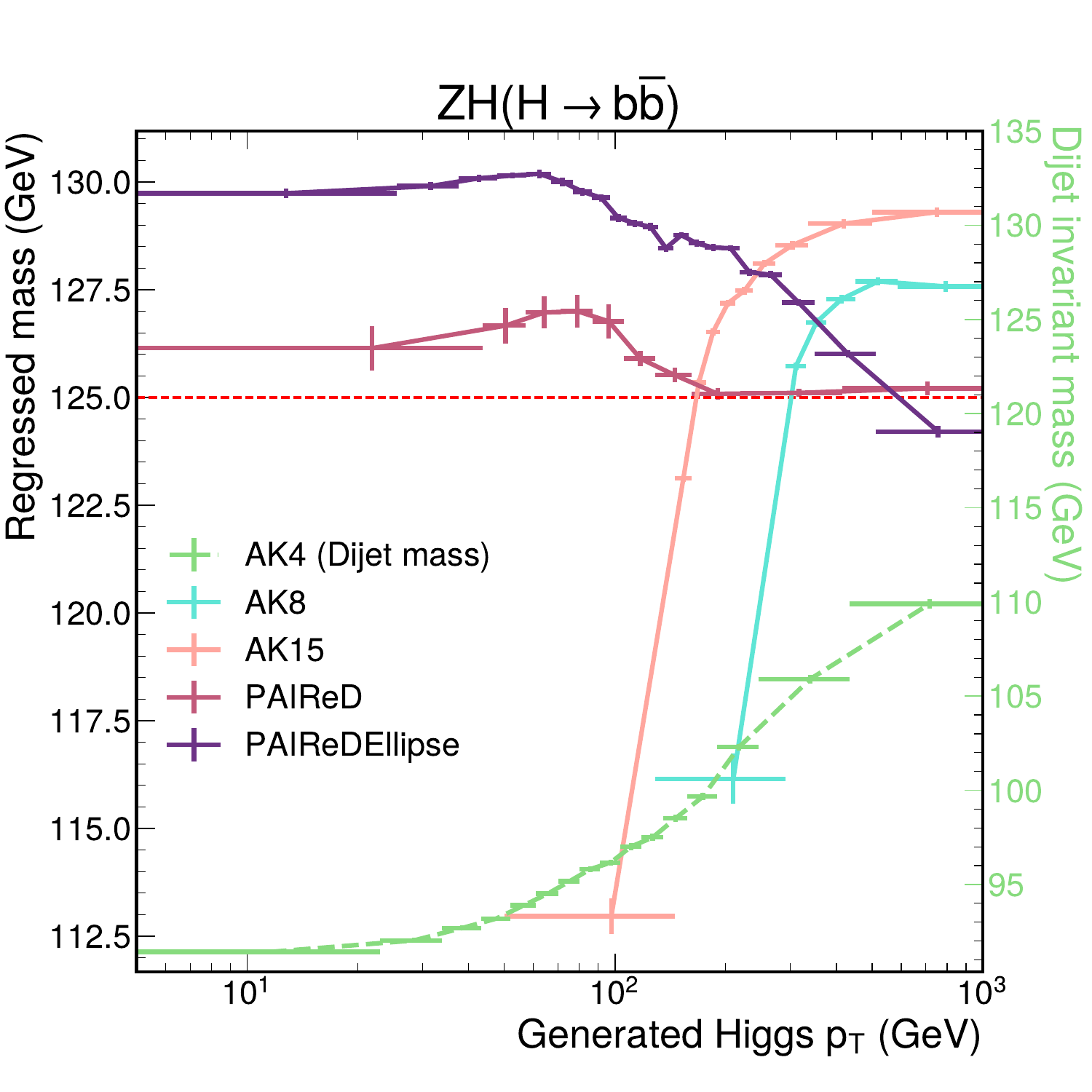}\includegraphics[width=0.5\textwidth]{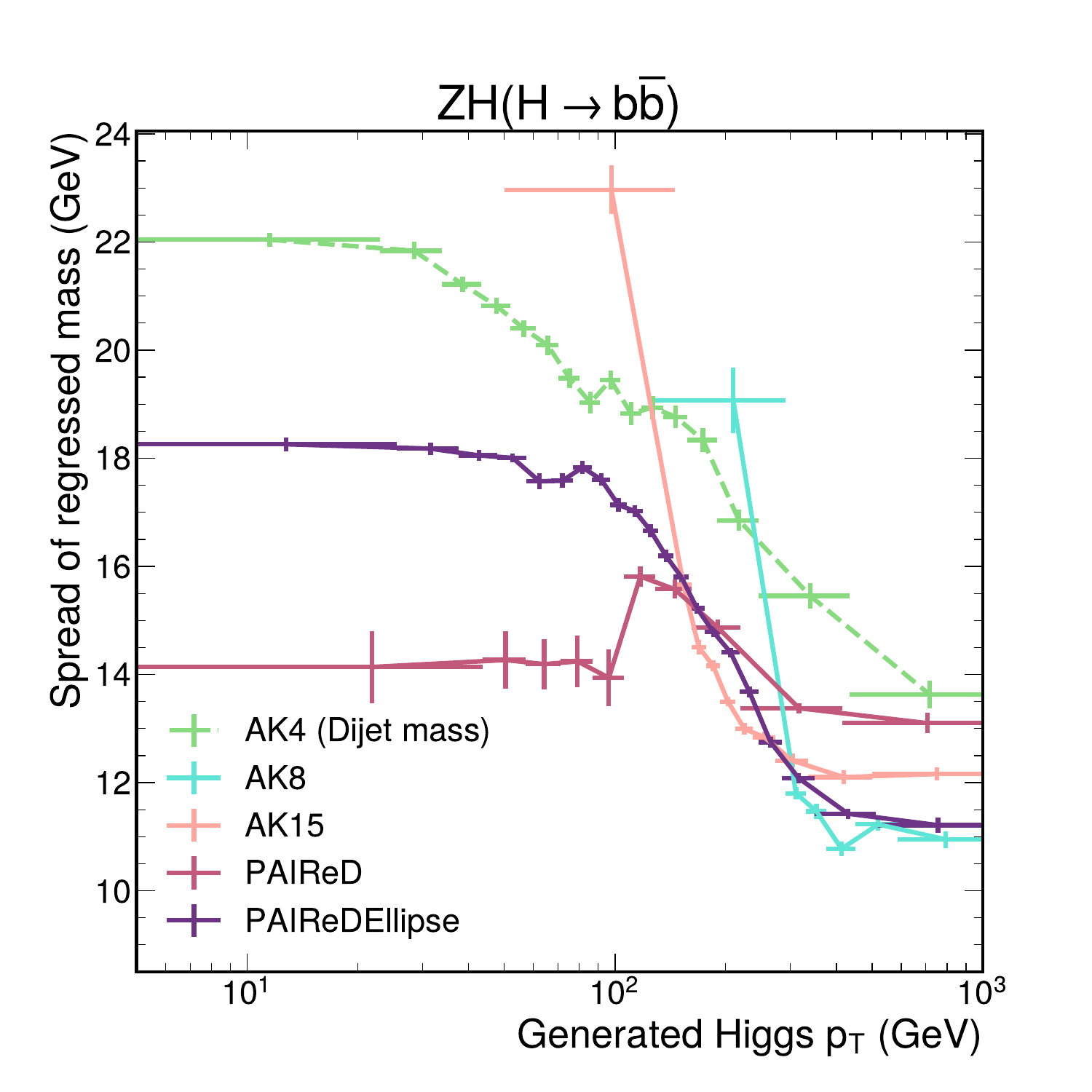}
\par\end{centering}
\begin{centering}
\includegraphics[width=0.5\textwidth]{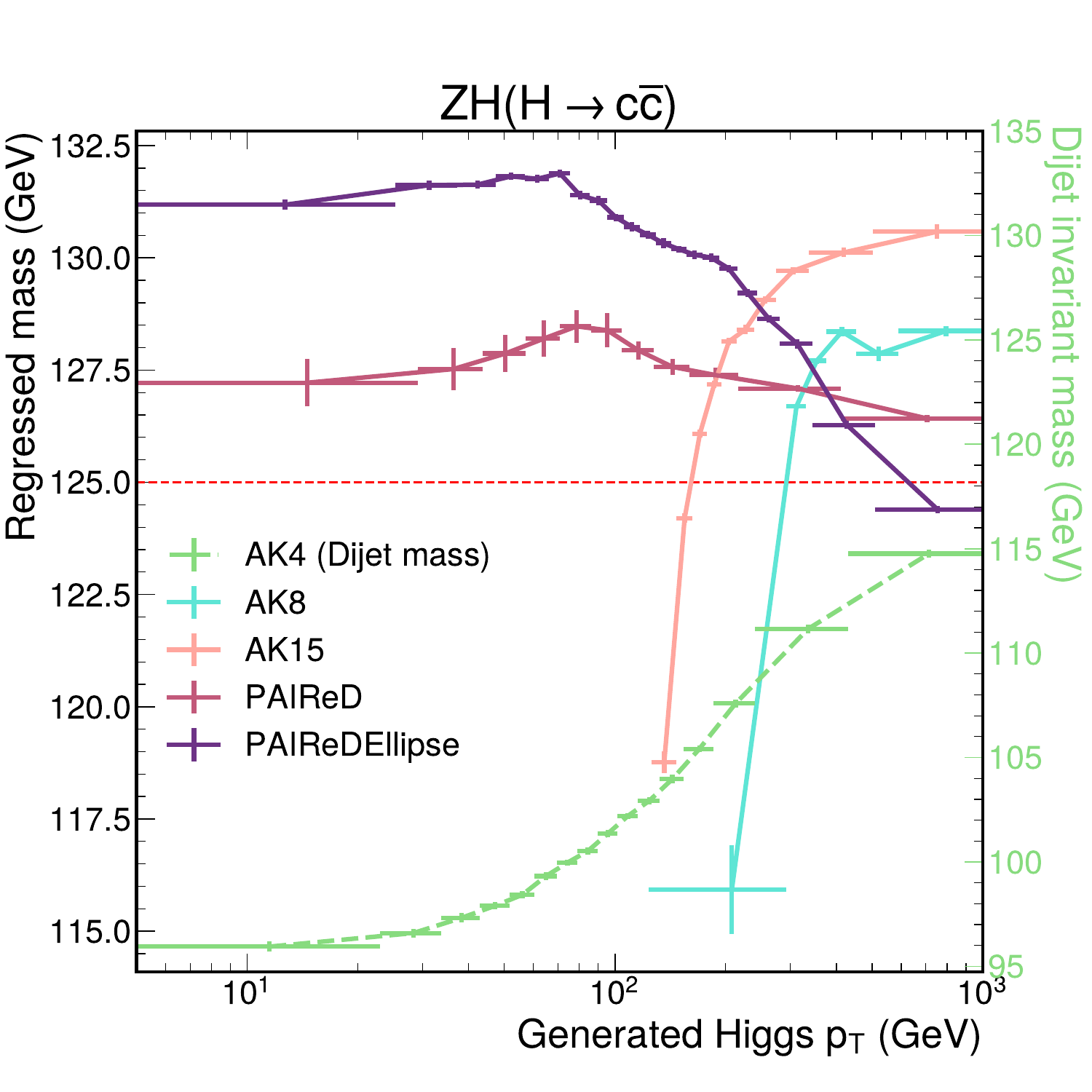}\includegraphics[width=0.5\textwidth]{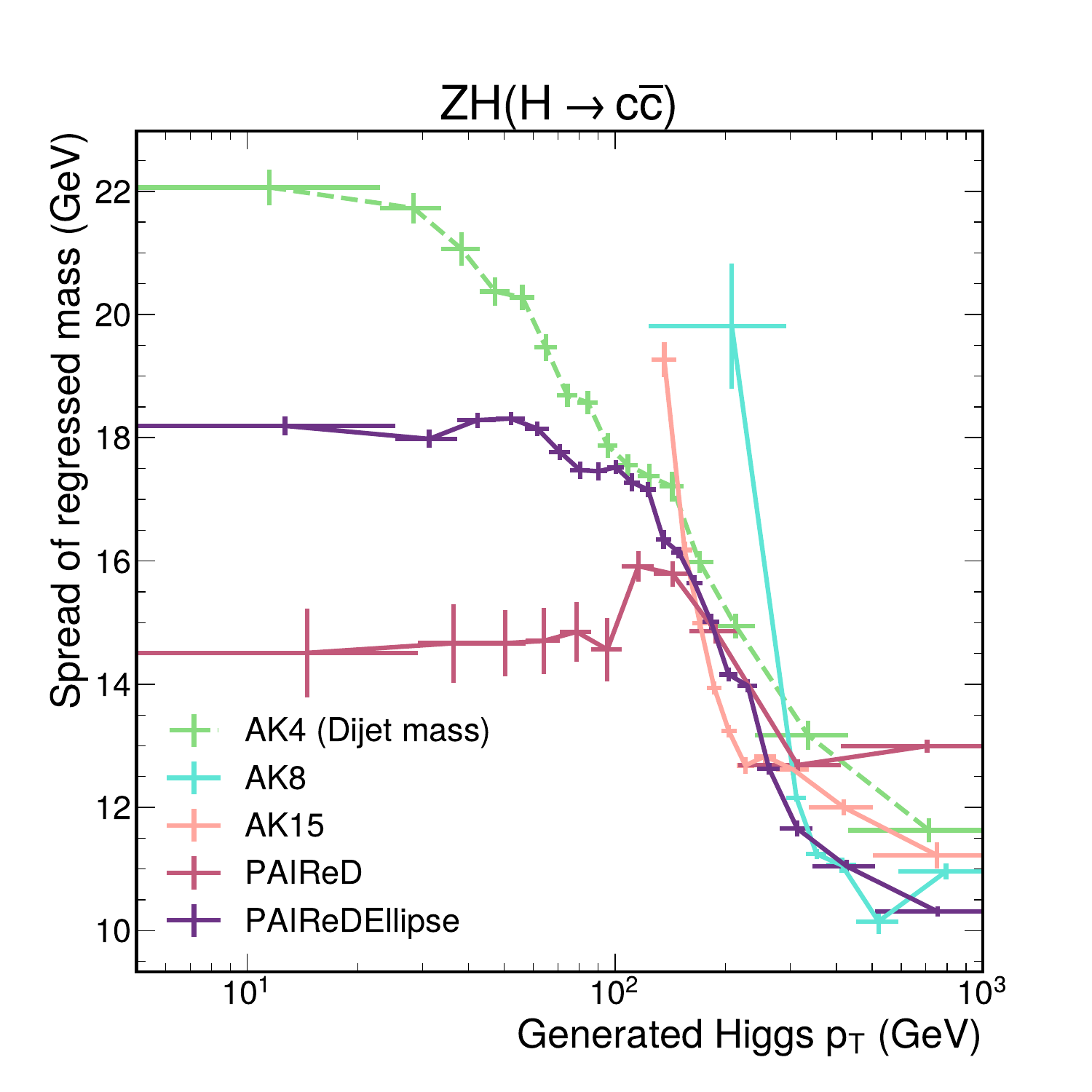}
\par\end{centering}
\caption{\label{fig:DCB}The mean (left) and spread (right) fit parameters
of the regressed mass as a function of the generated Higgs boson $p_{\text{T}}$
for different reconstruction strategies are shown. Uncertainty bands
indicate fit uncertainties. The binning for each strategy is chosen
in a way such that each bin contains approximately an equal number
of jets. The upper (lower) row shows results for bb (cc) jets taken
from ZH(H$\rightarrow\text{b\ensuremath{\bar{\text{b}}}}$) (ZH(H$\rightarrow\text{c}\bar{\text{c}}$))
processes. The fit results of the dijet invariant mass (the light-green
dashed lines) are also shown for the AK4-based reconstruction strategy.
The corresponding y-axis is shown on the right edge of the mean plot
(left) in light-green text.}
\end{figure}

It was observed that the mean and spread of the $m_{\text{reg}}$
distribution had a moderate dependence on the generated Higgs boson
$p_{\text{T}}$. The mass distributions were hence plotted for different
ranges of the generated Higgs boson $p_{\text{T}}$ and the DCB fit
was performed separately for every range. The mean and spread parameters
obtained from the fits are plotted as functions of the generated Higgs
boson $p_{\text{T}}$ in Fig. \ref{fig:DCB}. In general, a better
mass resolution is obtained at higher values of the Higgs boson $p_{\text{T}}$.
In all cases, the mean is compatible with the true mass of 125 GeV
within the respective resolutions.

Figure \ref{fig:DCB} also shows the fit results for the dijet invariant
mass in signal events using the AK4-based reconstruction strategy,
for reference. The AK4-based strategy exhibits a poorer mass resolution
compared to other strategies employing dedicated mass regression techniques.
However, a better mass resolution in AK4-based reconstruction is usually
achieved in physics analyses \citep{CMS:2018nsn,VHcc} using techniques
like b and c jet mass regression \citep{CMS:2019uxx} that evaluate
per-jet corrections. Such techniques have not been used in this paper.

The PAIReD approach achieves the best mass resolution at lower Lorentz-boosts
of the generated Higgs boson. The PAIReDEllipse approach has a slightly
poorer resolution, since PAIReDEllipse jets are expected to contain
a larger number of reconstructed particles from pileup and unrelated
hadronic activities compared to the PAIReD approach at high opening
angles. The AK8 and AK15 approaches either do not yield sufficient
jets or have a poor mass resolution at low boosts. This is because
they fail to encompass the full decay at low $p_{\text{T}}$'s. On
the other hand, the PAIReDEllipse, AK8, and AK15 approaches exhibit
similar mass resolutions at high Lorentz-boosts. 

\begin{figure}
\begin{centering}
\includegraphics[width=0.5\textwidth]{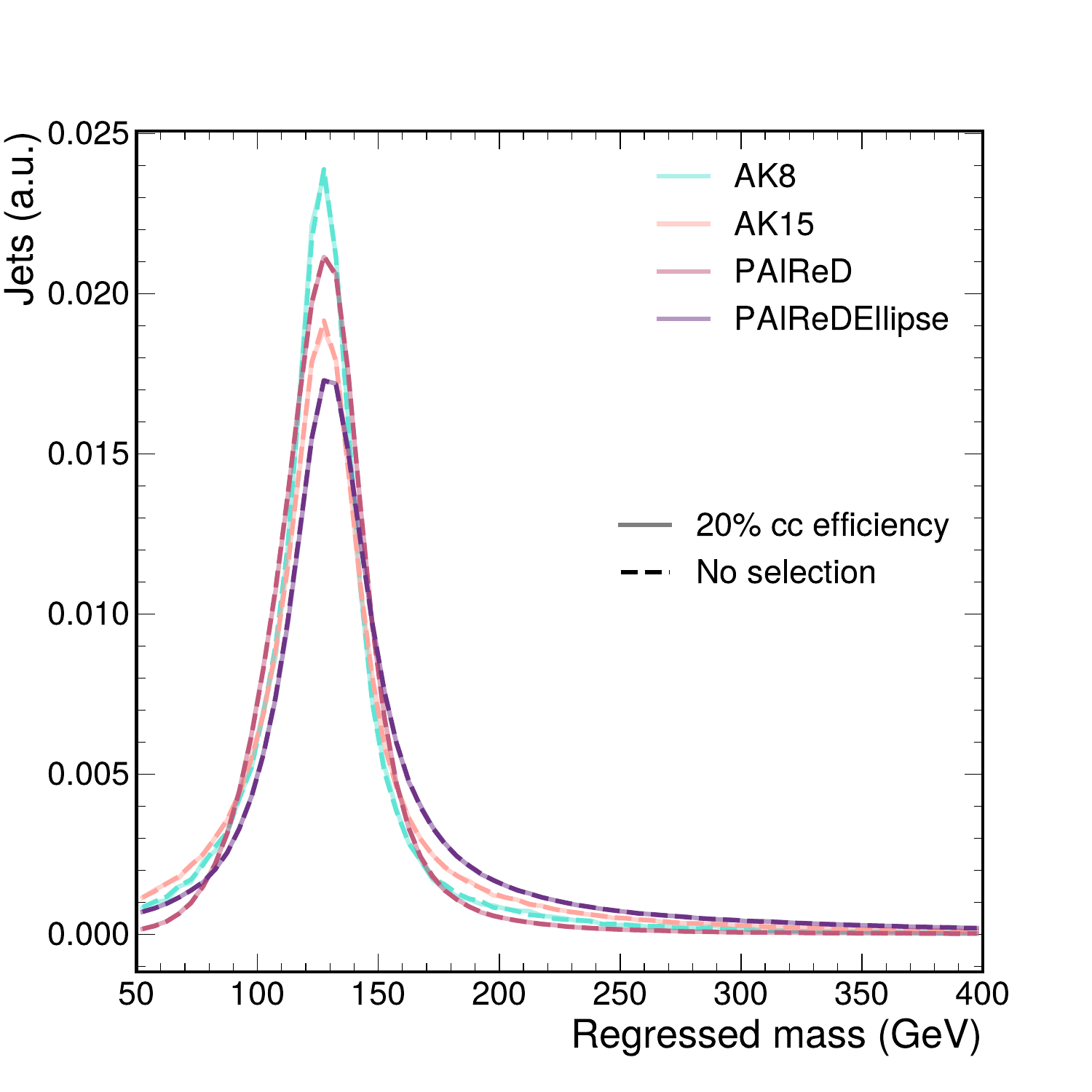}\includegraphics[width=0.5\textwidth]{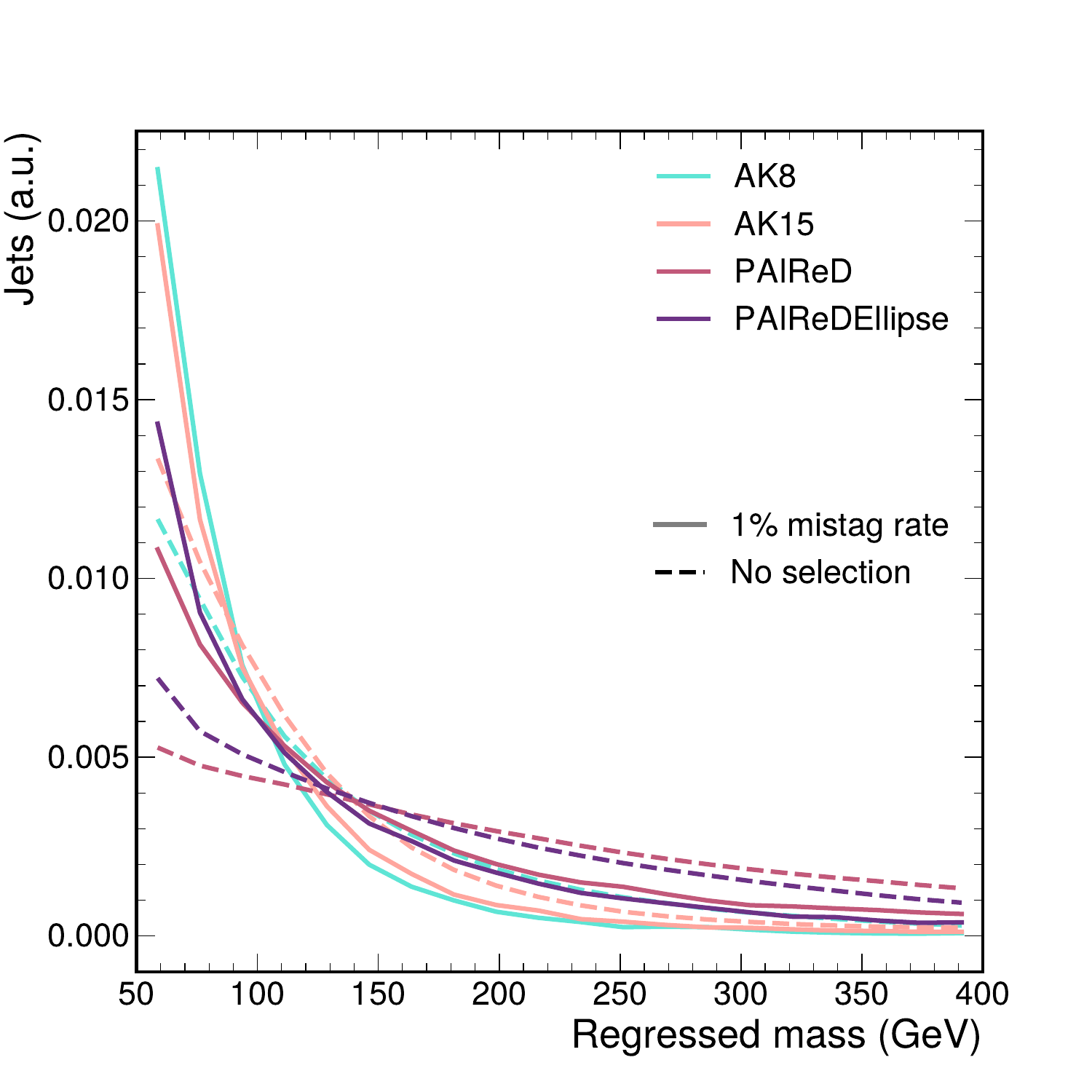}
\par\end{centering}
\caption{\label{fig:sculpt}The distribution of the regressed mass of the cc
signal (left) and ll background (right) jets before (the dashed lines)
and after (the solid lines) a selection criterion is applied on the
output scores of different classifiers. The selection criteria are
chosen such that exactly 20\% (1\%) of the signal (background) jets
are selected in each case. All curves are individually normalized
to unity. In the left plot, the pre- and post-selection curves for
the same kind of jet overlap.}
\end{figure}

To validate whether the jet flavor tagging classifiers trained in
this study introduce artificial peaks in the mass spectrum of background
jets---a phenomenon referred to as mass sculpting---the distribution
of $m_{\text{reg}}$ is plotted for background jets that satisfy a
tight selection criterion in the classifier output score. Figure \ref{fig:sculpt}
(right) shows the $m_{\text{reg}}$ spectrum of background jets corresponding
to various reconstruction strategies at a fixed ll mistag rate of
1\% achieved by the respective classifier, along with the corresponding
pre-selection distributions for reference. While the application of
a selection criterion on the tagger score moderately affects the overall
$m_{\text{reg}}$ distribution, it does not induce artificial peaks
in the background mass spectrum and the distribution continues to
have a smooth decline with increasing mass. On the other hand, the
signal jets shown in Fig. \ref{fig:sculpt} (left) have a mass spectrum
that sharply peaks at around the Higgs mass and remains unchanged
upon the application of tagger selections. Therefore, the signal and
background jets maintain a separation in their mass spectra even after
selections on classifier scores are applied. This ensures a fair comparison
among these strategies, as well as between these strategies and the
inherently mass-agnostic AK4-based reconstruction approach.

\begin{figure}
\begin{centering}
\includegraphics[width=0.5\textwidth]{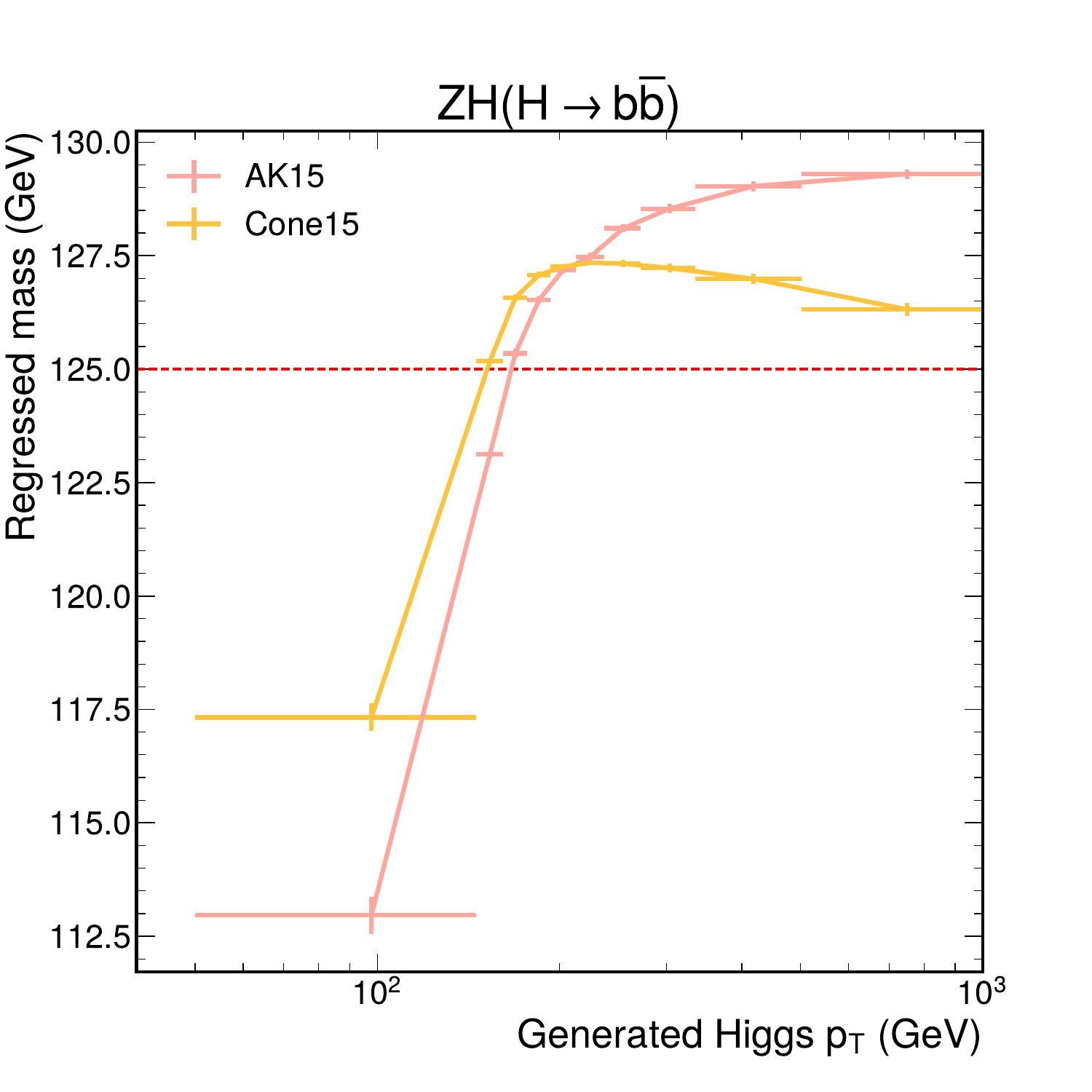}\includegraphics[width=0.5\textwidth]{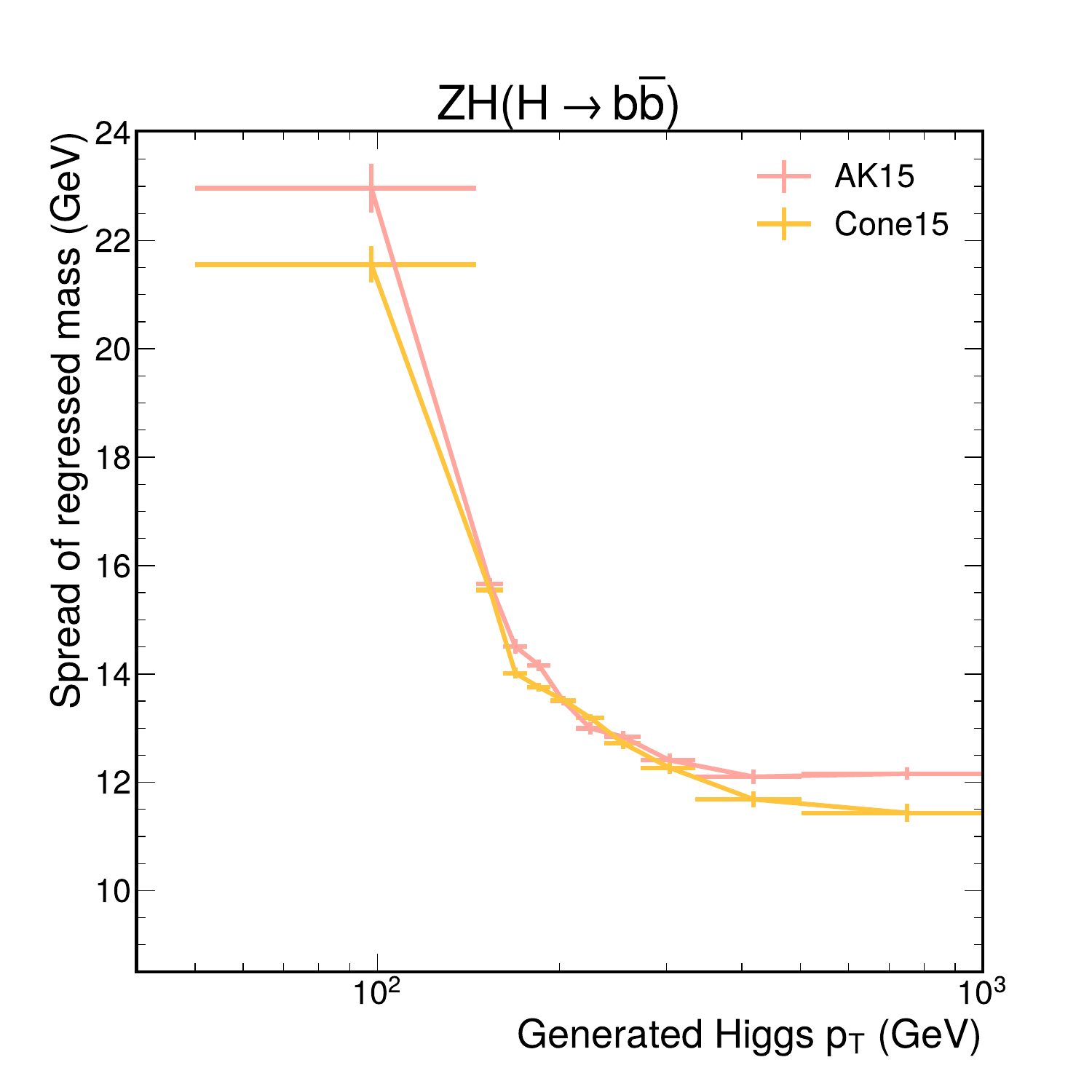}
\par\end{centering}
\begin{centering}
\includegraphics[width=0.5\textwidth]{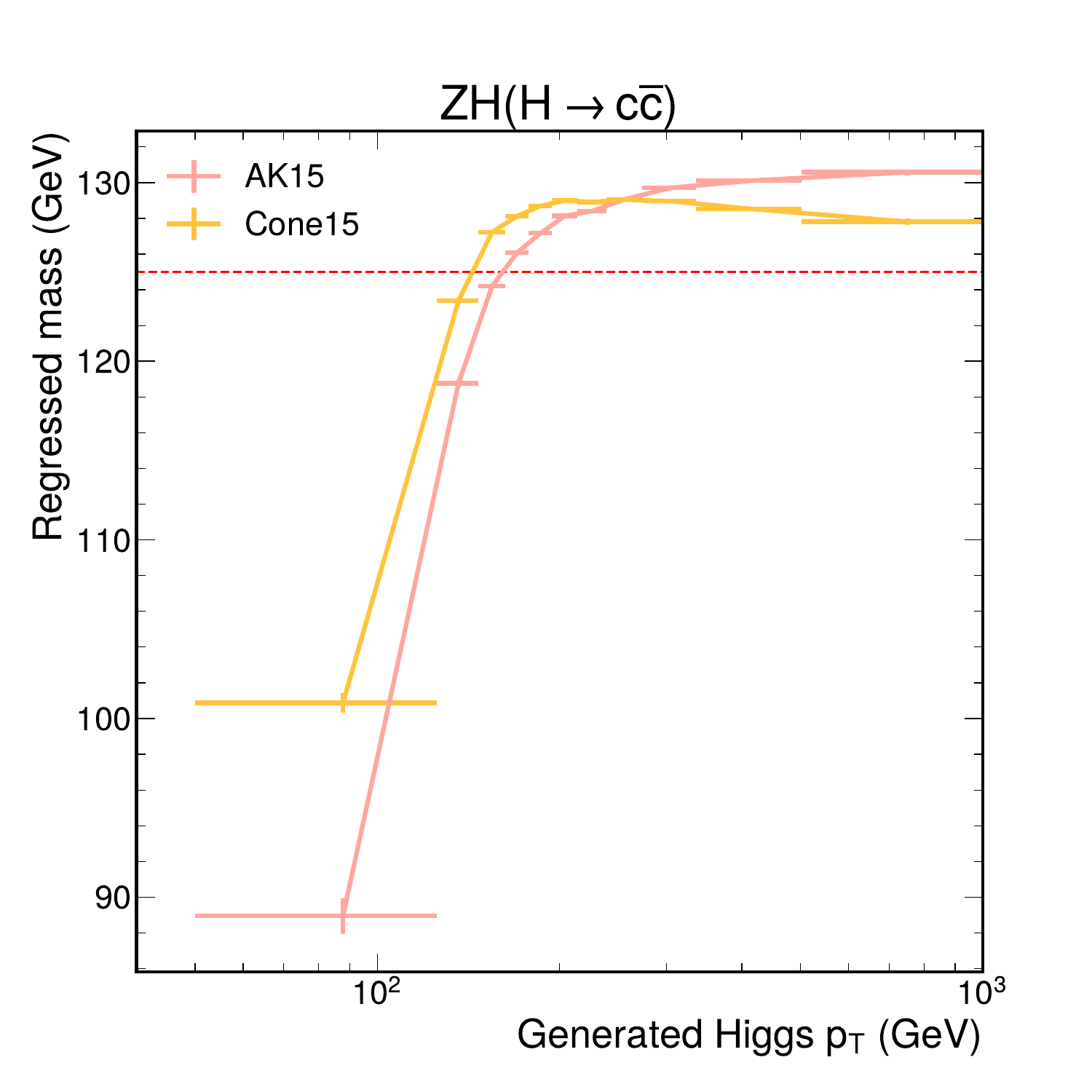}\includegraphics[width=0.5\textwidth]{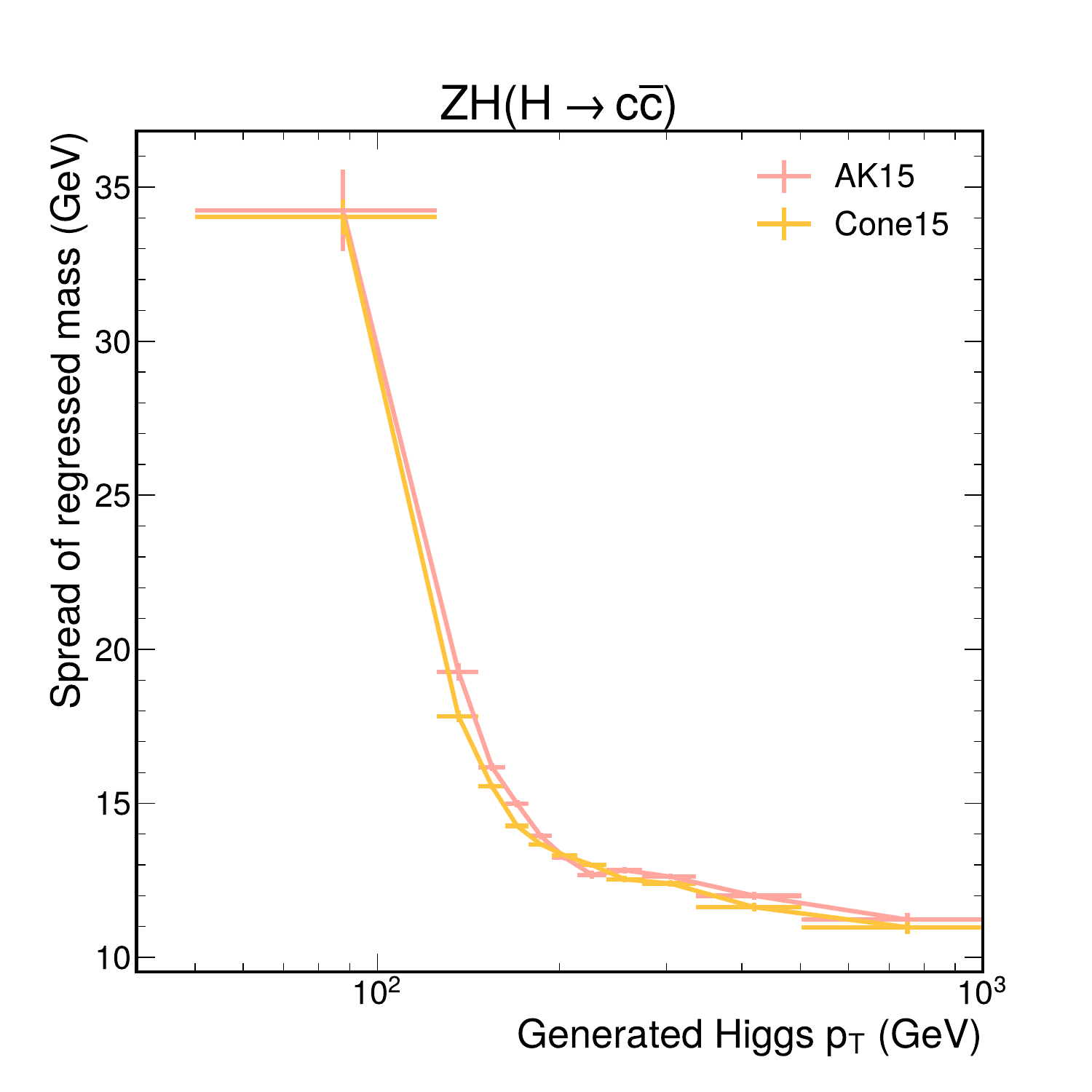}
\par\end{centering}
\caption{\label{fig:DCB-1}The mean (left) and spread (right) fit parameters
of the regressed mass as a function of the generated Higgs boson $p_{\text{T}}$
for AK15 and Cone15 jets are shown. Uncertainty bands indicate fit
uncertainties. The upper (lower) row shows results for bb (cc) jets
taken from ZH(H$\rightarrow\text{b\ensuremath{\bar{\text{b}}}}$)
(ZH(H$\rightarrow\text{c}\bar{\text{c}}$)) processes.}
\end{figure}

The mass regression technique was also applied on Cone15 jets (cf.
Appendix \ref{subsec:AK15-and-Cone15}). The results of the DCB fit
to the regressed masses for AK15 and Cone15 jets are shown in Fig.
\ref{fig:DCB-1}. The mass resolutions obtained from the two methods
are similar. Regression using the Cone15 approach, however, yields
means closer to the generated Higgs boson mass both at low and high
boosts of the Higgs boson.

\section{Differential signal sensitivity ($\nicefrac{S}{\sqrt{B}}$) in a
ZH(H$\rightarrow$$\text{c}\bar{\text{c}}$) analysis\label{sec:Differential-signal-sensitivity}}

We define a signal sensitivity metric as the ratio of the number of
selected signal events ($S$) to the square-root of the number of
passing background events ($B$) after all selection criteria and
classifier score requirements are applied (cf. Sec. \ref{sec:Constructing-a-mock}).
This can be evaluated at a target end-to-end signal efficiency value,
as well as across various reconstruction strategies. As discussed
in Sec. \ref{sec:Simulation}, only the major background (Z+jj) has
been considered in this study. To estimate the approximate signal
and background yields, the ZH(H$\rightarrow\text{c}\bar{\text{c}}$)
and Z+jj processes are assigned cross sections of 0.0027 and 270 pb,
respectively. The former value is taken from the LHC Higgs Cross Section
Working Group report \citep{HiggsXSec}, while the latter is obtained
from \textsc{Pythia8} after \textsc{MLM} matching is performed during
the sample generation. The simulations are scaled to an integrated
luminosity of 200 $\text{fb}^{-1}$, corresponding to the expected
data to be delivered by the LHC in Run-3.

\begin{figure}
\begin{centering}
\includegraphics[width=0.5\textwidth]{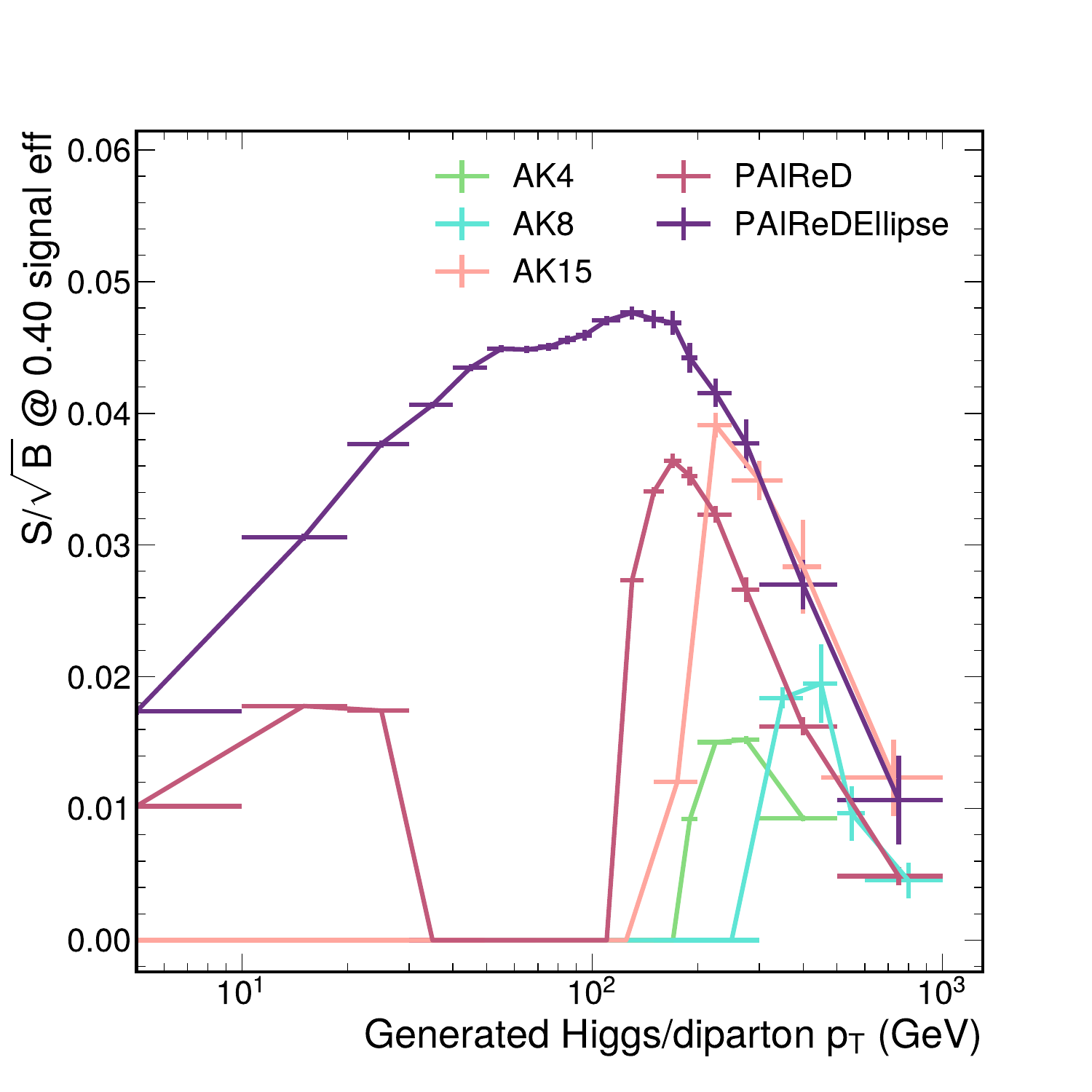}\includegraphics[width=0.5\textwidth]{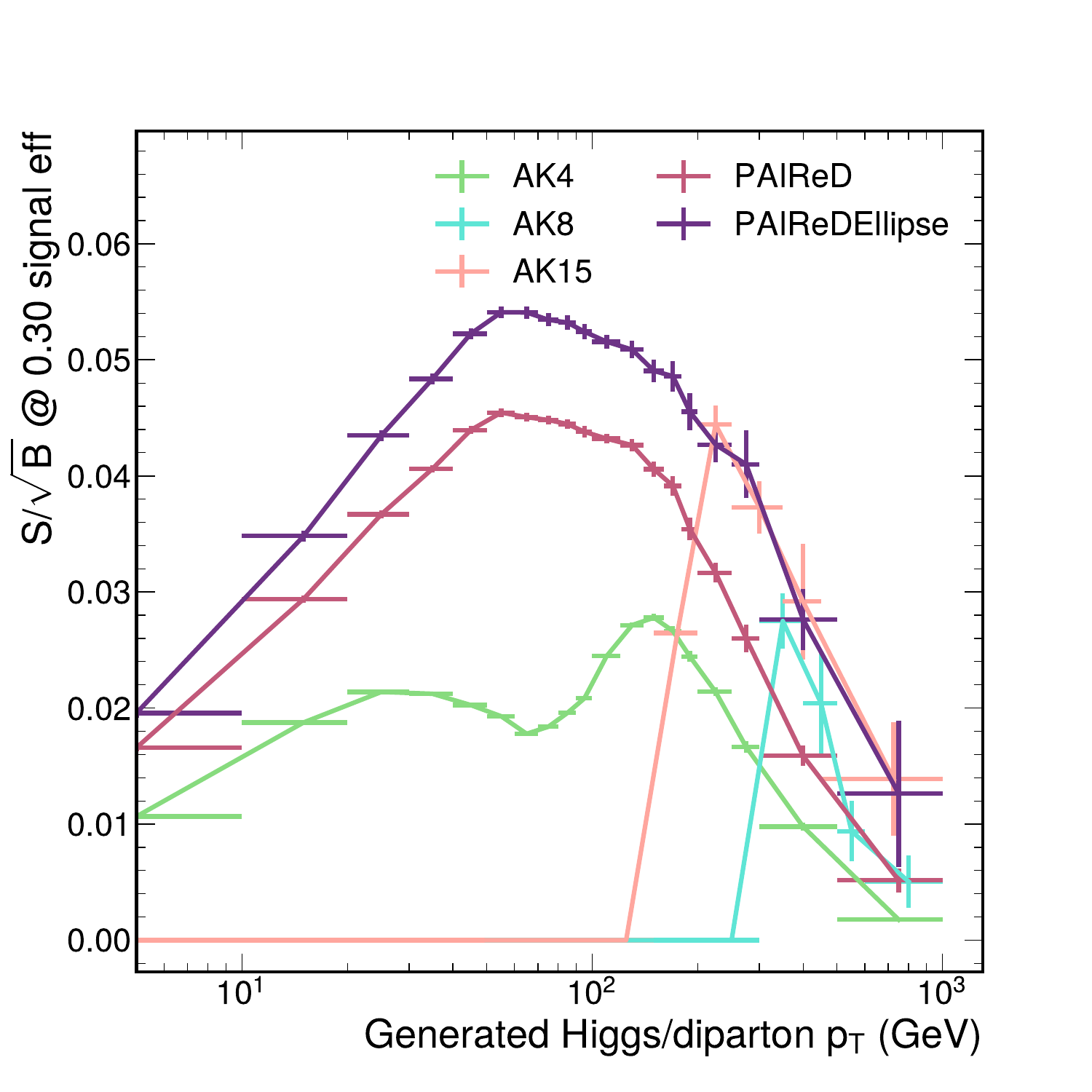}
\par\end{centering}
\begin{centering}
\includegraphics[width=0.5\textwidth]{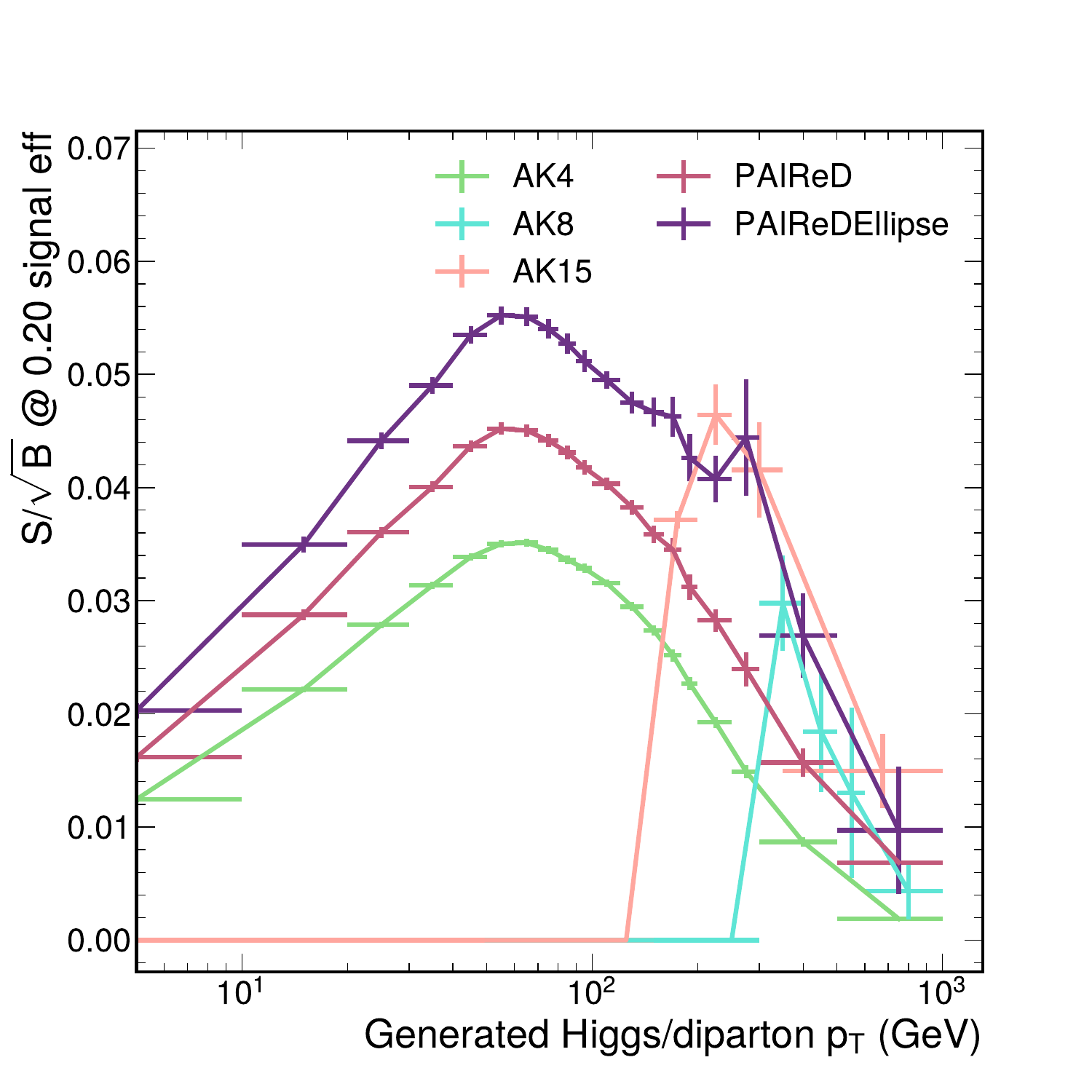}\includegraphics[width=0.5\textwidth]{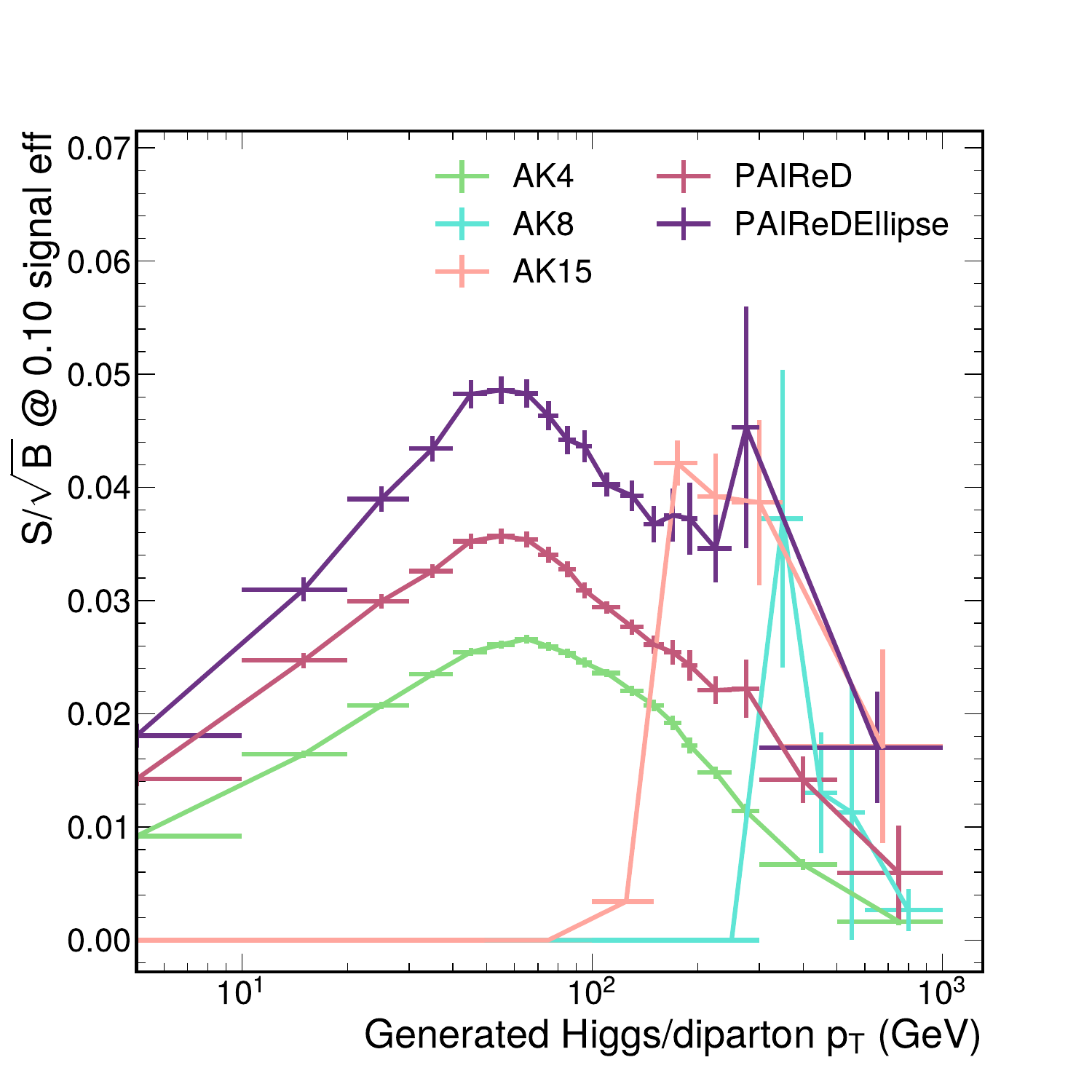}
\par\end{centering}
\caption{\label{fig:SoverB}The $\nicefrac{S}{\sqrt{B}}$ values at end-to-end
signal efficiencies of 0.4 (upper-left), 0.3 (upper-right), 0.2 (lower-left),
and 0.1 (lower-right) as a function of the generated Higgs boson $p_{\text{T}}$
(for signal) or diparton $p_{\text{T}}$ (for background), plotted
for various reconstruction strategies. Signal and background yields
in each bin are normalized to reflect a fixed Higgs/diparton $p_{\text{T}}$
range of 10 GeV. The target end-to-end signal efficiencies are calculated
per bin. The $\nicefrac{S}{\sqrt{B}}$ value is set to 0 wherever
the target signal efficiency cannot be achieved.}
\end{figure}

The $\nicefrac{S}{\sqrt{B}}$ values for four different target signal
efficiencies are shown in Fig. \ref{fig:SoverB}. The signal and background
yields in each bin are normalized to reflect a fixed Higgs/diparton
$p_{\text{T}}$ range, specifically a range of 10 GeV, so that the
$\nicefrac{S}{\sqrt{B}}$ value is independent of the choice of bin
width.

The trends observed in Fig. \ref{fig:SoverB} closely follow those
observed in Fig. \ref{fig:BkgRej}. At low Lorentz-boosts, the PAIReDEllipse
strategy achieves up to twice the signal sensitivity that the AK4-based
strategy achieves, while the PAIReD strategy has an intermediate performance.
At higher boosts, the PAIReDEllipse and AK15 strategies have similar
performances. The maximum $\nicefrac{S}{\sqrt{B}}$ is achieved in
Higgs boson $p_{\text{T}}$ ranges of 50--80 GeV at per-bin signal
efficiencies of 0.2--0.3, using the PAIReDEllipse strategy. This
reinforces the hypothesis that novel reconstruction and tagging strategies
employed at the resolved-jet regime can significantly improve the
overall sensitivity of such analyses. However, these studies rely
only on the major background simulation and do not consider any systematic
uncertainties, and can therefore be regarded as rough estimates only. 

\section{Event-level BDT performance at different Higgs $p_{\text{T}}$ and
jet multiplicities\label{sec:Event-level-BDT-performance}}

\begin{figure}
\begin{centering}
\includegraphics[width=0.5\textwidth]{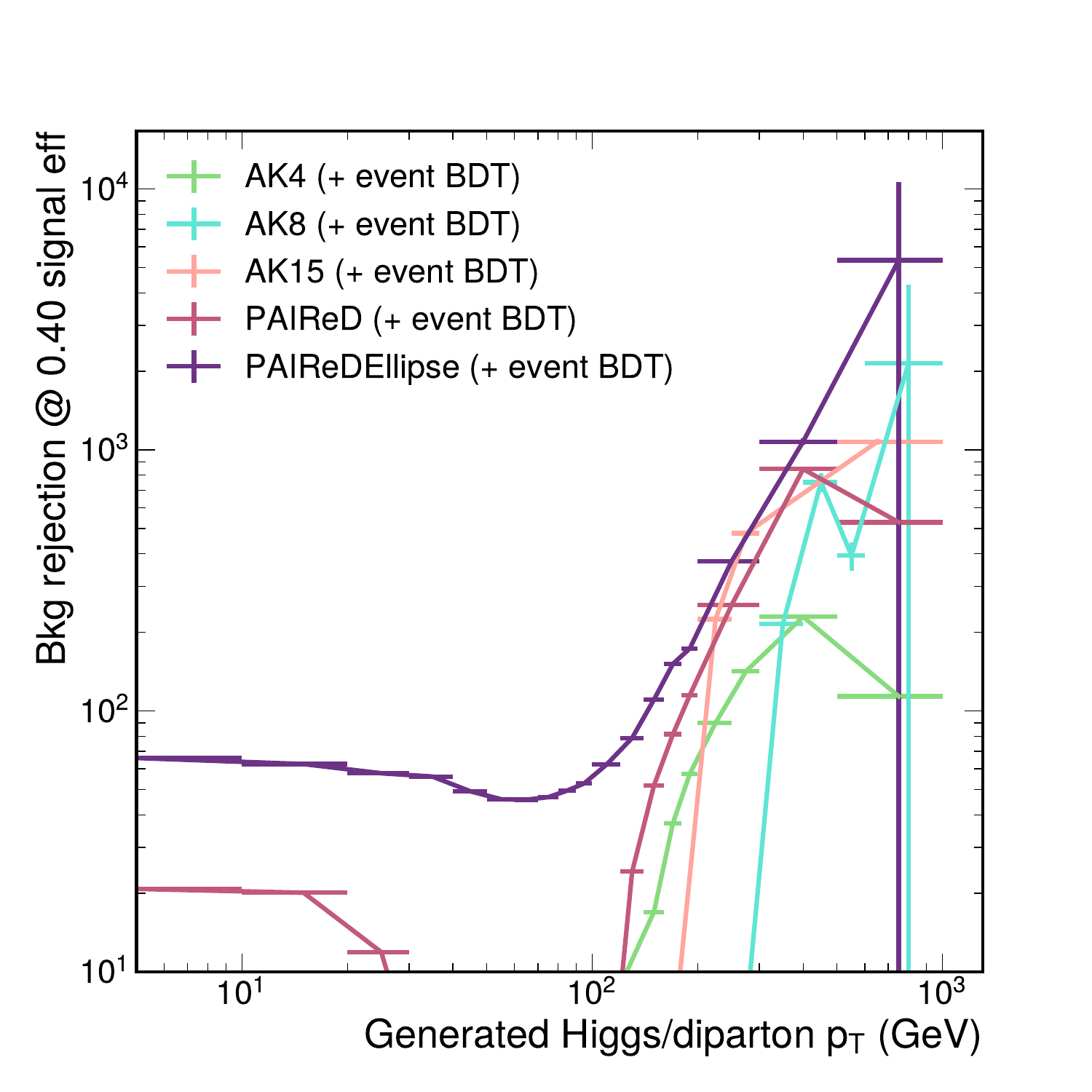}\includegraphics[width=0.5\textwidth]{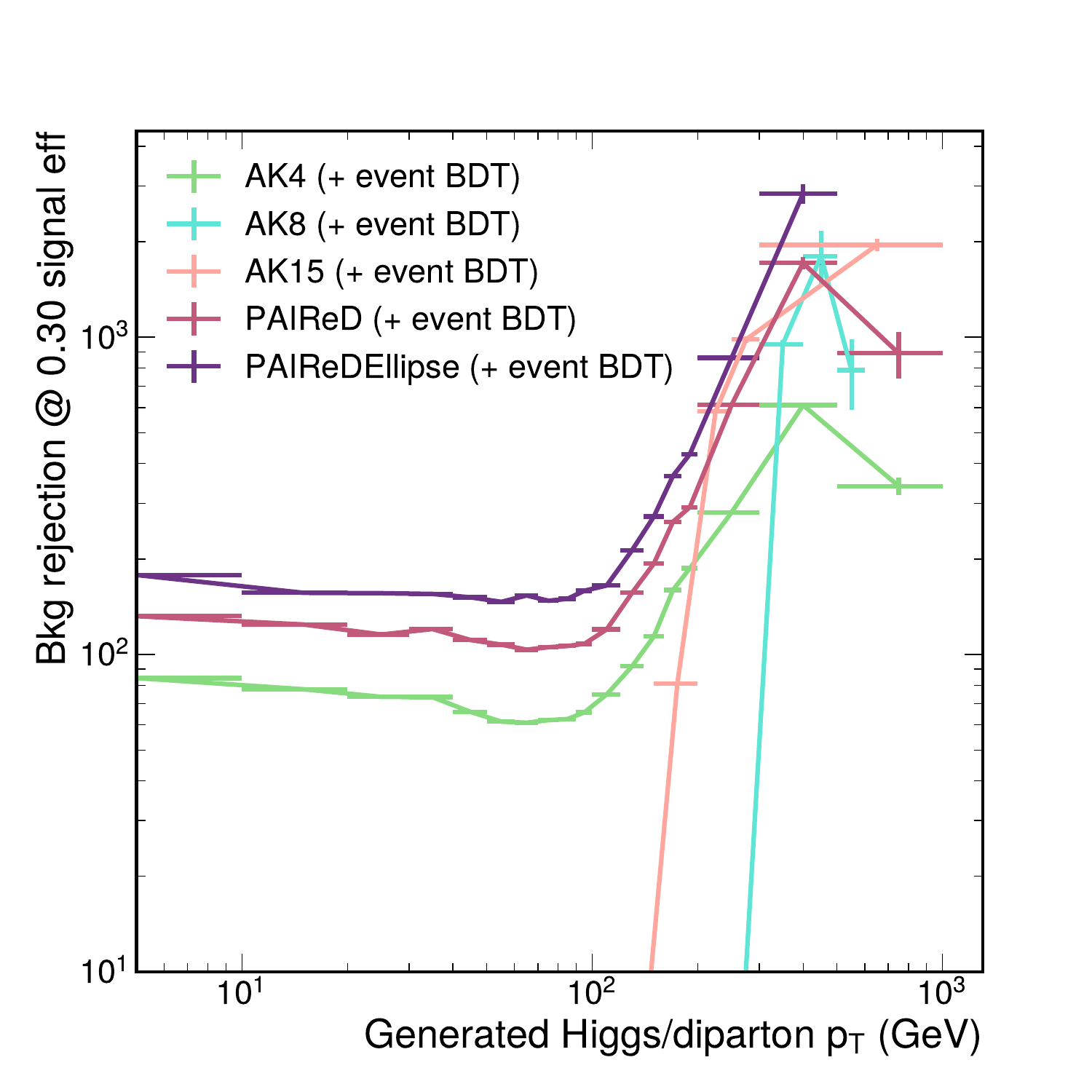}
\par\end{centering}
\begin{centering}
\includegraphics[width=0.5\textwidth]{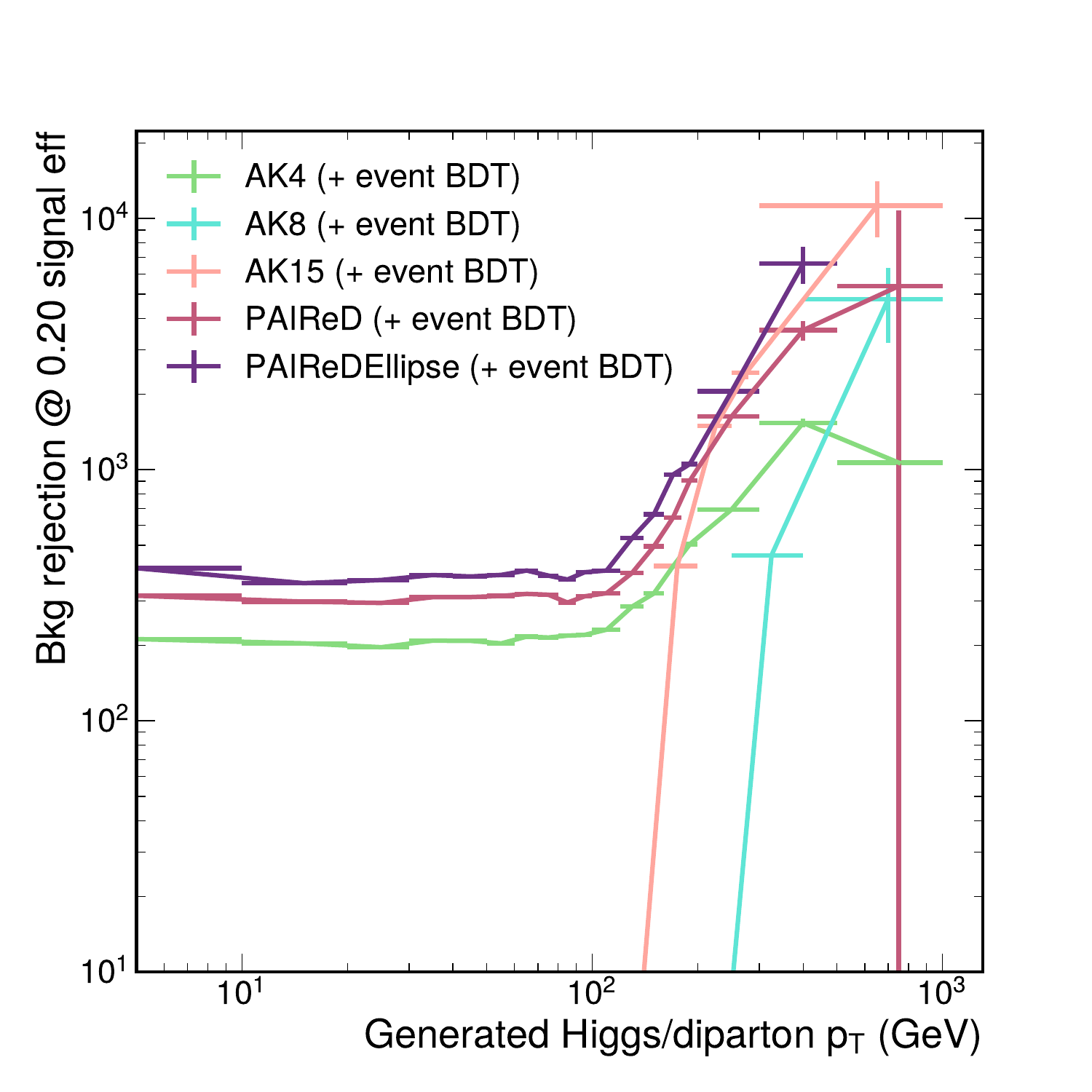}\includegraphics[width=0.5\textwidth]{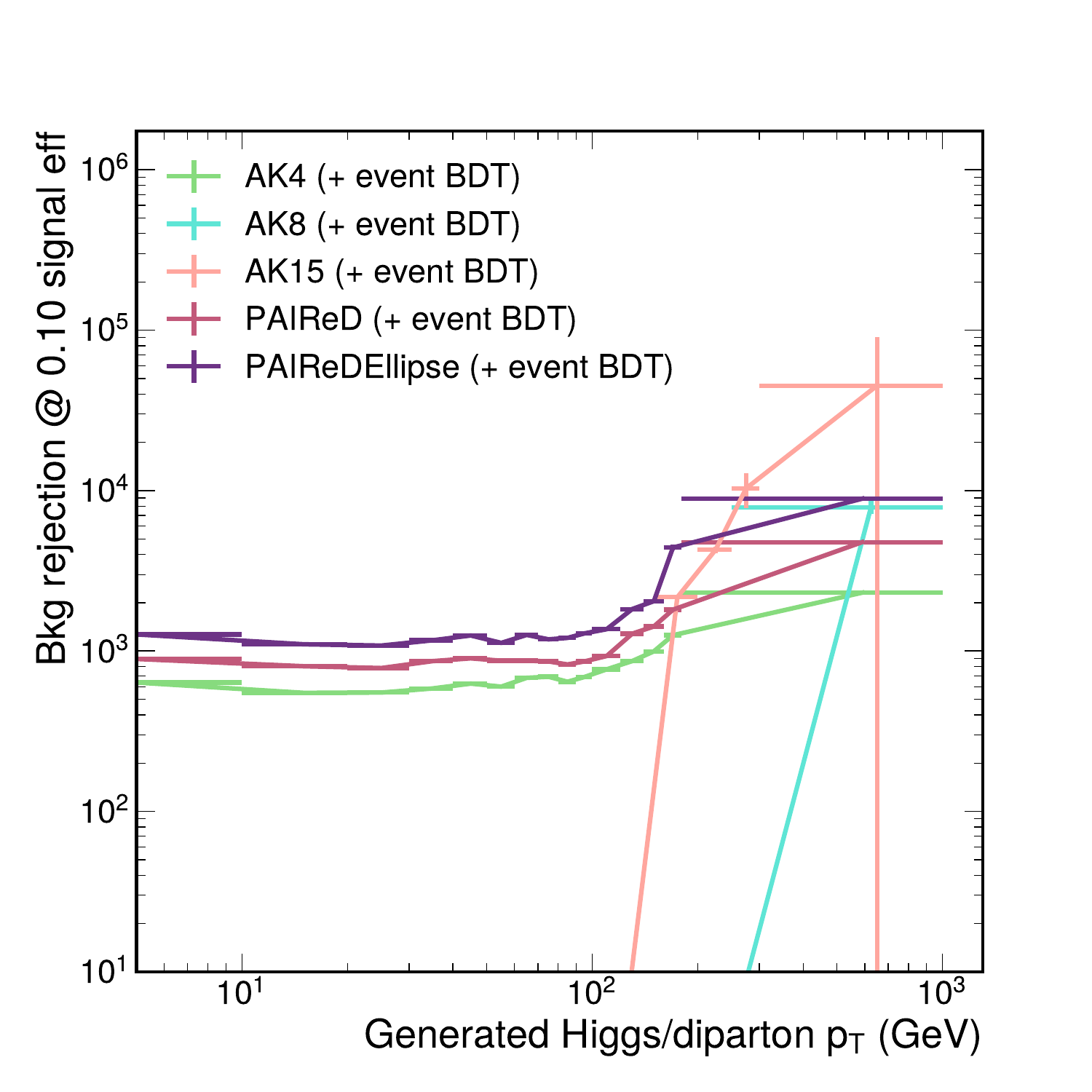}
\par\end{centering}
\caption{\label{fig:BkgRej-BDT}The end-to-end background rejection at end-to-end
signal efficiencies of 0.4 (upper-left), 0.3 (upper-right), 0.2 (lower-left),
and 0.1 (lower-right) as a function of the generated Higgs boson $p_{\text{T}}$
(for signal) or diparton $p_{\text{T}}$ (for background) achieved
with event-level BDTs (cf. Sec. \ref{subsec:Event-level-classification}),
plotted for various reconstruction strategies. End-to-end signal efficiencies
and background rejections are calculated per bin. The background rejection
is set to 0 wherever the target signal efficiency cannot be achieved.
Bins with very low yield of background events are either merged with
adjacent bins or are omitted. Compared to Fig. \ref{fig:BkgRej},
bin widths and statistical uncertainties are larger in this figure
as the number of events used is approximately half.}
\end{figure}

The performance of the event-level BDTs described in Sec. \ref{subsec:Event-level-classification}
is further evaluated as a function of the Higgs/diparton $p_{\text{T}}$.
The end-to-end background rejection rates achieved at end-to-end signal
efficiencies of 0.4, 0.3. 0.2, and 0.1 are shown in Fig. \ref{fig:BkgRej-BDT}. 

The overall background rejections achieved at a given signal efficiency
are higher compared to the corresponding values in Fig. \ref{fig:BkgRej}.
This is expected since the BDT is able to leverage more event-level
information in addition to the tagger scores. The trends across reconstruction
strategies, however, remain similar, with the BDT trained with PAIReDEllipse
jet scores rejecting about 2--2.5 times more background events compared
to the BDT trained with AK4 jet scores (as opposed to a 2.5--4 times
improvement reported in the case of tagger-only discrimination in
Sec. \ref{subsec:End-to-end-efficiencies}). This reflects the fact
that an event-level BDT in case of the AK4 jet-based reconstruction
approach can partially account for correlations between two jets by
leveraging the tagger scores of the two jets (4 additional inputs,
as shown in Tab. \ref{tab:BDT inputs}). However, this still falls
short of the performance achievable using event-level BDTs with the
PAIReD(Ellipse) approaches, as the latter leverage additional information
and correlations at the jet constituent level that are inaccessible
to AK4-based taggers. 

\begin{figure}
\begin{centering}
\includegraphics[width=0.5\textwidth]{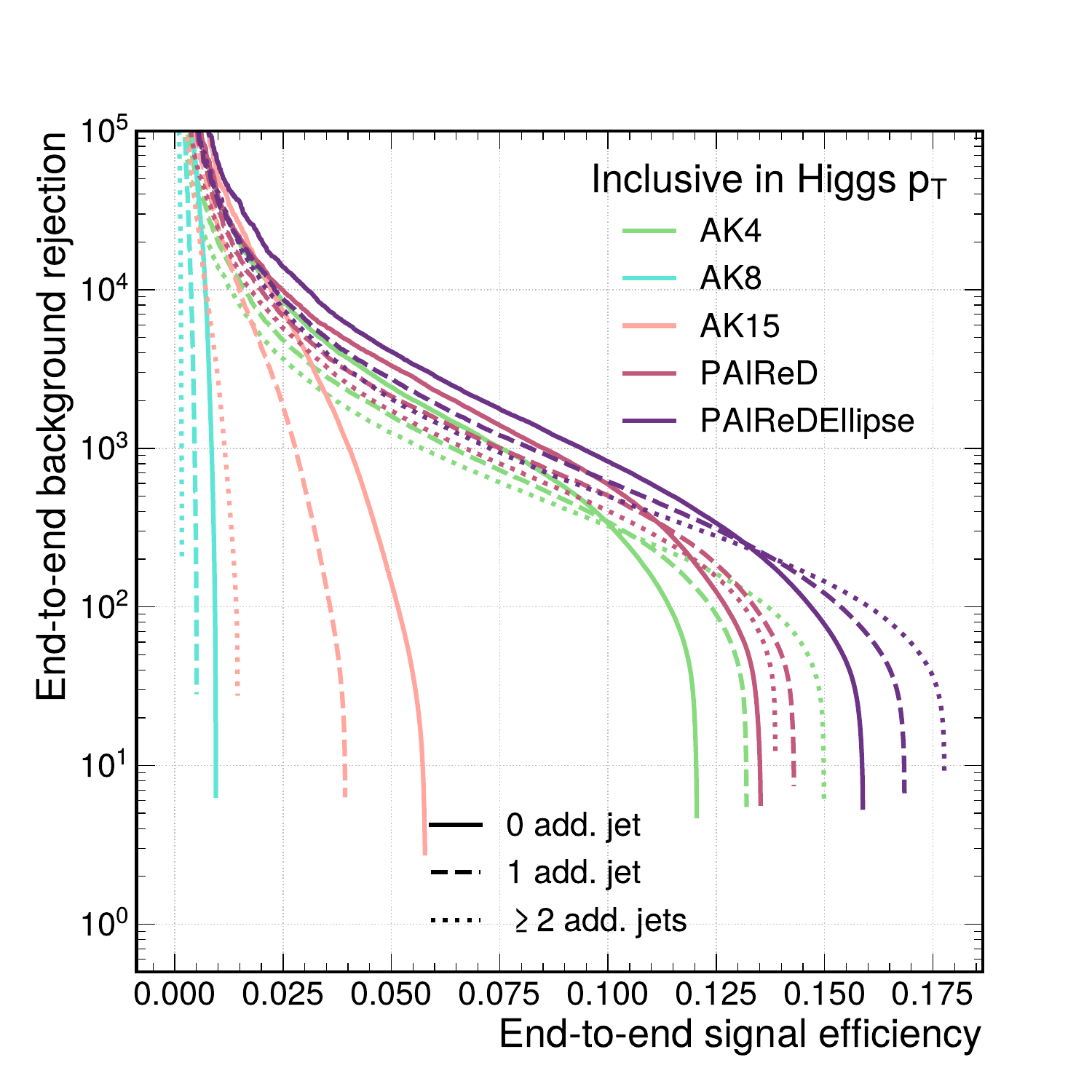}\includegraphics[width=0.5\textwidth]{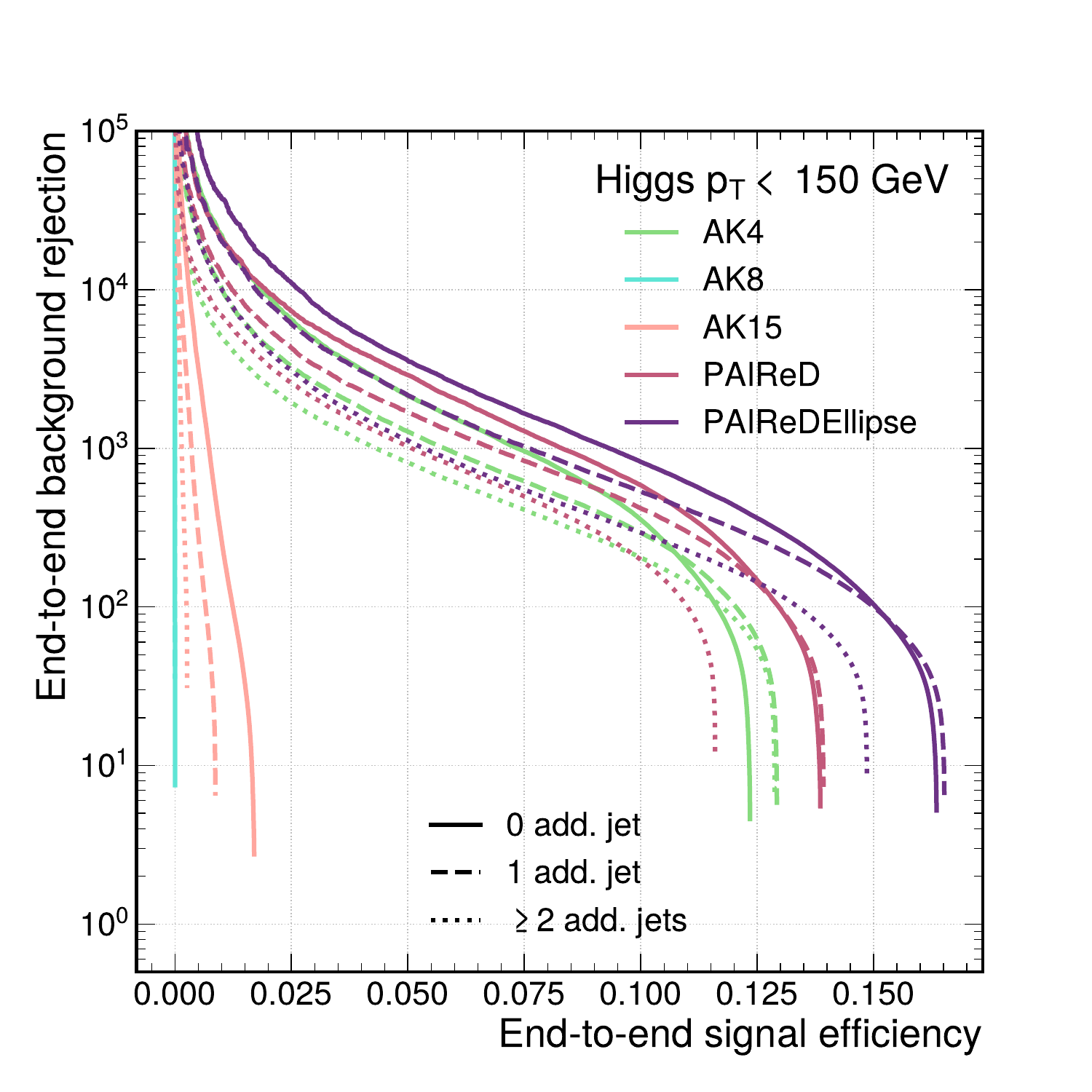}
\par\end{centering}
\begin{centering}
\includegraphics[width=0.5\textwidth]{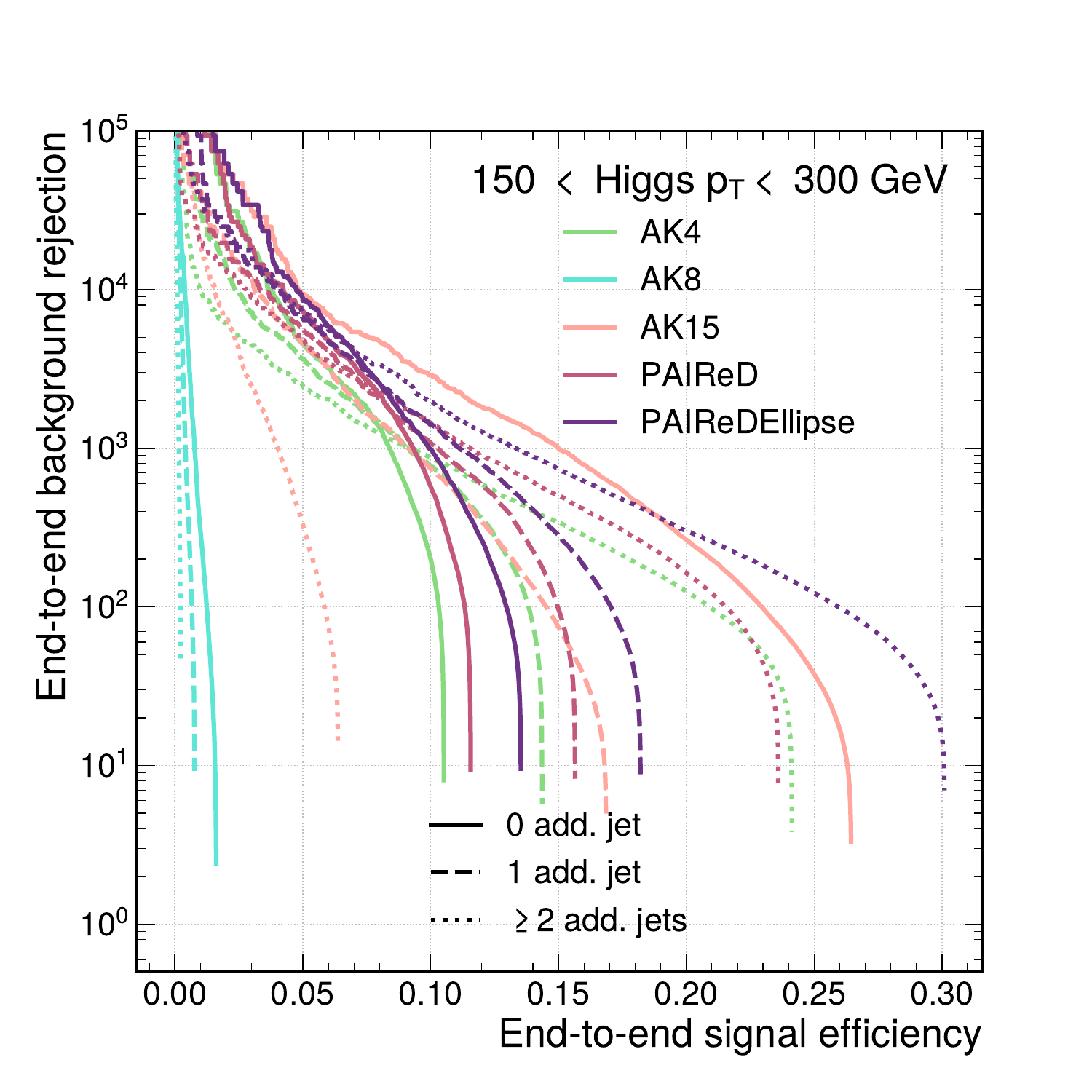}\includegraphics[width=0.5\textwidth]{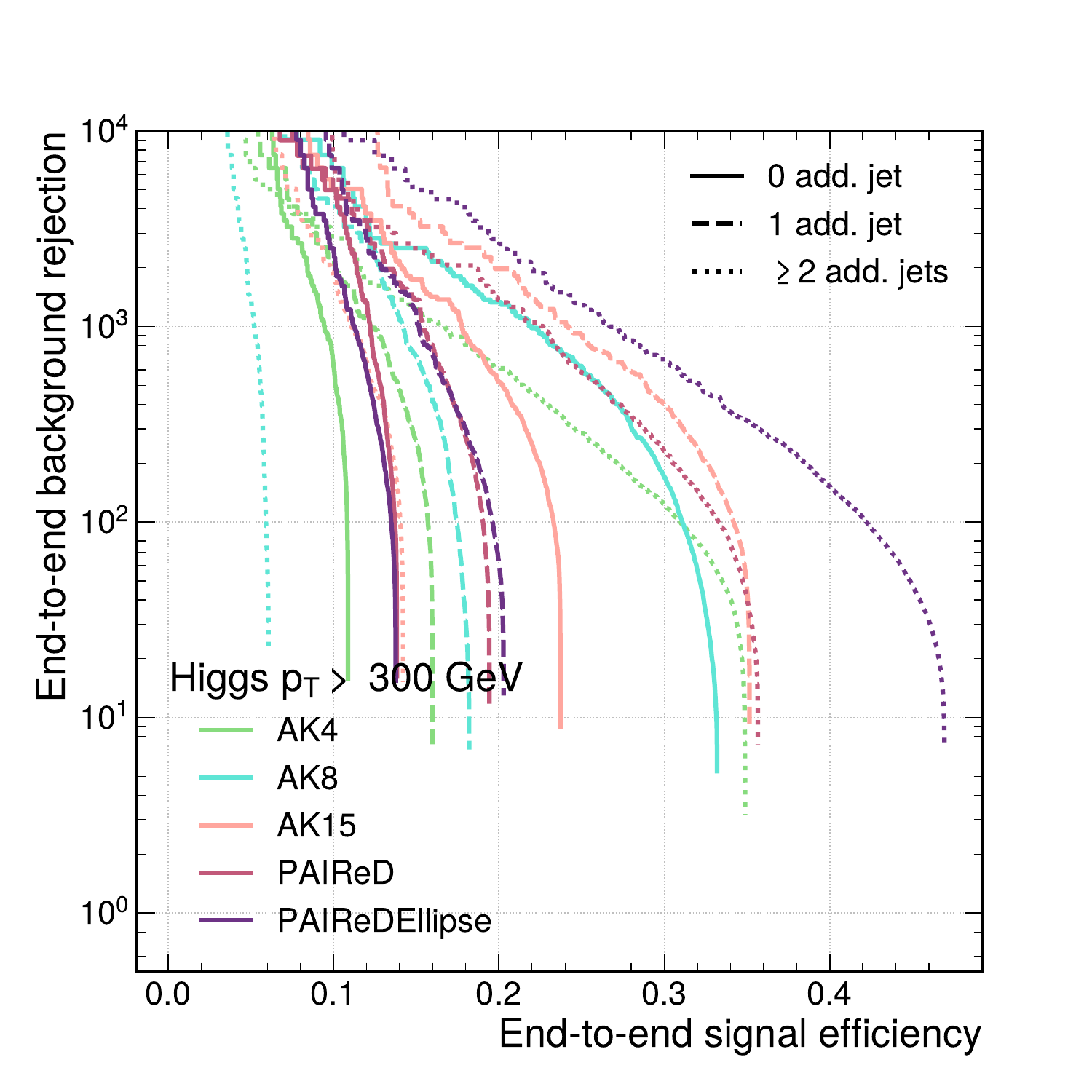}
\par\end{centering}
\caption{\label{fig:BDT-ROC-njet}The end-to-end background event rejection
rate as a function of the end-to-end signal efficiency inclusively
in Higgs boson (or diparton) $p_{\text{T}}$ (upper-left), at low
($p_{\text{T}}$ < 150 GeV) $p_{\text{T}}$ ranges (upper-right),
at medium (150 < $p_{\text{T}}$ < 300 GeV) $p_{\text{T}}$ ranges
(lower-left), and at high ($p_{\text{T}}$ > 300 GeV) $p_{\text{T}}$
ranges (lower-right), across different jet multiplicities, as predicted
by the event-level BDT. The solid lines represent events with exactly
2 AK4 jets, 1 AK8/AK15 jet, or 1 PAIReD/PAIReDEllipse jet. The dashed
lines represent events with exactly 3 AK4 jets, 2 AK8/AK15 jets, or
3 PAIReD/PAIReDEllipse jets. The dotted lines represent events with
at least 4 AK4 jets, 3 AK8/AK15 jets, or 6 PAIReD/PAIReDEllipse jets.
The highest value along the x-axis reached by each curve represents
the maximum signal reconstruction efficiency (w.r.t. the total number
of generated events, irrespective of the number of reconstructable
jets) achievable using the respective event reconstruction strategy
(cf. Sec. \ref{subsec:Signal-reconstruction-efficiency}) with the
respective number of reconstructed jets.}
\end{figure}

The performance of the BDTs are also evaluated for different jet multiplicities.
The corresponding ROC curves are shown in Fig. \ref{fig:BDT-ROC-njet}.
It is observed that the classification performance using PAIReD(Ellipse)
jets does not drop significantly at low boosts despite the higher
combinatorial complexity. In fact, events with high small-radius jet
multiplicities (and consequently high PAIReD(Ellipse) jet multiplicities)
play a significant role especially at high boosts of the Higgs boson.

\section{Validation with ZZ(Z$\rightarrow$$\text{c}\bar{\text{c}}$) analysis\label{sec:Validation-with-ZZ(Z)}}

Additional ZZ(Z$\rightarrow$$\text{c}\bar{\text{c}}$) processes
are simulated as per the description in Sec. \ref{sec:Simulation},
with the difference that the Z$\rightarrow$$\text{c}\bar{\text{c}}$
decay is performed in MG5 instead of \textsc{Pythia8}. A mock ZZ(Z$\rightarrow$$\text{c}\bar{\text{c}}$)
analysis is designed in a way similar to the ZH(H$\rightarrow$$\text{c}\bar{\text{c}}$)
analysis described in Sec. \ref{sec:Constructing-a-mock}. 

\begin{figure}[!b]
\begin{centering}
\includegraphics[width=0.5\textwidth]{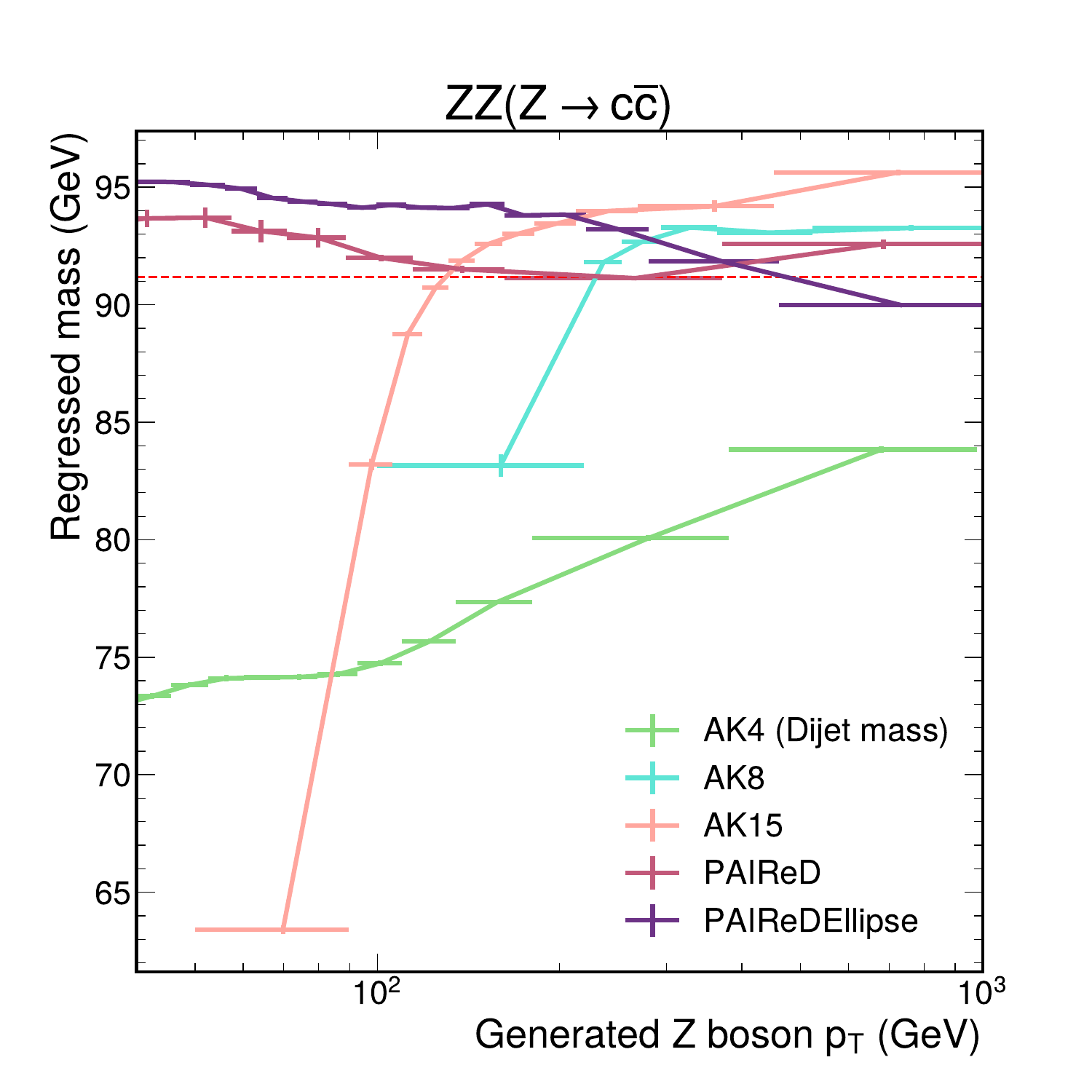}\includegraphics[width=0.5\textwidth]{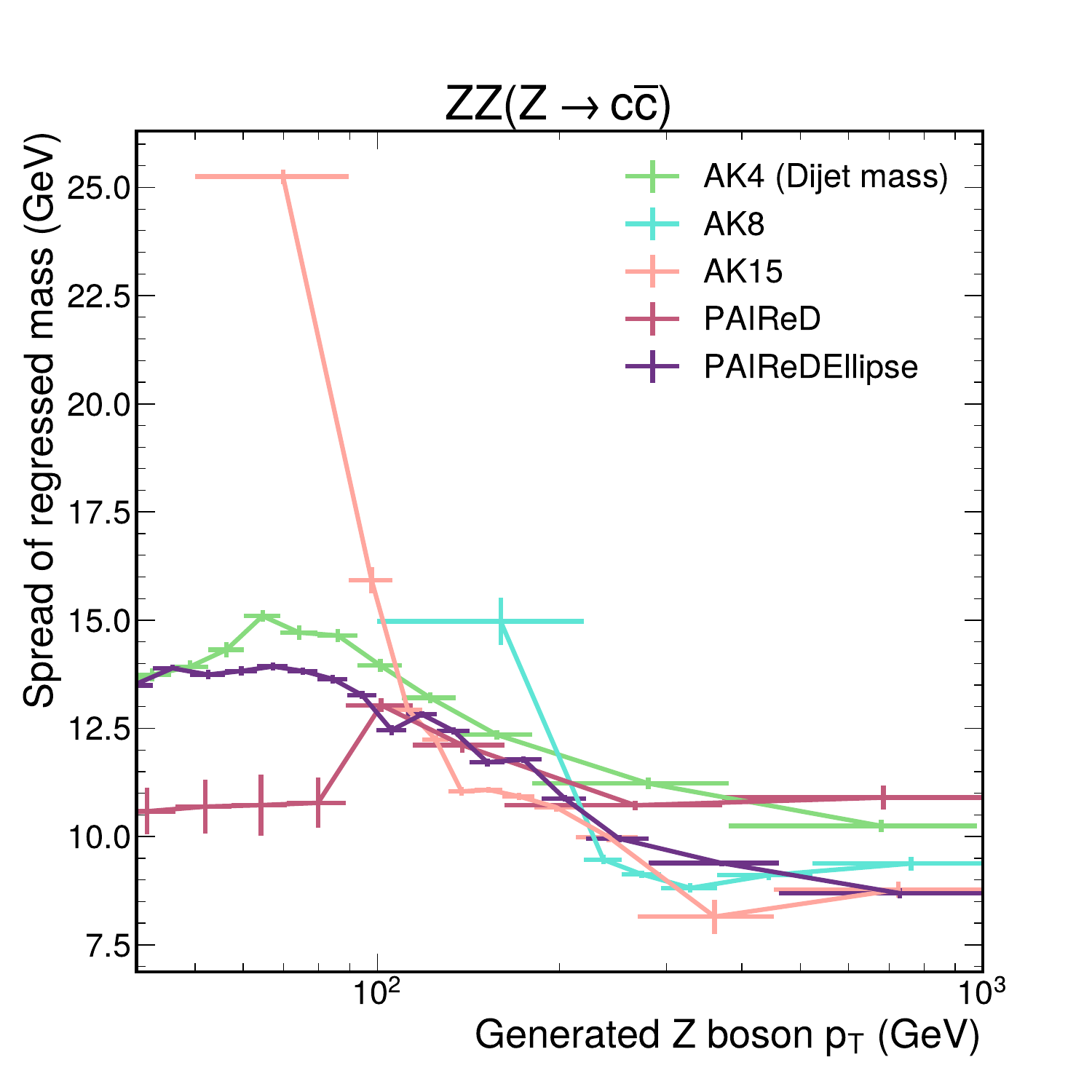}
\par\end{centering}
\caption{\label{fig:DCB-2}The mean (left) and spread (right) fit parameters
of the regressed mass as a function of the generated Z boson $p_{\text{T}}$
for different reconstruction strategies are shown. Uncertainty bands
indicate fit uncertainties. The binning for each strategy is chosen
in a way such that each bin contains approximately an equal number
of jets. The cc jets from the ZZ(Z$\rightarrow\text{c}\bar{\text{c}}$)
processes are used. The y-axis indicates the dijet invariant mass
for the AK4-based reconstruction strategy.}
\end{figure}

The mass regression trainings obtained from using the Higgs boson
samples are used to predict the regressed mass values of the hadronically-decaying
Z boson in the ZZ(Z$\rightarrow$$\text{c}\bar{\text{c}}$) samples.
The regressed mass and its spread for different reconstruction strategies
are shown in Fig. \ref{fig:DCB-2}. The decay width of the Z boson
($\Gamma_{\text{Z}}\approx2.5$ GeV \citep{PDG2022}) is neglected
as it is small compared to the best mass resolution ($\sigma_{\text{Z}}\approx8$
GeV) achieved using this method ($\Gamma_{\text{Z}}^{2}\ll\text{\ensuremath{\sigma_{\text{Z}}^{2}}}$).
The relative mass resolution ($\nicefrac{\sigma}{\bar{x}}$) values
are consistent with the ones obtained for the Higgs boson sample (Fig.
\ref{fig:DCB}). 

\begin{figure}
\begin{centering}
\includegraphics[width=0.5\textwidth]{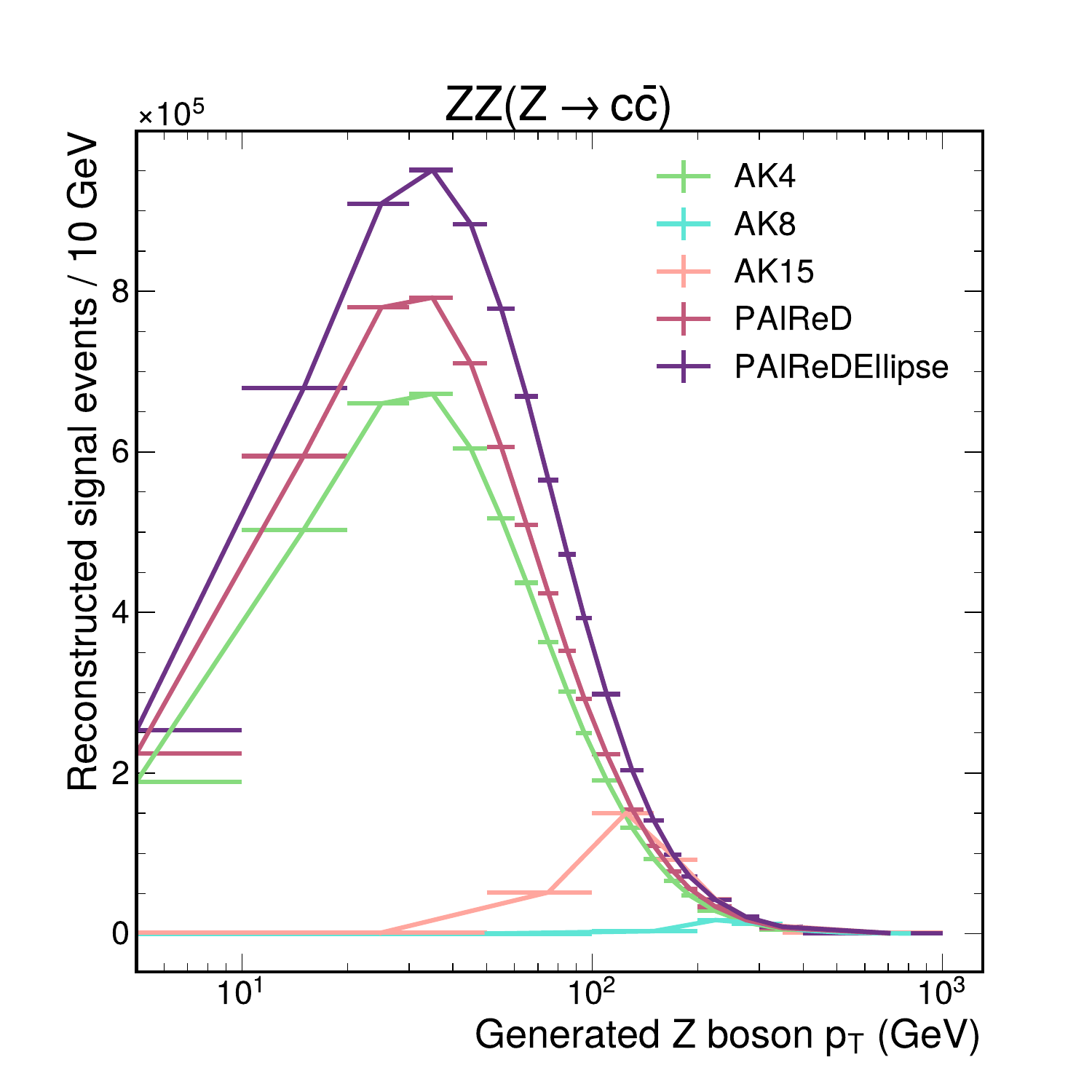}\includegraphics[width=0.5\textwidth]{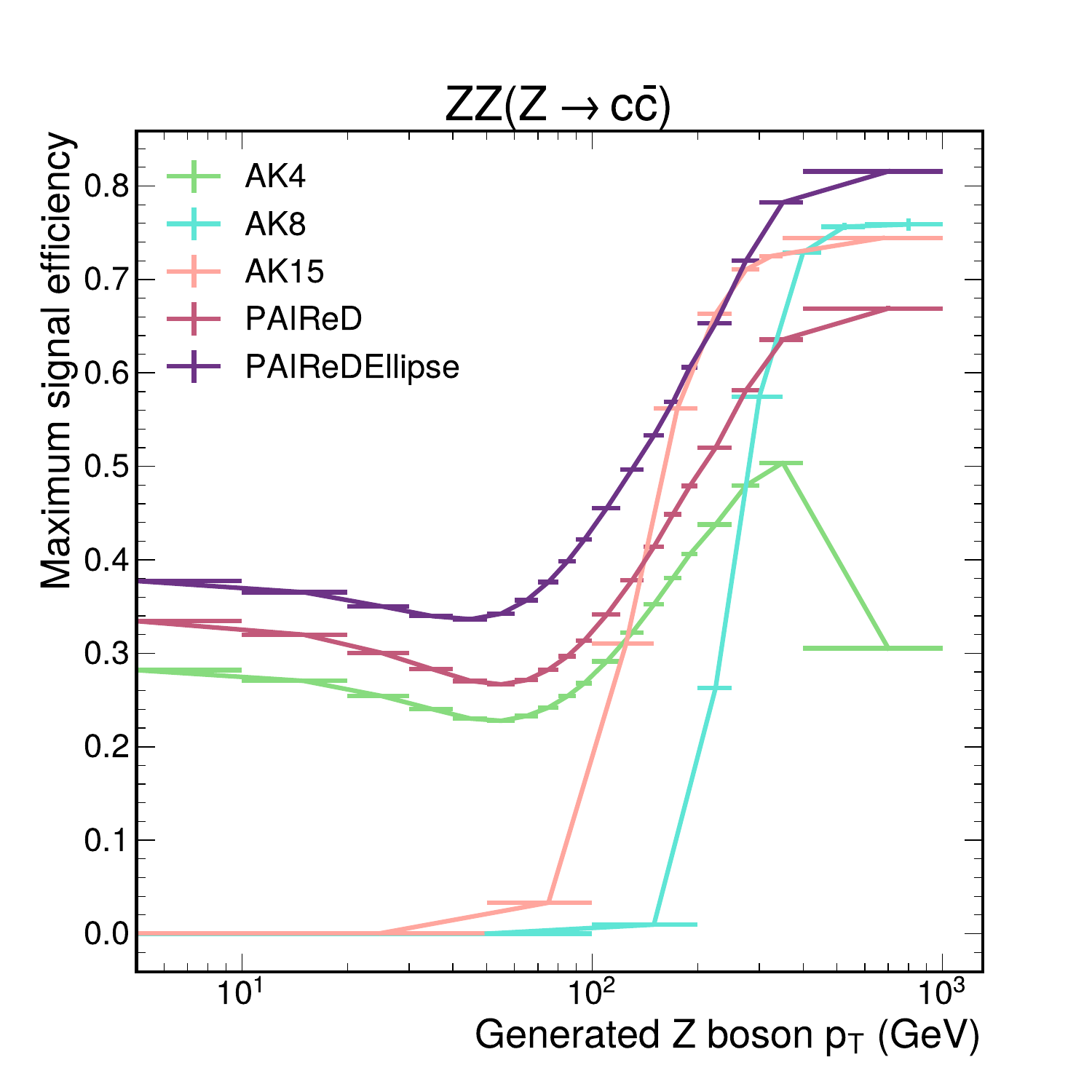}
\par\end{centering}
\begin{centering}
\includegraphics[width=0.5\textwidth]{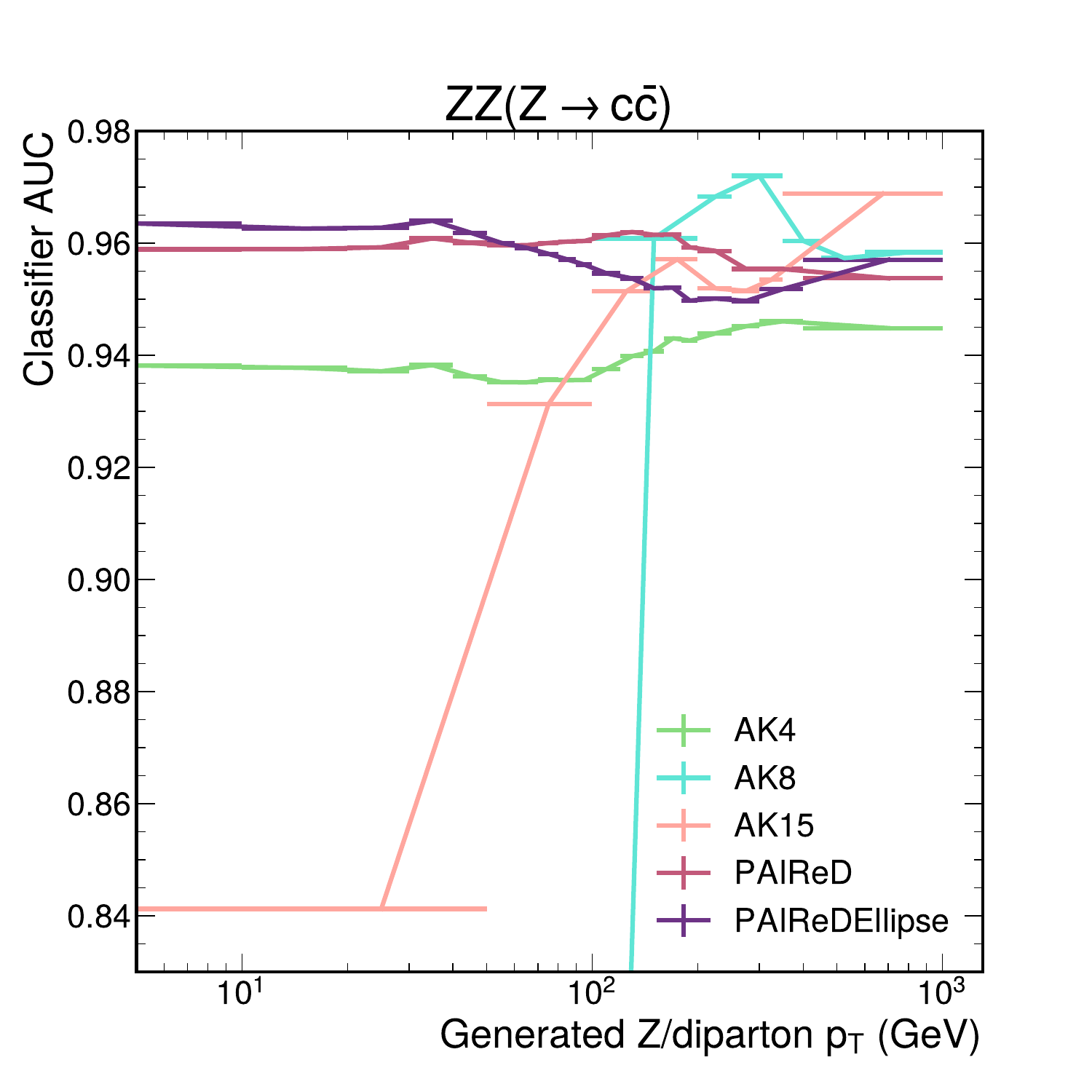}
\par\end{centering}
\caption{\label{fig:MaxSigEff-1}The number of reconstructed ZZ(Z$\rightarrow\text{c}\bar{\text{c}}$)
signal events out of the \textasciitilde 11M signal events generated
(upper left) and the maximum signal efficiency or the signal reconstruction
efficiency (upper right) as functions of the generated Z boson $p_{\text{T}}$,
for various reconstruction strategies. The lower plot shows the area
under the ROC curves (AUC) obtained when classifiers corresponding
to various reconstruction strategies are evaluated on the signal and
background events passing the respective selections, plotted as a
function of the Z boson $p_{\text{T}}$ (for signal events) or the
diparton $p_{\text{T}}$ (for background events). }
\end{figure}

\begin{figure}
\begin{centering}
\includegraphics[width=0.5\textwidth]{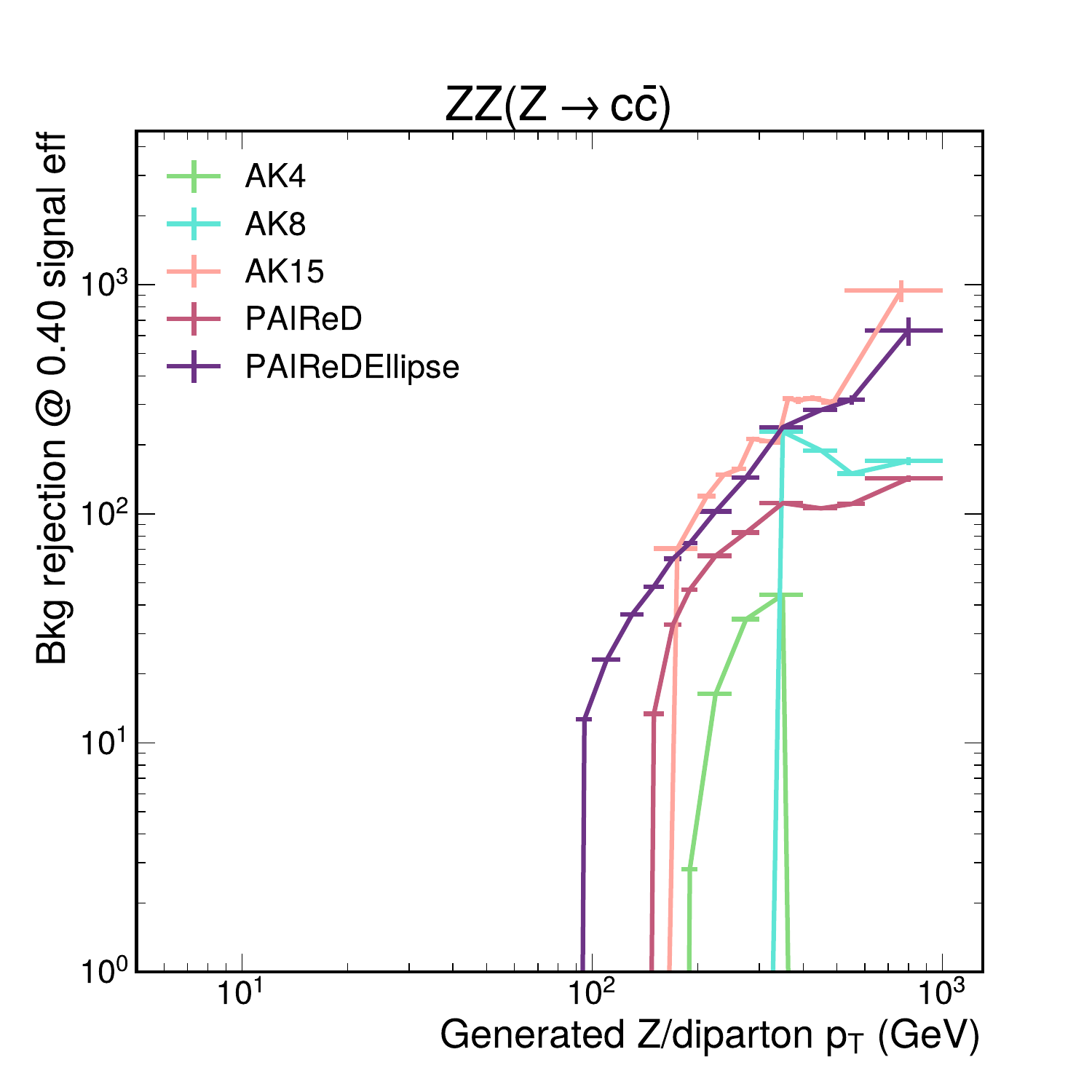}\includegraphics[width=0.5\textwidth]{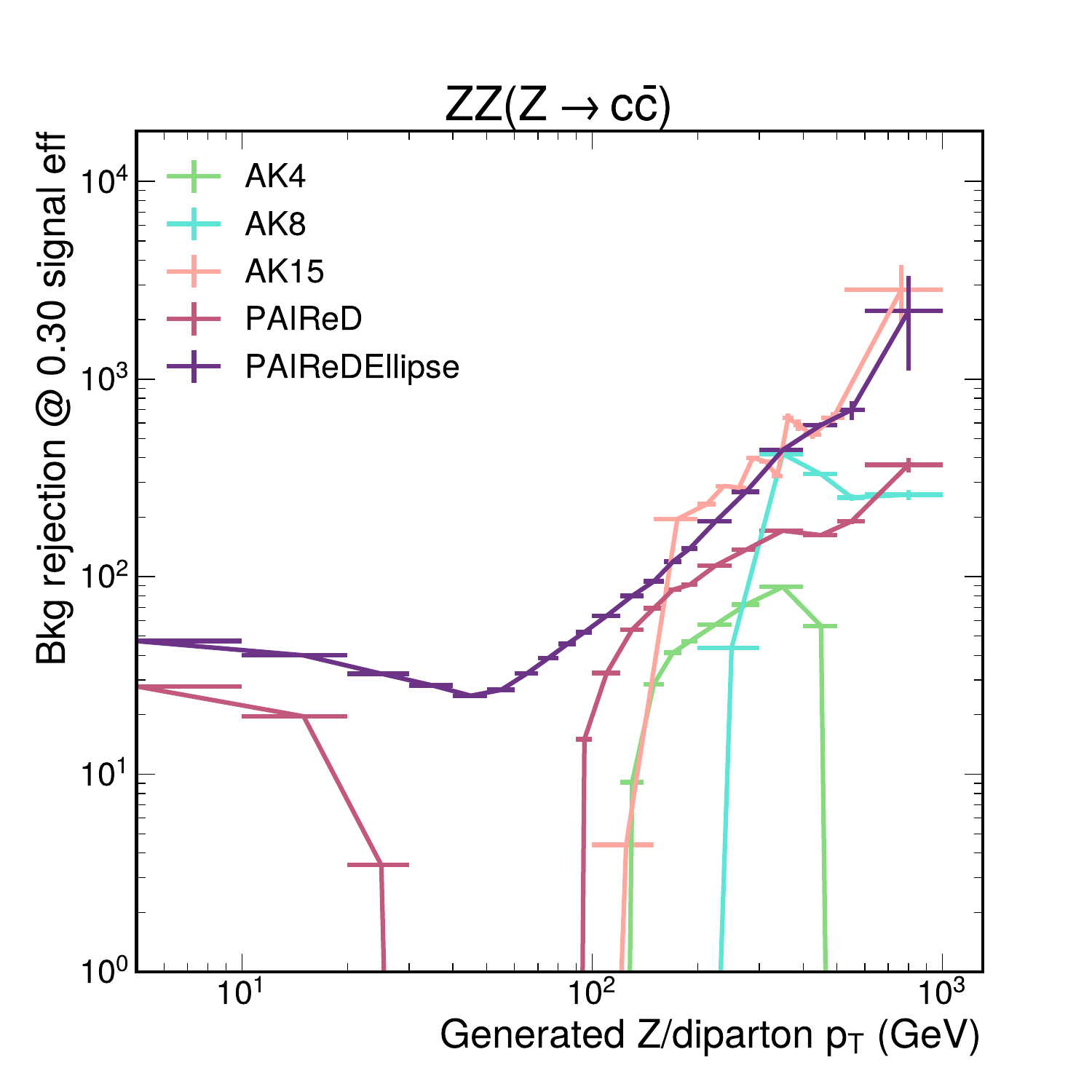}
\par\end{centering}
\begin{centering}
\includegraphics[width=0.5\textwidth]{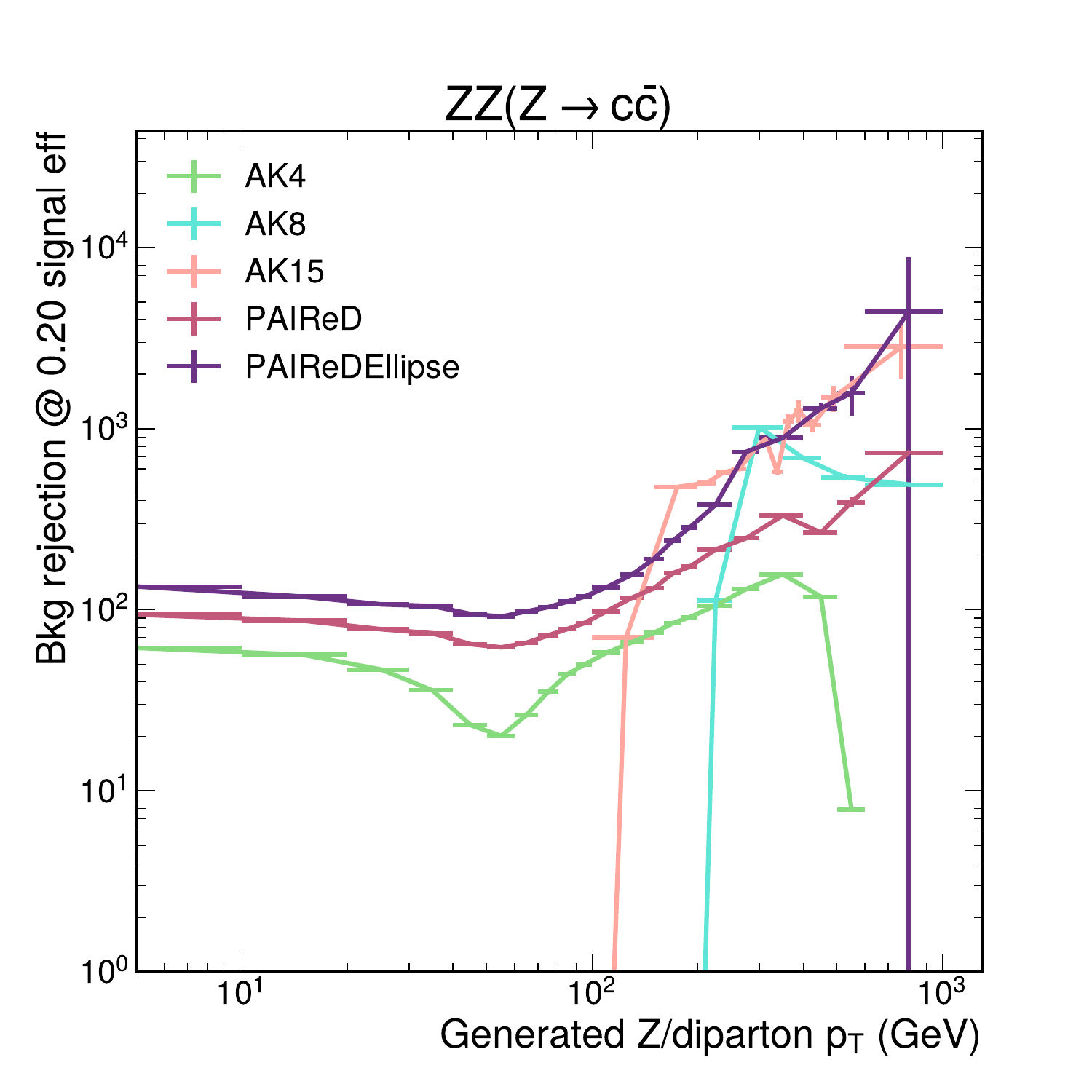}\includegraphics[width=0.5\textwidth]{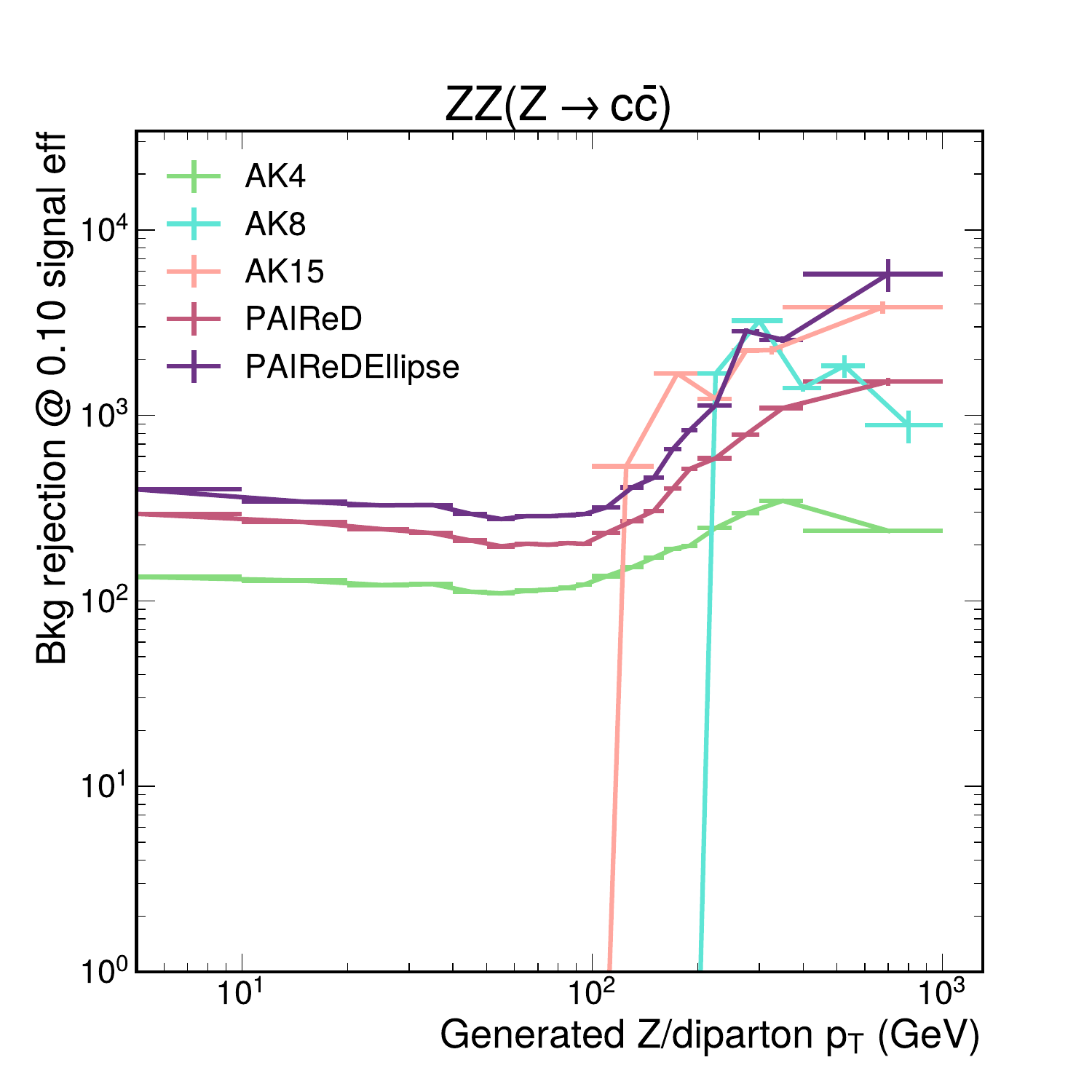}
\par\end{centering}
\caption{\label{fig:BkgRej-1}The end-to-end background rejection at end-to-end
signal efficiencies of 0.4 (upper-left), 0.3 (upper-right), 0.2 (lower-left),
and 0.1 (lower-right) as a function of the generated Z $p_{\text{T}}$
(for signal) or diparton $p_{\text{T}}$ (for background), plotted
for various reconstruction strategies. End-to-end signal efficiencies
and background rejections are calculated per bin. The background rejection
is set to 0 wherever the target signal efficiency cannot be achieved.
Bins with very low yield of background events are either merged with
adjacent bins or are omitted.}
\end{figure}

\begin{figure}
\begin{centering}
\includegraphics[width=0.5\textwidth]{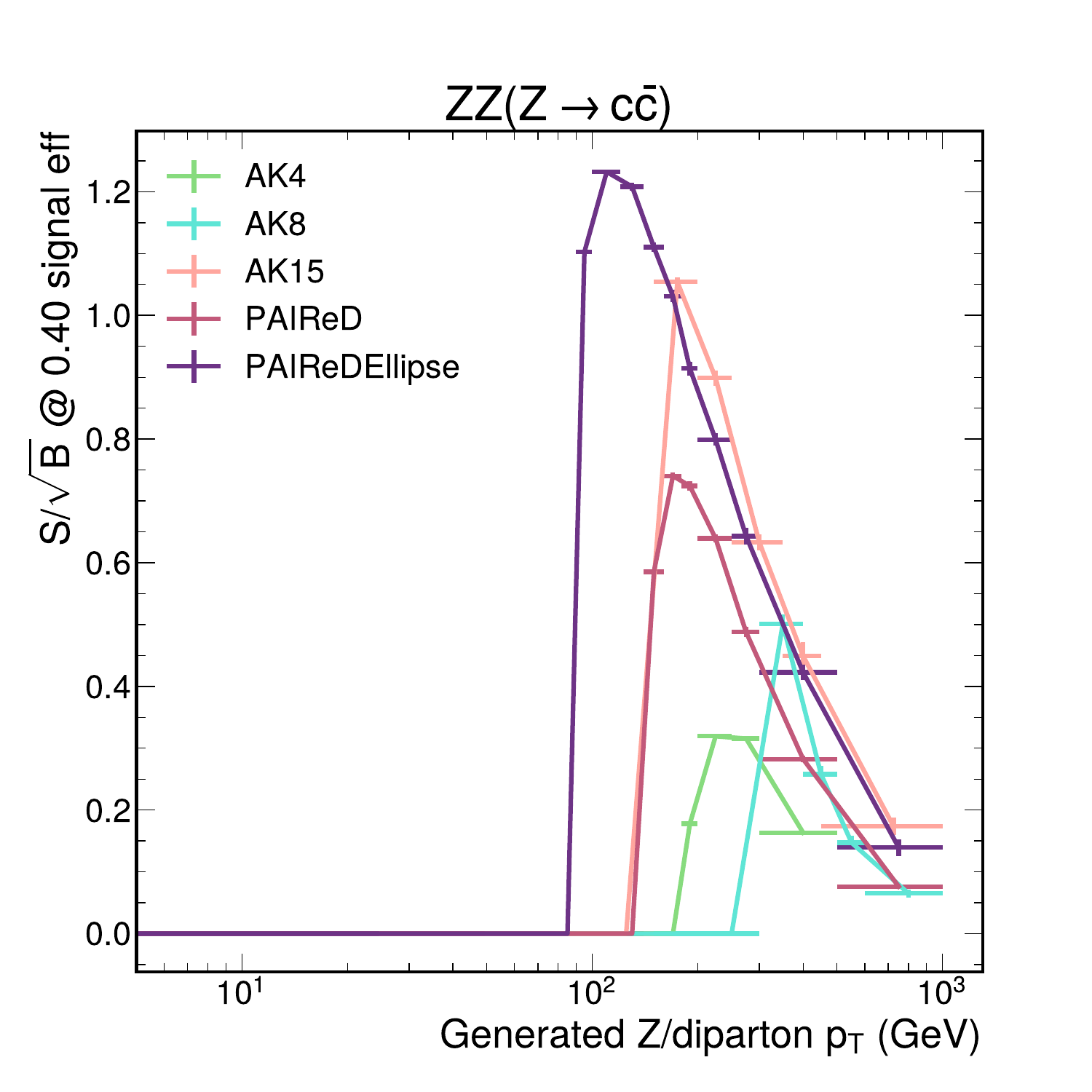}\includegraphics[width=0.5\textwidth]{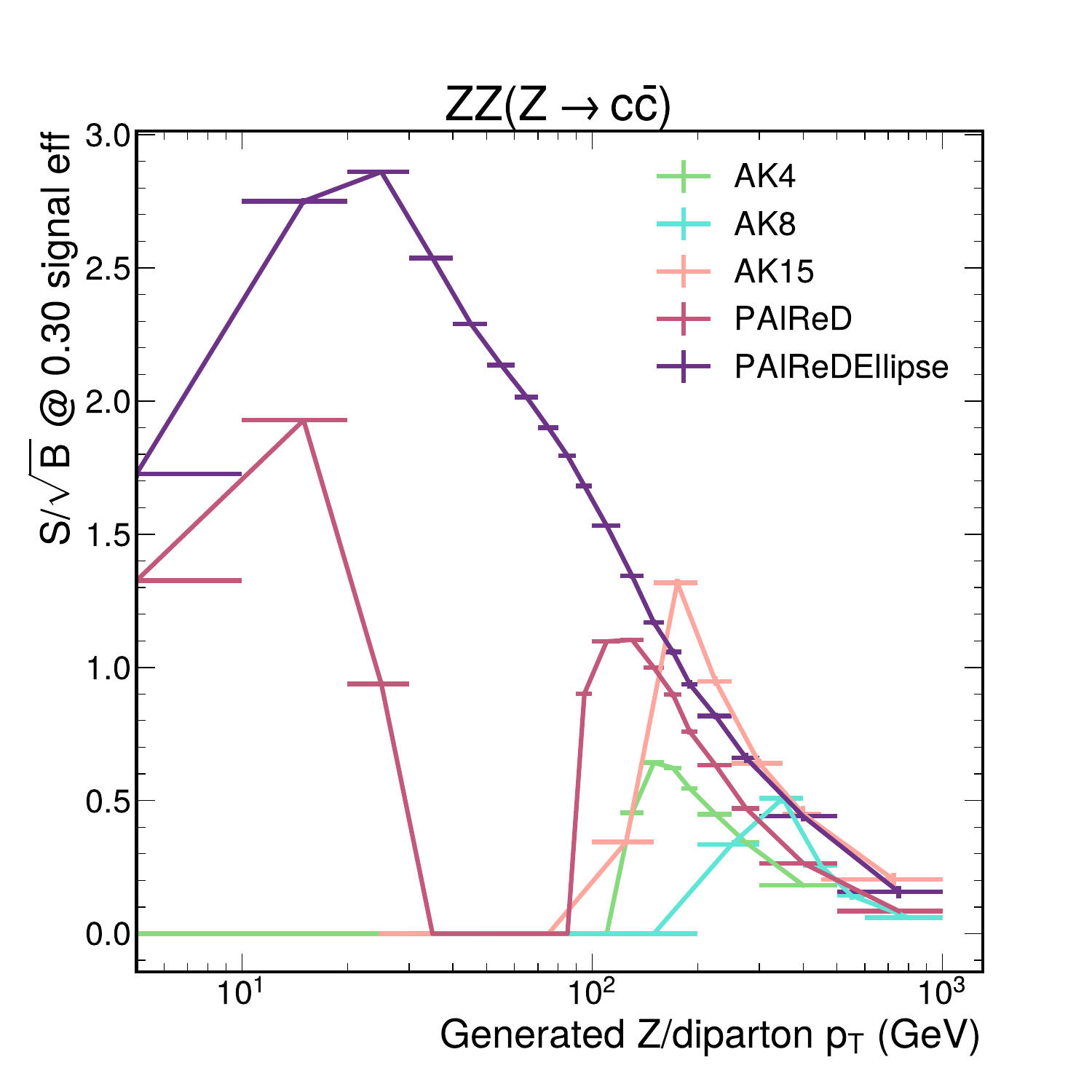}
\par\end{centering}
\begin{centering}
\includegraphics[width=0.5\textwidth]{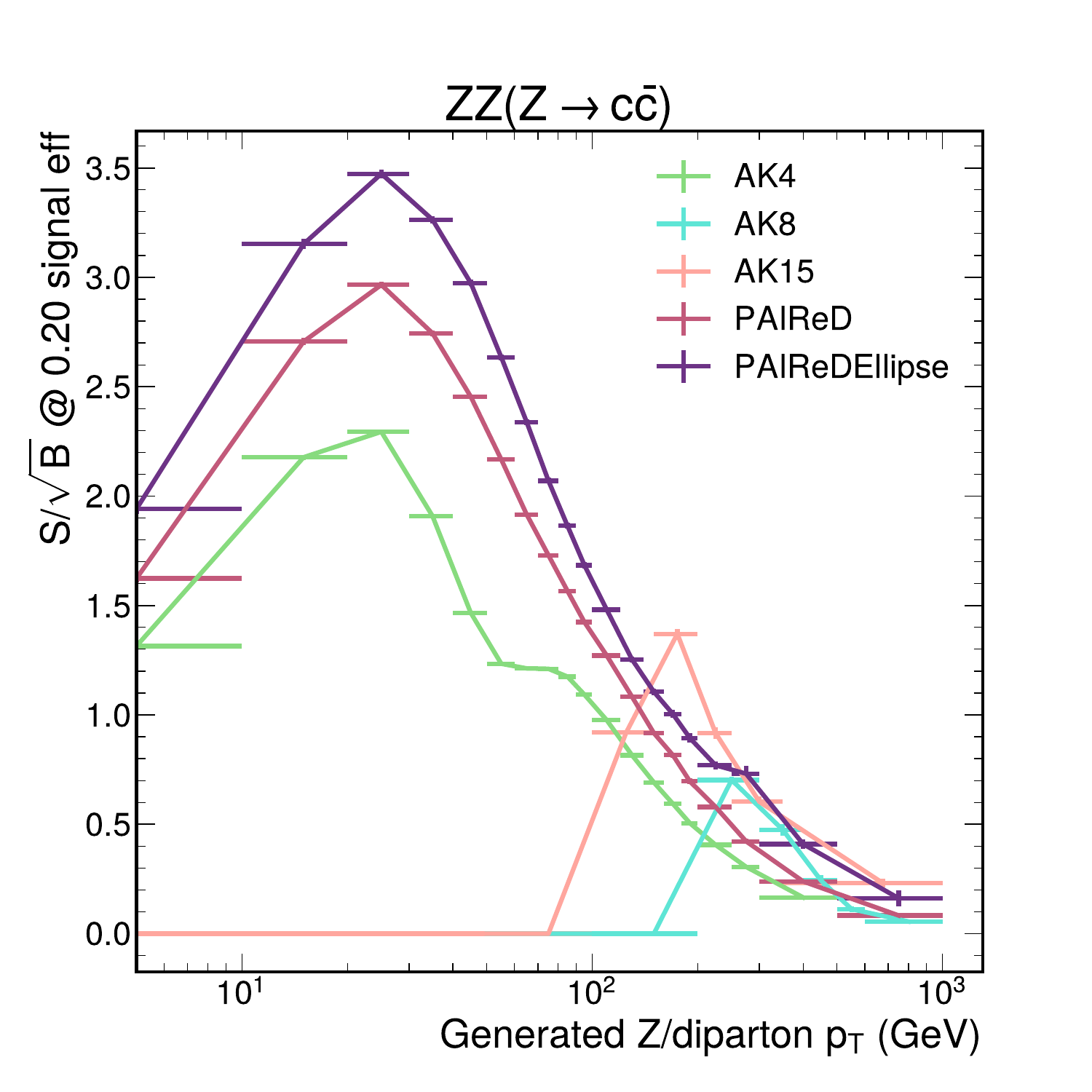}\includegraphics[width=0.5\textwidth]{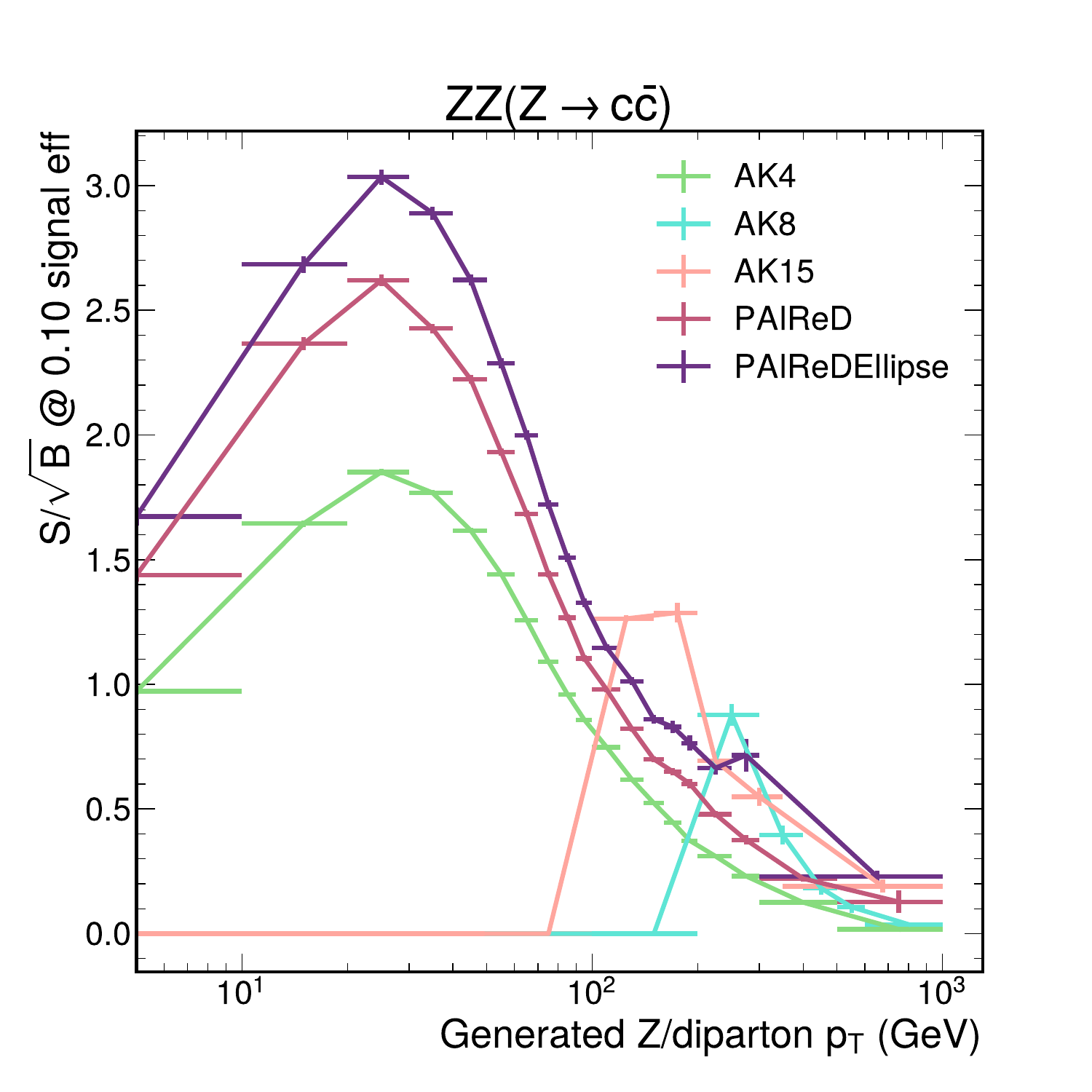}
\par\end{centering}
\caption{\label{fig:SoverB-1}The $\nicefrac{S}{\sqrt{B}}$ values at end-to-end
signal efficiencies of 0.4 (upper-left), 0.3 (upper-right), 0.2 (lower-left),
and 0.1 (lower-right) as a function of the generated Z boson $p_{\text{T}}$
(for signal) or diparton $p_{\text{T}}$ (for background), plotted
for various reconstruction strategies. Signal and background yields
in each bin are normalized to reflect a fixed Z/diparton $p_{\text{T}}$
range of 10 GeV. The target end-to-end signal efficiencies are calculated
per bin. The $\nicefrac{S}{\sqrt{B}}$ value is set to 0 wherever
the target signal efficiency cannot be achieved.}
\end{figure}

Similarly, the existing classifier trainings are used to perform the
ZZ(Z$\rightarrow$$\text{c}\bar{\text{c}}$) analysis. The signal
reconstruction efficiencies and the classifier AUCs as a function
of the hadronic Z boson (or diparton) $p_{\text{T}}$ are shown in
Fig. \ref{fig:MaxSigEff-1}. The background rejection and differential
signal sensitivity estimates ($\nicefrac{S}{\sqrt{B}}$) at different
values of the end-to-end signal efficiencies are shown in Figs. \ref{fig:BkgRej-1}
and \ref{fig:SoverB-1}, respectively. A cross section of 0.155 pb
is used for the ZZ(Z$\rightarrow$$\text{c}\bar{\text{c}}$) process.
This cross section is predicted by \textsc{Pythia8} after \textsc{MLM}
jet matching is performed.

The trends observed in the ZZ(Z$\rightarrow$$\text{c}\bar{\text{c}}$)
analysis closely follow those observed in the ZH(H$\rightarrow$$\text{c}\bar{\text{c}}$)
analysis. The overall signal efficiency is lower in case of the ZZ(Z$\rightarrow$$\text{c}\bar{\text{c}}$)
analysis owing to the regressed mass (or dijet invariant mass) requirement
of $m_{\text{reg}}\in[50,200]$ GeV, tuned for the Higgs boson analysis.
However, the observations highlight the generalizability of the reconstruction
algorithms and classifier trainings to a wider set of final states.

\section{Performance against $\text{t}\bar{\text{t}}$ background\label{sec:Performance-with-ttbar}}

The ``1-lepton'' channel in the VH(H$\rightarrow\text{c}\bar{\text{c}}$)
searches \citep{Aad:2022aa,VHcc} identify WH(H$\rightarrow\text{c}\bar{\text{c}}$)
with the W boson decaying leptonically as the signal process. The
major backgrounds in this channel are the W+jj and semileptonic $\text{t}\bar{\text{t}}$
processes. As far as hadronic reconstruction is concerned, it is expected
that the performance of discriminating WH(H$\rightarrow\text{c}\bar{\text{c}}$)
from W+jj events will be similar to the one for discriminating ZH(H$\rightarrow\text{c}\bar{\text{c}}$)
from Z+jj background (cf. Sec. \ref{sec:Constructing-a-mock}). However,
since $\text{t}\bar{\text{t}}$ has a very different signature compared
to Z+jj, we perform a dedicated study of the performance of the various
tagging strategies with $\text{t}\bar{\text{t}}$ as the background.
We use the same simulated ZH(H$\rightarrow\text{c}\bar{\text{c}}$)
signal as a proxy for the WH(H$\rightarrow\text{c}\bar{\text{c}}$)
process to avoid simulating additional processes. This is a valid
comparison since only the hadronic parts of the processes are relevant.
About 10 million semileptonic $\text{t}\bar{\text{t}}$ events are
simulated following the methods detailed in Sec. \ref{sec:Simulation}.
Of these, 5 million (\emph{test dataset}) events are used as the background
in the following test and the remaining are defined as the \emph{training
dataset}.

The same regression and classification models that were trained using
ZH(H$\rightarrow\text{b}\bar{\text{b}}$), ZH(H$\rightarrow\text{c}\bar{\text{c}}$),
and Z+jj samples are used in this study. The respective models are
evaluated on AK4, PAIReD, and PAIReDEllipse jets reconstructed in
the $\text{t}\bar{\text{t}}$ sample. Similar to Sec. \ref{subsec:Event-level-classification},
event-level BDTs are trained to discriminate ZH(H$\rightarrow\text{c}\bar{\text{c}}$)
events from $\text{t}\bar{\text{t}}$ events for the three different
reconstruction strategies, using the tagger scores, jet kinematics,
and Higgs candidate mass (cf. Table \ref{tab:BDT inputs}) as inputs. 

\begin{figure}
\begin{centering}
\includegraphics[width=0.5\textwidth]{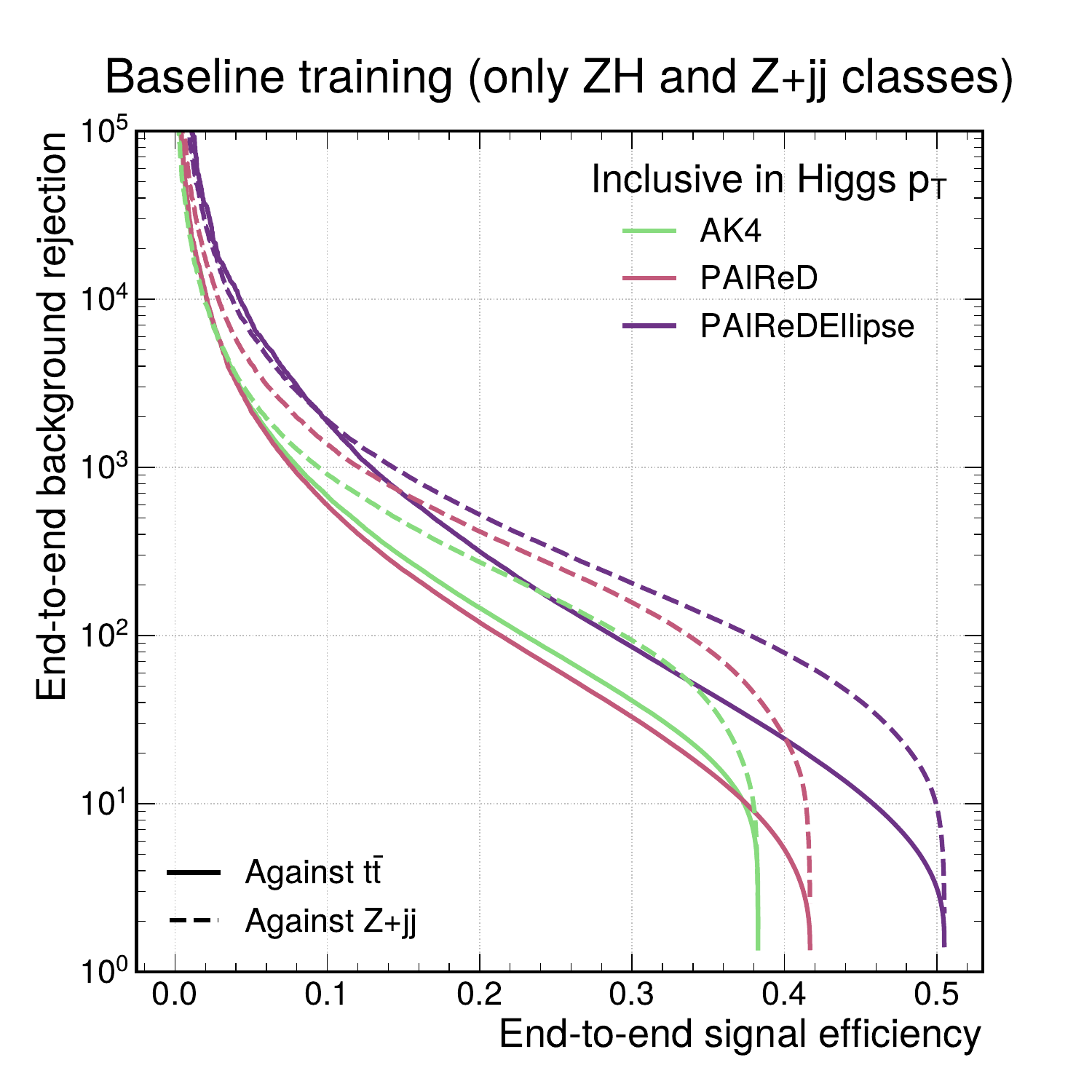}\includegraphics[width=0.5\textwidth]{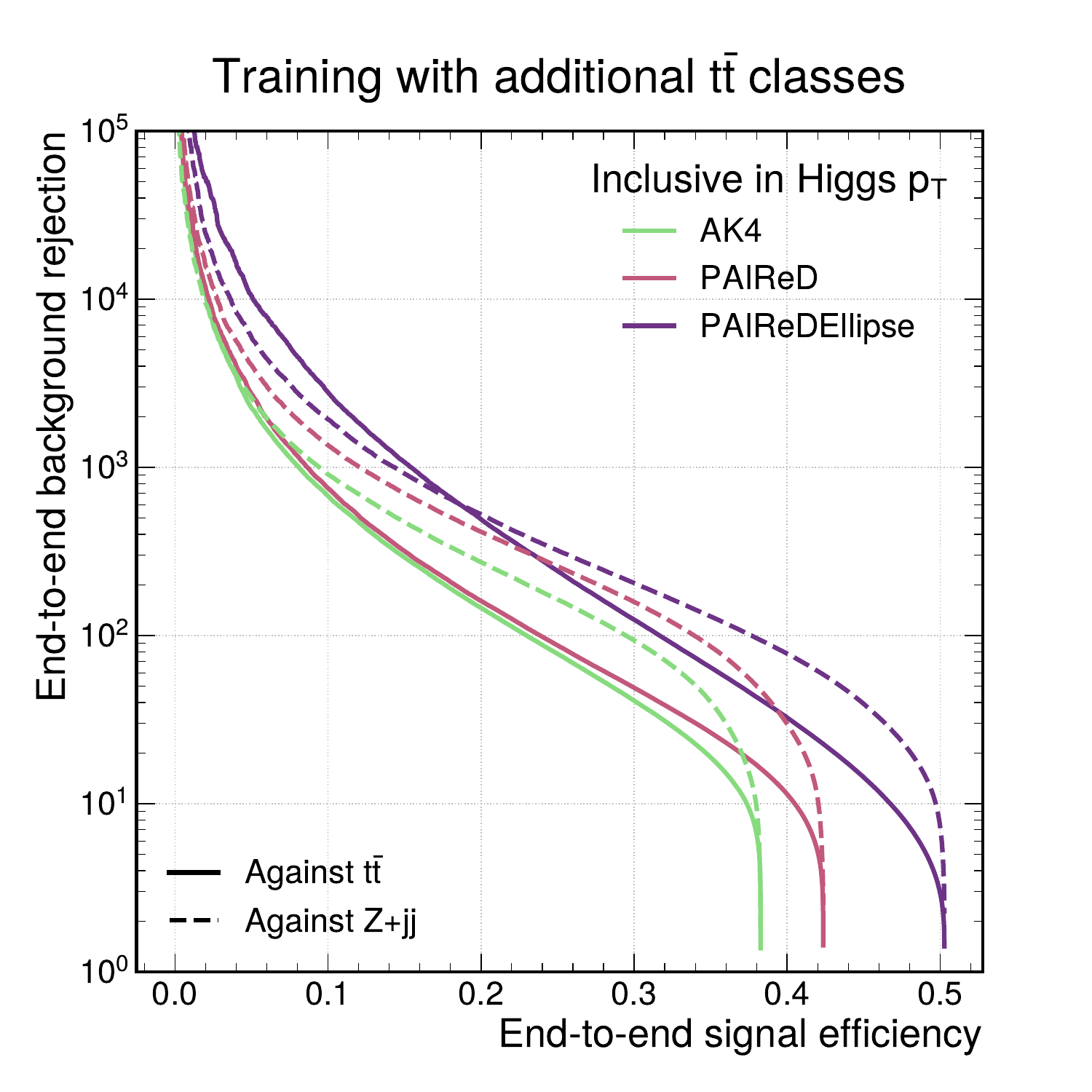}
\par\end{centering}
\caption{\label{fig:ttbar}The end-to-end $\text{t}\bar{\text{t}}$ (the solid
lines) and Z+jj (the dashed lines) background event rejection rates
as a function of the end-to-end signal efficiency, as predicted by
the signal-versus-$\text{t}\bar{\text{t}}$ and signal-versus-Z+jj
event-level BDTs, respectively. The left plot shows the baseline training
with only Z+jj samples represented in the trainings, while the right
plot shows results of PAIReD(Ellipse) trainings performed with additional
$\text{t}\bar{\text{t}}$ classes. The AK4 trainings are identical
in the two plots.}
\end{figure}

The performance of the signal-versus-$\text{t}\bar{\text{t}}$ event-level
BDT is shown in Fig. \ref{fig:ttbar} (left), along with a comparison
with the signal-versus-Z+jj BDT (duplicated from Fig. \ref{fig:BDT}
(upper-left)). It can be seen that the discrimination against $\text{t}\bar{\text{t}}$
is weaker compared to that against Z+jj across all reconstruction
scenarios. This can be attributed to the fact that the $\text{t}\bar{\text{t}}$
event topology is inherently more similar to the signal event topology
than the Z+jj topology is, owing to an abundance of heavy-flavor jets
in the former. The PAIReDEllipse-based BDT achieves higher maximum
signal efficiency compared to the other approaches and achieves a
similar gain (factor of \textasciitilde 2) in background rejection
compared to the AK4 approach at a given signal efficiency. However,
the PAIReD approach falls short of the AK4 approach between signal
efficiencies of \textasciitilde 0.08 to \textasciitilde 0.37, which
was not the case for Z+jj background. It can be noted that the classification
models used here were trained with only Z+jj as the background for
the purpose of demonstrating the strategy. 

To perform a fairer comparison, we train extended models of the PAIReD-
and PAIReDEllipse-based taggers to incorporate additional classes
corresponding to $\text{t}\bar{\text{t}}$ backgrounds. In addition
to the bb, cc, and ll classes in the baseline (3-class) model, we
define bXtt and cXtt classes. PAIReD(Ellipse) jets containing at least
one b quark from a $\text{t}\bar{\text{t}}$ event in their area are
labelled as bXtt. This definition includes non-resonant $\text{b}\bar{\text{b}}$
pairs in $\text{t}\bar{\text{t}}$ events. PAIReD(Ellipse) jets containing
no b quarks but at least 1 c quark in a $\text{t}\bar{\text{t}}$
event are labelled cXtt. All other PAIReD(Ellipse) jets are labelled
ll. New classifier trainings are performed using ZH(H$\rightarrow\text{b}\bar{\text{b}}$),
ZH(H$\rightarrow\text{c}\bar{\text{c}}$), Z+jj, and $\text{t}\bar{\text{t}}$
samples with 5 classes in the output layer. The $\text{t}\bar{\text{t}}$
\emph{training dataset }consisting of 5 million semileptonic $\text{t}\bar{\text{t}}$
events is used for this purpose. The same regression models trained
using ZH(H$\rightarrow\text{b}\bar{\text{b}}$) and ZH(H$\rightarrow\text{c}\bar{\text{c}}$)
samples are used to define the regressed mass of PAIReD(Ellipse) jets. 

Once again, two separate event-level BDTs are trained to discriminate
ZH(H$\rightarrow\text{c}\bar{\text{c}}$) events from Z+jj and $\text{t}\bar{\text{t}}$
backgrounds, respectively, using half of the \emph{test datasets}.
The CCvsLL and CCvsBB inputs to these BDTs are redefined to include
the cXtt and bXtt scores in the denominators, respectively, using
predictions from the taggers with extended classes (5-class models).
The performances of the retrained signal-versus-$\text{t}\bar{\text{t}}$
and signal-versus-Z+jj BDTs are evaluated on the remaining half of
the \emph{test dataset} and are compared with the results from the
baseline AK4-based approach. The comparison is shown in Fig. \ref{fig:ttbar}
(right). It can be seen that the PAIReD approach achieves a better
performance in rejecting $\text{t}\bar{\text{t}}$ compared to the
AK4 approach when the 5-class model is used. The PAIReDEllipse approach
exhibits an even stronger $\text{t}\bar{\text{t}}$ rejection compared
to the corresponding 3-class model. The rejection rates of Z+jj backgrounds
are altered only marginally. Therefore, extending the trainings to
include jets from various topologies and physics processes appears
to improve the generalization capabilities of the taggers. Future
studies and experimental implementations may consider including jets
from $\text{t}\bar{\text{t}}$, QCD multijet, and other similar processes
as background jets in the training. 

\section{Effects of clustering algorithms\label{sec:Effects-of-(not)}}

The PAIReD and PAIReDEllipse jets used throughout this paper are constructed
without using any sequential clustering algorithms. Their areas are
defined using the centers of clustered jets (in this case, AK4 jets)
as seeds, and all reconstructed particles lying in the defined area
are considered constituents of the jet. This is akin to non-sequential
cone algorithms, and in contrast to the commonly-used sequential jet
clustering algorithms like the AK and Cambridge-Aachen \citep{Dokshitzer:1997in,Wobisch:1998wt}
algorithms in which particles are added to the jet one at a time starting
from a seed.

Jet clustering algorithms like AK offer some advantages such as IRC
safety \citep{Banfi:2004yd}, and less sensitivity to pileup and unrelated
hadronic activities. Infrared safety means that the presence of soft
radiation does not affect the results of the algorithm. Collinear
safety means that the results of the algorithm remain unaffected after
introducing a collinear splitting of one of the inputs. These advantages
are lacking when we define PAIReD and PAIReDEllipse jets. On the other
hand, previous studies \citep{Lai:2021ckt,Lu:2022cxg,Athanasakos:2023fhq}
have shown that classifiers leveraging IRC-unsafe information outperform
classifiers with only IRC-safe inputs in a variety of jet classification
tasks. It has been hypothesized that the gain comes from the facts
that IRC-unsafe classifiers leverage very-soft radiations from jets,
which IRC-safe approaches typically do not have access to, and IRC-unsafe
classifiers take exact position of the constituents as inputs, as
opposed to IRC-safe approaches taking only relative positions as inputs
\citep{Athanasakos:2023fhq}. The presence of pileup and UE, however,
mitigates the gains \citep{Lai:2021ckt}. In the following subsections,
we investigate this phenomenon in the context of PAIReD jets and AK15
jets in the presence of pileup.

\subsection{PAIReD jets with AK-clustered particles only\label{subsec:PAIReD-jets-with}}

In order to understand the extent of the performance improvement gained
by using unclustered particles, we define a new variant of the PAIReD
jet approach containing only the reconstructed particles that are
clustered inside the seed AK4 jets. This variant, referred to as PAIReDClustered
henceforth, does not contain unclustered particles that lie within
an angular distance of $\Delta R=0.4$ around the centers of the seed
AK4 jets like PAIReD does, and hence may be considered as an ablated
variant of the baseline PAIReD jet defined in this paper. The information
available to each tagging strategy is summarized in Table \ref{tab:Clustered}.

\begin{table}
\begin{centering}
\setlength\extrarowheight{8pt}
\begin{tabular}{>{\raggedright}m{0.4\textwidth}>{\centering}m{0.12\textwidth}>{\centering}m{0.12\textwidth}>{\centering}m{0.12\textwidth}>{\centering}m{0.12\textwidth}}
\hline 
 & AK4 & PAIReD\-Clustered & PAIReD & PAIReD\-Ellipse\tabularnewline
\hline 
Tagging strategy & Individual jet & Jet pairs & Jet pairs & Jet pairs\tabularnewline
Low-level features of constituents of a thin-radius jet & $\checkmark$ & $\checkmark$ & $\checkmark$ & $\checkmark$\tabularnewline
Correlations between constituents of two different jets &  & $\checkmark$ & $\checkmark$ & $\checkmark$\tabularnewline
Soft radiations/unclustered particles around seed jet centers &  &  & $\checkmark$ & $\checkmark$\tabularnewline
Unclustered particles lying spatially between two seed jets &  &  &  & $\checkmark$\tabularnewline
\hline 
\end{tabular}
\par\end{centering}
\caption{\label{tab:Clustered}Summary of the information available to the
various tagging approaches. }
\end{table}

\begin{figure}
\begin{centering}
\includegraphics[width=0.65\textwidth]{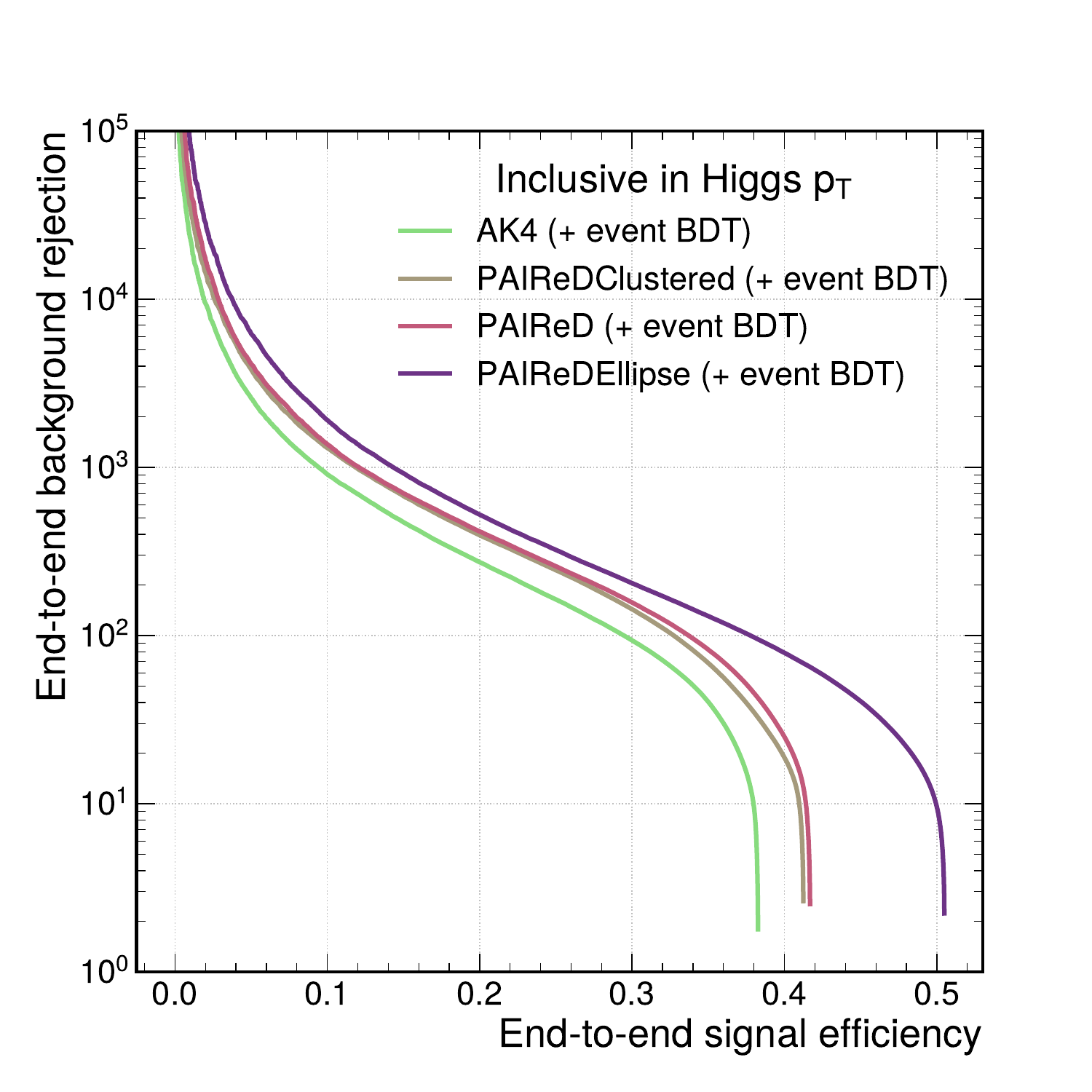}
\par\end{centering}
\caption{\label{fig:Clustered}The end-to-end background event rejection rate
as a function of the end-to-end signal efficiency inclusively in Higgs
boson (or diparton) $p_{\text{T}}$ as predicted by event-level BDTs.}
\end{figure}

Dedicated mass regression and classification networks are trained
for PAIReDClustered jets with the same samples, labels, and techniques
described in Secs. \ref{sec:Simulation}--\ref{sec:Neural-network-training}.
An event level BDT to discriminate ZH(H$\rightarrow\text{c}\bar{\text{c}}$)
signal events from Z+jj backgrounds is trained following the prescription
discussed in Sec. \ref{subsec:Event-level-classification} and using
the same input variables corresponding to the PAIReD jet column in
Table \ref{tab:BDT inputs}. The end-to-end background event rejection
rate is plotted as a function of the end-to-end signal efficiency
for the AK4 approach and the three variants of the PAIReD approach,
as is shown in Fig. \ref{fig:Clustered}. The rates are also summarized
in Table \ref{tab:Clustered-1} for three different signal efficiencies.

\begin{table}
\begin{centering}
\setlength\extrarowheight{8pt}
\begin{tabular}{>{\centering}m{0.15\textwidth}>{\centering}m{0.12\textwidth}>{\centering}m{0.12\textwidth}>{\centering}m{0.12\textwidth}>{\centering}m{0.12\textwidth}}
\hline 
Signal efficiency & AK4 & PAIReD\-Clustered & PAIReD & PAIReD\-Ellipse\tabularnewline
\hline 
10 & 908.7 & 1305 & 1373 & 1910\tabularnewline
20 & 273.0 & 394.9 & 415.7 & 523.8\tabularnewline
30 & 93.58 & 143.4 & 157.3 & 205.0\tabularnewline
\hline 
\end{tabular}
\par\end{centering}
\caption{\label{tab:Clustered-1}End-to-end background event rejection rates
for different values of end-to-end signal efficiencies (first column)
for various algorithms, evaluated with an event-level BDT, inclusively
in Higgs boson (or diparton) $p_{\text{T}}$.}
\end{table}

It can be seen that the performance of the PAIReDClustered tagging
approach is marginally worse than that of the PAIReD approach, but
significantly better than that of the AK4 approach. On the one hand,
this shows that the improvement in the PAIReD approach over traditional
algorithms arises almost exclusively from leveraging correlations
between low-level features of two different jets arising from the
same heavy particle. An additional, small improvement is gained from
using soft, unclustered particles around the seed jet centers; this
observation is in agreement with those in Refs. \citep{Lai:2021ckt,Lu:2022cxg,Athanasakos:2023fhq}.
On the other hand, this also shows that it is possible to use an alternative,
ablated variant of the PAIReD approach and still improve over traditional
approaches, in case the PS, UE, and pileup---and hence the unclustered
particles used by the baseline PAIReD variant---cannot be modeled
correctly in simulations.

\subsection{AK15 and Cone15 jets\label{subsec:AK15-and-Cone15}}

To examine the effect of unclustered particles in tagging large-radius
jets, we additionally define a cone-based jet (Cone15) with radius
1.5 units, for every AK15 jet reconstructed in the simulated events.
The axis of every Cone15 jet is defined to be at the center of the
corresponding AK15 jet and all particles in a radius of $\Delta R=1.5$
units around the jet axis are considered its constituents. This makes
the Cone15 jets geometrically similar to the AK15 jets with the difference
that the former does not use the AK algorithm, except for defining
the jet center. The regression and classification trainings (Sec.
\ref{sec:Neural-network-training}) and the mock ZH(H$\rightarrow\text{c}\bar{\text{c}}$)
analysis (Sec. \ref{sec:Constructing-a-mock}) are repeated replacing
AK15 jets with Cone15 jets.

\begin{figure}
\begin{centering}
\includegraphics[width=0.5\textwidth]{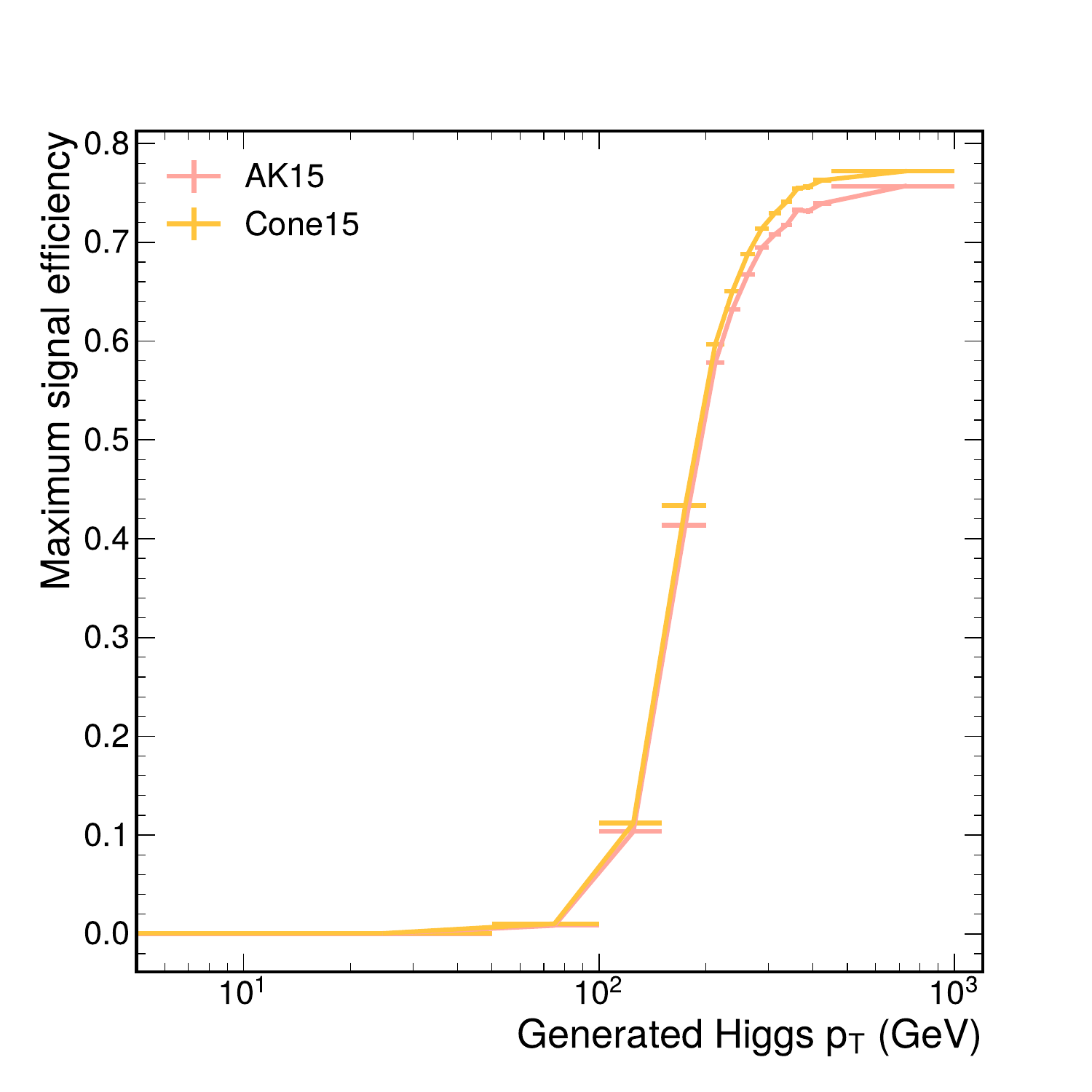}\includegraphics[width=0.5\textwidth]{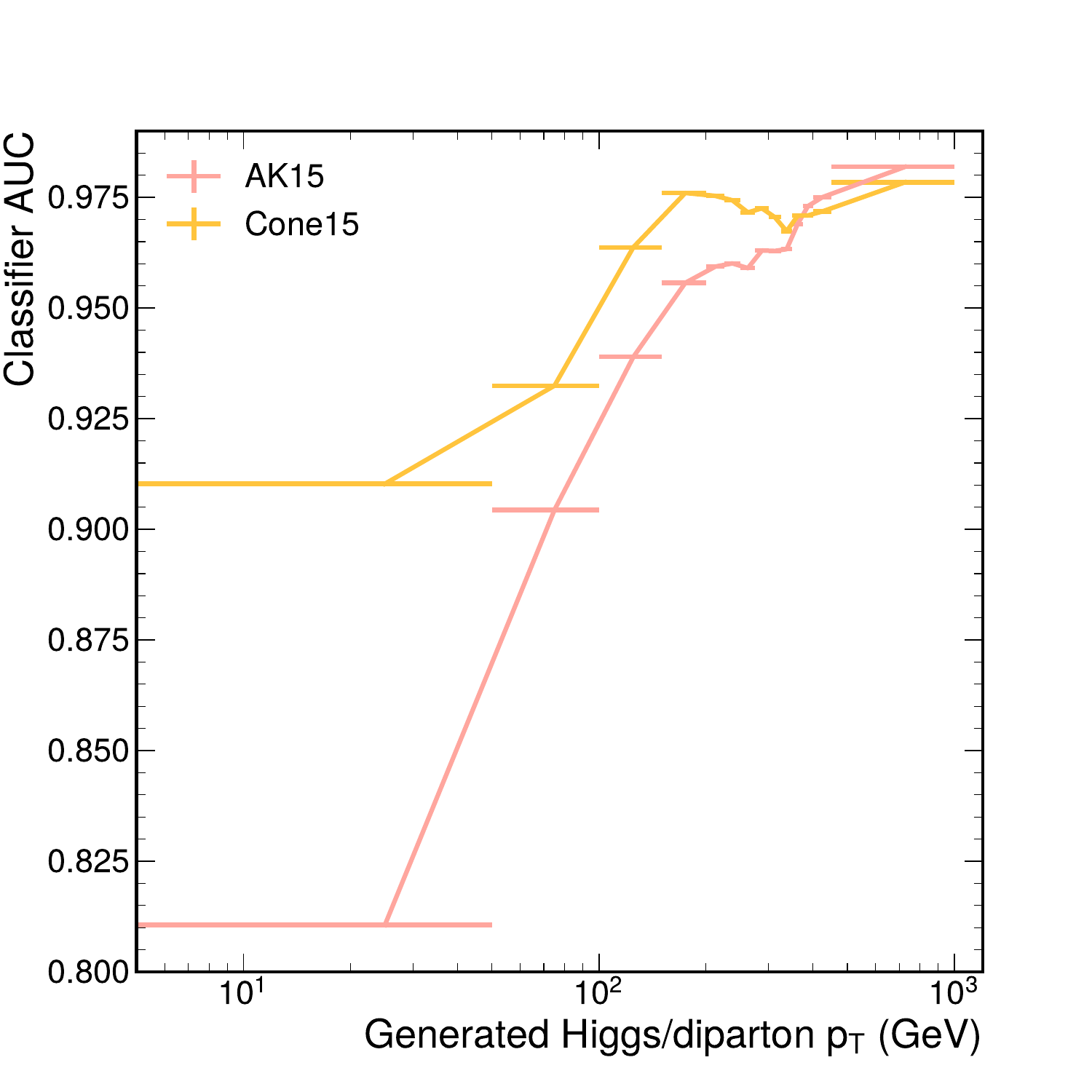}
\par\end{centering}
\begin{centering}
\includegraphics[width=0.5\textwidth]{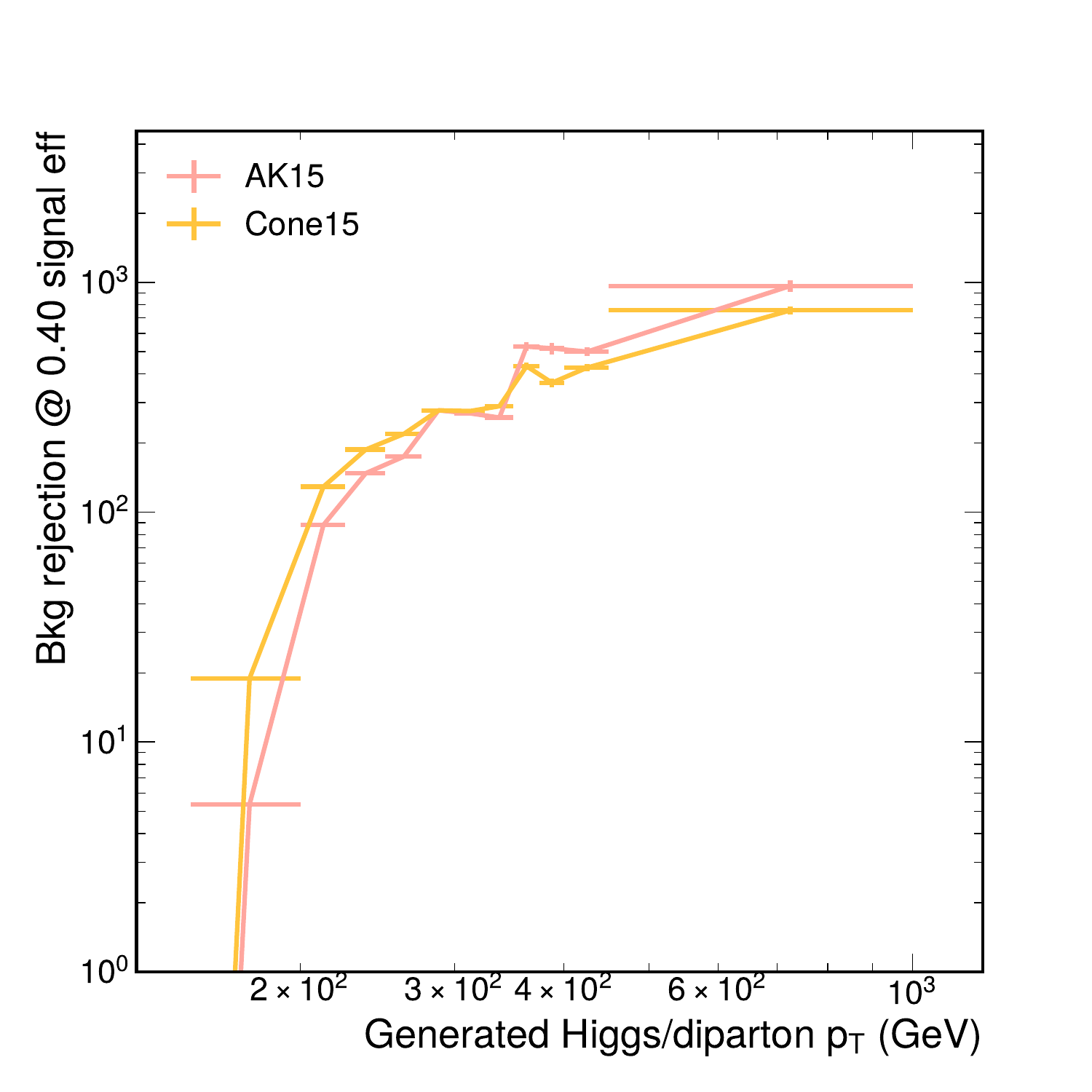}\includegraphics[width=0.5\textwidth]{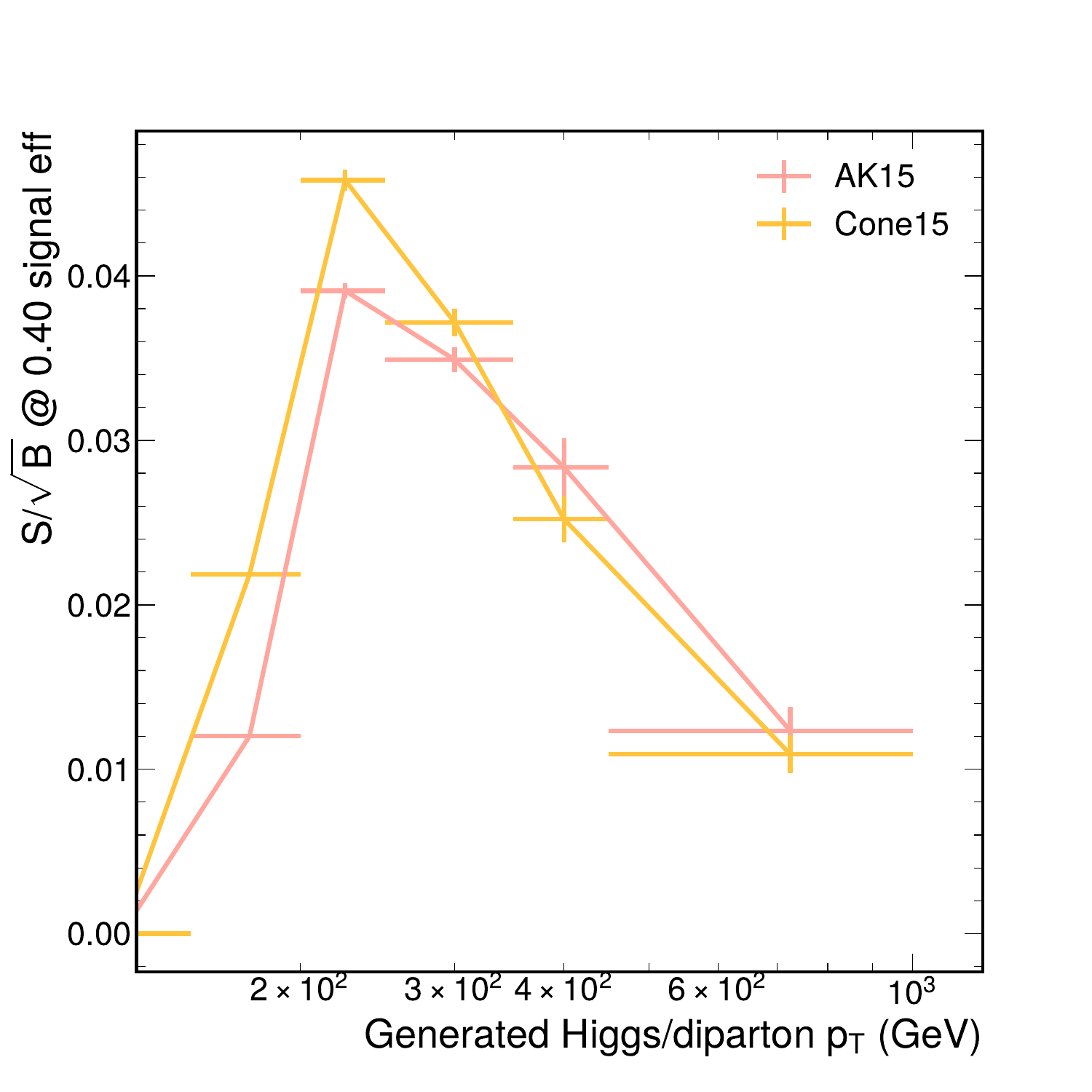}
\par\end{centering}
\caption{\label{fig:Cone15Class}The maximum signal efficiency (upper-left),
the classifier AUCs (upper-right), and the end-to-end background rejection
rate (lower-left) and signal sensitivity (lower-right) at an end-to-end
signal efficiency of 0.4 are plotted as a function of Higgs boson
$p_{\text{T}}$ (or diparton $p_{\text{T}}$) comparing the AK15-
and Cone15-based approaches.}
\end{figure}

The maximum signal efficiencies, classifier AUCs, end-to-end background
rejections, and signal sensitivity estimates in the context of a ZH(H$\rightarrow\text{c}\bar{\text{c}}$)
analysis are shown in Fig. \ref{fig:Cone15Class}. The Cone15 and
AK15 have similar signal reconstruction efficiencies, except for at
high Higgs boson boosts where the Cone15 approach reconstructs about
2\% more signal events. The Cone15 approach selects the correct jet
as the Higgs boson candidate in a slightly higher number of events,
resulting in this observed gain in signal reconstruction efficiency.
However, this gain is offset by the classifier performance in the
next step, where the AK15-based classifier performs better only at
high boosts ($p_{\text{T}}\apprge$ 350 GeV) of the Higgs boson. The
Cone15-based classifier has a significantly better classification
performance at lower boosts. The combined effect can be seen in the
end-to-end background rejection and signal sensitivity plots, where
the two approaches show similar performances, with the Cone15 performing
marginally better at low boosts ($\apprle$ 300 GeV), while the AK15-based
approach performs better at higher Higgs boson $p_{\text{T}}$'s ($\apprge$
350 GeV).

The improvement in the classification performance at low Higgs boson
boosts using the Cone15 approach suggests that the presence of particles
(mostly soft radiations) in addition to the ones clustered by the
AK algorithm within the jet radius helps improve the classifier's
accuracy at higher angular separation of the partons. However, the
difference in the final end-to-end background rejection is marginal,
as few AK15 jets are reconstructed at low boosts. On the other hand,
a similar performance between the two approaches at higher boosts
suggests that a sequential clustering algorithm may be more crucial
at low angular separation of the partons, where the hadronization
products of the two partons have a larger overlap.

The study in this subsection is limited to only one example and the
conclusions drawn here are empirical. A detailed description and a
theoretical analysis of the synergistic effects of various jet clustering
methods and attention-based deep learning networks are beyond the
scope of this paper. Nevertheless, the performance of the PAIReDEllipse
strategy may potentially be improved at high boosts in the future
by utilizing advanced jet grooming and pileup mitigation techniques
\citep{Bertolini:2014bba,Komiske:2017ubm,Hansen:2018osj,ArjonaMartinez:2018eah,Alipour-fard:2023yjz},
and potentially at all boosts by replacing the AK4 seeds with jets
constructed with newer algorithms, such as XCone \citep{Stewart:2015aa,Thaler:2015aa},
dynamic-radius algorithms \citep{refId0,Mukhopadhyaya:2023aa}, SIFT
\citep{larkoski2023jet}, or supervised jet clustering \citep{Ju:2020tbo}.
These comparisons could be investigated in future studies.
\end{document}